\documentclass[trackchanges,twocolumn]{aastex7}
\usepackage{comment}
\usepackage{booktabs}
\usepackage{multirow}
\usepackage{amsmath}
\usepackage[normalem]{ulem}
\usepackage[export]{adjustbox} 
\newcommand{\hco}{HCO$^+$}
\newcommand{\co}{$^{12}$CO}
\newcommand{\ico}{$^{13}$CO}

\newcommand{\kms}{km\,s$^{-1}$}

\shortauthors{W. Yang et al.}
\submitjournal{ApJ}

\begin{document}
\title{Molecular signatures of shocked gas and enhanced cosmic-ray ionization in supernova remnant W49B revealed by ALMA}

\author[orcid=0000-0002-3599-6608,sname='Yang']{Wenjin Yang} 
\affiliation{School of Astronomy \& Space Science, Nanjing University, 163 Xianlin Avenue, Nanjing 210023, People's Republic of China}
\affiliation{Key Laboratory of Modern Astronomy and Astrophysics, Nanjing University, Ministry of Education, Nanjing 210023, People's Republic of China}
\email[show]{wjyang@nju.edu.cn}

\author[0000-0002-5683-822X,sname='Zhou']{Ping Zhou} 
\affiliation{School of Astronomy \& Space Science, Nanjing University, 163 Xianlin Avenue, Nanjing 210023, People's Republic of China}
\affiliation{Key Laboratory of Modern Astronomy and Astrophysics, Nanjing University, Ministry of Education, Nanjing 210023, People's Republic of China}
\email[show]{pingzhou@nju.edu.cn}

\author[0000-0002-7299-2876,sname='Zhang']{Zhi-Yu Zhang} 
\affiliation{School of Astronomy \& Space Science, Nanjing University, 163 Xianlin Avenue, Nanjing 210023, People's Republic of China}
\affiliation{Key Laboratory of Modern Astronomy and Astrophysics, Nanjing University, Ministry of Education, Nanjing 210023, People's Republic of China}
\email{zzhang@nju.edu.cn}

\author[0000-0002-0747-0078,sname='Zhang']{Gao-yuan Zhang} 
\affiliation{Research Center for Astronomical Computing, Zhejiang Laboratory, Hangzhou, People's Republic of China}
\email{gaoyuan.zhang@zhejianglab.org}

\author[0000-0002-2755-1879,sname='Koo']{Bon-Chul Koo} 
\affiliation{Department of Physics and Astronomy, Seoul National University, Seoul 08826, Republic of Korea}
\email{bckoo@snu.ac.kr}

\author[0000-0002-3866-414X,sname='Gong']{Yan Gong} 
\affiliation{Purple Mountain Observatory and Key Laboratory of Radio Astronomy, Chinese Academy of Sciences, 10 Yuanhua Road, Nanjing, 210033, People's Republic of China}
\email{ygong@pmo.ac.cn}

\author[0000-0003-1576-0961,sname='Liu']{Ruo-Yu Liu} 
\affiliation{School of Astronomy \& Space Science, Nanjing University, 163 Xianlin Avenue, Nanjing 210023, People's Republic of China}
\affiliation{Key Laboratory of Modern Astronomy and Astrophysics, Nanjing University, Ministry of Education, Nanjing 210023, People's Republic of China}
\email{ryliu@nju.edu.cn}

\author[0000-0003-2062-5692,sname='Sano']{Hidetoshi Sano} 
\affiliation{Faculty of Engineering, Gifu University, 1-1 Yanagido, Gifu 501-1193, Japan}
\affiliation{Department of Intelligence Science and Engineering, Graduate School of Natural Science and Technology, Gifu University, 1-1 Yanagido, Gifu, 501-1193, Japan}
\affiliation{Center for Space Research and Utilization Promotion (c-SRUP), Gifu University, 1-1 Yanagido, Gifu 501-1193, Japan}
\email{sano.hidetoshi.w4@f.gifu-u.ac.jp}

\author[0000-0002-9776-5610,sname='Tu']{Tian-Yu Tu} 
\affiliation{School of Astronomy \& Space Science, Nanjing University, 163 Xianlin Avenue, Nanjing 210023, People's Republic of China}
\affiliation{Key Laboratory of Modern Astronomy and Astrophysics, Nanjing University, Ministry of Education, Nanjing 210023, People's Republic of China}
\email{tianyu.tu1105@outlook.com}

\author[0000-0002-4753-2798,sname='Chen']{Yang Chen} 
\affiliation{School of Astronomy \& Space Science, Nanjing University, 163 Xianlin Avenue, Nanjing 210023, People's Republic of China}
\affiliation{Key Laboratory of Modern Astronomy and Astrophysics, Nanjing University, Ministry of Education, Nanjing 210023, People's Republic of China}
\email{ygchen@nju.edu.cn}

\author[0000-0002-6488-8227,sname='Le']{Le Ngoc Tram} 
\affiliation{Leiden Observatory, Leiden University, PO Box 9513, 2300, RA, Leiden, The Netherlands}
\email{nle@strw.leidenuniv.nl}

\author[0000-0001-8926-2215,sname='Lin']{Liang-hao Lin} 
\affiliation{Max-Planck-Institut f{\"u}r Radioastronomie, Auf dem H{\"u}gel 69, D-53121 Bonn, Germany}
\email{lhlin@mpifr-bonn.mpg.de}

\author[sname='Wyrowski']{Friedrich Wyrowski} 
\affiliation{Max-Planck-Institut f{\"u}r Radioastronomie, Auf dem H{\"u}gel 69, D-53121 Bonn, Germany}
\email{wyrowski@mpifr-bonn.mpg.de}


\begin{abstract}
Galactic supernova remnants (SNRs) interacting with their surrounding molecular clouds (MCs) provide important laboratories for investigating how supernova feedback is transferred to the interstellar medium. 
We performed observations of the \hco~$J$=1--0, 3--2, and 4--3 lines, as well as the \co~$J$=3--2 line, toward the eastern and western regions of SNR W49B using the Atacama Large Millimeter/submillimeter Array (ALMA) and the Atacama Pathfinder Experiment (APEX) telescope. 
In W49B west, we identify twelve regions with broad-line profiles in both \hco\ and \co\ emission, with FWHM of 25--60~\kms, providing clear evidence of SNR–MC interaction. These emissions are compact (at $\sim$5\arcsec\ resolution) and spatially coincident with 2.12~$\mu$m H$_2$ emission, but generally offset from the 1.64~$\mu$m [Fe~II] and radio continuum peaks.
From non-LTE analyses, we derive that the shocked gas has densities of (0.5--1.6) $\times$ 10$^{5}$~cm$^{-3}$, and high $N_{\rm HCO^+}$/$N_{\rm CO}$ abundance ratios of ($4.2$--$14.5) \times10^{-4}$ compared with those reported in other SNRs. 
Reproducing such high abundance ratios in shocked MCs requires high cosmic ray ionization rates (CRIRs) of a few 10$^{-14}$--10$^{-13}$~s$^{-1}$, about 3--4 orders of magnitude above the standard Galactic level. 
Some regions exhibit enhanced HCO$^+$/CO intensity ratios in their high-velocity line wings, possibly due to multiple unresolved shocks spanning a range of velocities and gas densities, spatial variations in the CRIR, or both. By contrast, the ratio remains nearly constant with velocity in other regions. For these regions, we further explore the propagation of low-energy CRs through the post-shock layer. 
A subsequent energy-budget analysis of a representative shocked clump indicates that the shock-associated kinetic-energy reservoir substantially exceeds the estimated cumulative CR and X-ray heating inputs, with CR heating exceeding X-ray heating.

\end{abstract}


\section{Introduction}

Supernovae (SNe) are among the most energetic events in galaxies, playing a key role in regulating star formation and galaxy evolution \citep{2018MNRAS.473.4077P}. 
However, observational constraints on SN feedback remain challenging, and one of the reasons is the uncertainty in the specific roles of different feedback mechanisms, including shocks, radiation, and cosmic rays (CRs), in shaping the surrounding molecular clouds (MCs) \citep{2017ARA&A..55...59N}. Among these feedback channels, shocks have been the most widely studied heating mechanism in supernova remnants (SNRs),  with their properties depending on both the evolutionary stage of the remnant and the local ambient conditions \citep{2026ApJ..1004..117U}, while X-ray radiation and low-energy ($<$1~GeV) CRs can penetrate into MCs shielded from UV radiation and play important roles in the heating, ionization and chemistry \citep[e.g.,][]{1978ApJ...222..881G, padovani09, gabici22,2022ARA&A..60..247W}.
Notably, about 1/4 of Galactic SNRs are known to interact with MCs \citep{2010ApJ...712.1147J, 2014IAUS..296..170C}, and tens of SNRs emit X-ray and $\gamma$-ray radiation. In such environments, MCs can efficiently slow down SNR shocks, absorb radiation, and are affected by CR propagation, leading to significant variations in their physical and chemical conditions.
Thus, Galactic SNRs and their molecular environments (i.e., SNR-MC associations) are prime targets for quantifying SN feedback through multi-wavelength observations \citep{2020pesr.book.....V}.

W49B (G43.3$-$0.2) is the most luminous Galactic SNR in the X-ray band \citep{2005ApJ...632L..99I}, and remarkably bright in the $\gamma$-ray band \citep{2010ApJ...722.1303A,2018A&A...612A...5H}, with the latter originating from $\pi^0$ decay due to interactions between accelerated high-energy CRs and nearby dense materials. 
It is also the most luminous Galactic SNR in near-IR 2.12~$\mu$m H$_2$ emission \citep{2007ApJ...654..938K, 2019AJ....157..123L}.
These properties indicate that W49B is still undergoing vigorous energy release into its surroundings, making it an excellent target for studying the different channels of SN feedback.
 With an estimated age of $\sim$5--6 kyr, W49B is evolving in a complex and inhomogeneous environment where shocks of different velocities coexist. X-ray observations suggest a fast forward shock \citep[$\sim$ 500~\kms,][]{2018A&A...615A.150Z} propagating through relatively diffuse gas with densities of a few to a few tens of cm$^{-3}$, whereas molecular observations reveal interactions with denser gas, into which much slower shocks are driven  \citep[e.g.,][]{2021ApJ...919..123S,2022ApJ...931..144Z}.


Based on IRAM 30~m observations, \cite{2022ApJ...931..144Z} reported unusually high \hco/\co\,$J$=1--0 intensity ratios (0.2--1) in the shocked gas toward the southwestern region of W49B (hereafter W49B west), and in quiescent clouds located outside the southeastern boundary of the SNR (hereafter W49B east).
These ratios are 1--2 orders of magnitude larger than those found in typical MCs. 
Interestingly, in both regions where the \hco/\co\,$J$=1--0 ratios are enhanced, the spectral line profiles show markedly different behaviors: the HCO$^+$ $J$=1--0 line profile in W49B east is narrow, whereas in W49B west it is significantly broadened. This may suggest that this SNR influences the surrounding molecular clouds through different mechanisms in the two regions.
In addition, the complex velocity structure of molecular gas along the line of sight toward W49B has complicated the identification of the clouds physically associated with the remnant, and hence its distance \citep[e.g.,][]{2014ApJ...793...95Z,2021ApJ...919..123S,2022ApJ...931..144Z}.
In this work, we adopt a systemic local-standard-of-rest (LSR) velocity of 61--65~\kms\ and a distance of 7.9$\pm$0.6~kpc for W49B \citep{2022ApJ...931..144Z}, based on the velocity range of the shocked molecular gas and the approximately constant $\sim$64~\kms\ velocity of the 2.12~$\mu$m H$_2$ emission across the SNR \citep{2020AJ....160..263L}.

The strongly enhanced HCO$^+$/CO ratios near W49B could be explained by additional ionization from low-energy CRs and X-ray radiation from the SNR \citep{1996ApJ...466..561M,2014A&A...568A..50V}, which enhance the abundance of H$^+_3$ via reaction chains with H$_2$ molecules. 
HCO$^+$ molecules are primarily formed via the gas-phase ion-neutral reaction H$^+_3$ + CO $\to$ HCO$^+$ + H$_2$. 
Moreover, 
CRs and intense X-ray irradiation can modify the CO chemistry by promoting CO dissociation and ionization, thereby reducing the CO abundances near SNRs
\citep[e.g.,][]{2015ApJ...803...37B,2017ApJ...839...90B,2019MNRAS.486.1094M}. 
Therefore, variations in the HCO$^+$/CO abundance ratio provide a useful diagnostic of SNR-driven ionization feedback, consistent with predictions from state-of-the-art astrochemical simulations \citep[e.g.,][]{2023MNRAS.519..729B}.

However, single-dish observations \citep{2022ApJ...931..144Z} did not resolve the HCO$^+$ and CO emission, limiting a detailed multi-wavelength comparison with the infrared and X-ray images on small scales. In addition, the excitation conditions of HCO$^{+}$ and CO gas near W49B, including temperature, density, and their column densities, are poorly constrained, as only a single transition $J$=1--0 has been mapped.
To overcome these limitations, we present new Atacama Large Millimeter/submillimeter Array (ALMA) 12~m array and Atacama Compact Array (ACA) observations, targeting three HCO$^{+}$ lines and one \co\ line toward W49B east and west. 
These data allow us to resolve the molecular emission spatially and to perform detailed multi-wavelength comparisons, thereby probing the multiphase gas structure and the variations in the chemical and physical properties of molecular clouds at different distances from the energy sources.

The observations are described in Sect.~\ref{Sec:obs}, and the results are presented in Sect.~\ref{Sec:results}. We discuss our interpretation in Sect.~\ref{Sec:discuss}, and provide a summary in Sect.~\ref{Sec:sum}.



\begin{figure}[htbp]
\center
\includegraphics[width=0.5\textwidth]{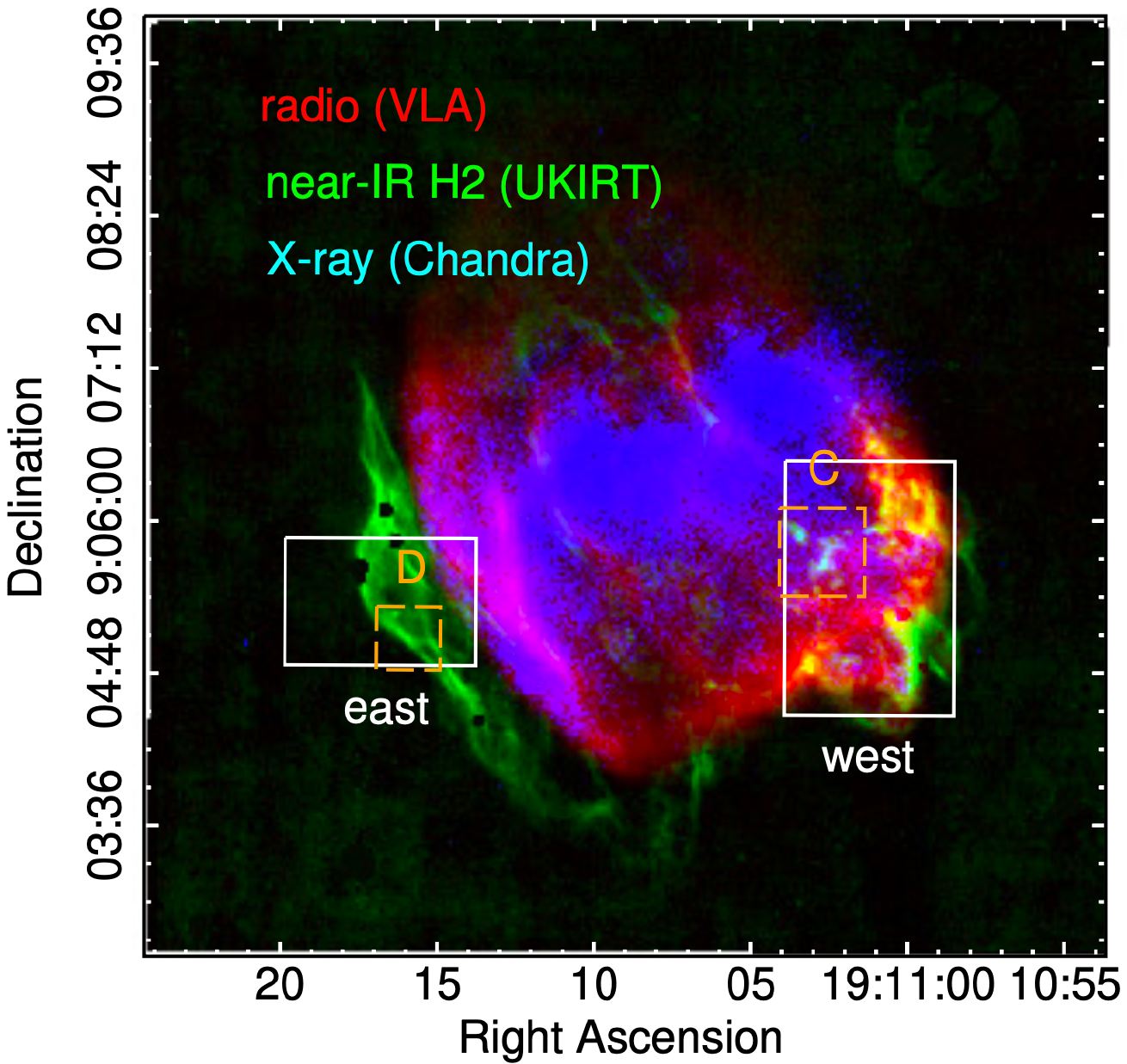}
\caption{Multi-wavelength composite image of SNR W49B, combining 327 MHz radio continuum emission (red, \citealt{2001ApJ...559..954L}), 2.12~$\mu$m H$_2$ emission (green, \citealt{2019AJ....157..123L}), and 0.3--10 keV X-ray emission (blue, \citealt{2018A&A...615A.150Z}). The orange dashed boxes (regions C and D) mark the areas reported by \cite{2022ApJ...931..144Z} as exhibiting the most prominent high \hco/CO ratios.
The white solid boxes indicate the east and west regions that are the focus of this work.
\label{fig:w49b_overview}}
\end{figure}

\section{Observations and data reduction}\label{Sec:obs}

\begin{deluxetable*}{rlccccccc}[t]
\tablecaption{Summary of the ALMA and/or ACA data.} \label{Tab:lines}
\centering
\tablecolumns{7}
\tablenum{1}
\centering
\tablehead{
\colhead{} & \colhead{} & \colhead{}  & \multicolumn{2}{c}{East} & \multicolumn{2}{c}{West} \\
\cmidrule(lr){4-5} \cmidrule(lr){6-7}
\colhead{Transition} & \colhead{Frequency\tablenotemark{a}} & \colhead{$n_{\rm crit}$\tablenotemark{b}}  & \colhead{$\theta_{\rm maj} \times \theta_{\rm min}$, PA} & \colhead{$\sigma$\tablenotemark{c}}   & \colhead{$\theta_{\rm maj} \times \theta_{\rm min}$, PA} & \colhead{$\sigma$\tablenotemark{c}} \\
\colhead{} & \colhead{(GHz)} & \colhead{(cm$^{-3}$)} & \colhead{(\arcsec $\times$ \arcsec, $^\circ$)}  & \colhead{(mJy beam$^{-1}$)} & \colhead{(\arcsec $\times$ \arcsec, $^\circ$)}  & \colhead{(mJy beam$^{-1}$)}  } 
\startdata
\hco~$J$=1--0 & 89.1885247  & 2.3$\times$10$^4$  & 3.89 $\times$ 2.87, $-$56.6 & 5 & 4.15 $\times$ 2.85, $-$57.0 & 5 \\
\hco~$J$=3--2 & 267.5576259 & 8.1$\times$10$^5$  & 6.28 $\times$ 4.06, $-$72.5 & 53 & 6.68 $\times$ 4.02, $-$71.2 & 57 \\
\hco~$J$=4--3 & 356.734223  & 2.0$\times$10$^6$  & 4.54 $\times$ 3.35, $-$85.7  & 139  & 4.78 $\times$ 3.31, $-$80.3 & 160 \\
\co~$J$=3--2  & 345.7959899 & 7.5$\times$10$^3$  & 5.15 $\times$ 3.31, +87.1 & 150  & 5.34 $\times$ 3.33, $-$88.1 & 150 \\
\co~$J$=2--1  & 230.538     & 2.1$\times$10$^3$  & 8.09 $\times$ 4.50, $-$74.9 & 85 & 7.56 $\times$ 4.65, $-$79.8  & 75 \\
\ico~$J$=2--1 & 220.3986842 & 1.9$\times$10$^3$  & 7.87 $\times$ 4.82, $-$80.0 & 84 & 7.95 $\times$ 4.85, $-$79.1  & 75 \\
\hline
Continuum   & 101.5 & --- & 3.40 $\times$ 2.38, $-$57.2 & 0.06 & 3.57 $\times$ 2.33, $-$57.1 & 0.06\\
\enddata
\tablenotetext{a}{The rest frequencies of the molecular transitions are taken from the CDMS database \citep{2005JMoSt.742..215M}. 
The 3~mm continuum data have a representative frequency of 101.5~GHz, with a frequency coverage of 99.7--103.2~GHz.}
\tablenotetext{b}{The critical densities are calculated assuming optically thin emission and a gas kinetic temperature of 100~K, following the method described by \citet{2015PASP..127..299S}.}
\tablenotetext{c}{The noise levels for all lines were measured at a channel spacing of 0.6~\kms.}
\end{deluxetable*}

\subsection{ALMA observations}\label{Sec:alma_obs}

The ALMA observations toward W49B east and west were conducted in Cycle 9 (project id: 2022.1.01608.S, PI: G. Zhang).
We carried out observations of \hco~$J$=1--0, 3--2, 4--3, and \co~$J$=3--2 lines, together with continuum emission in ALMA bands 3 ($\sim$3~mm), 6 ($\sim$1~mm), and 7 ($\sim$0.87~mm). 
Using a mosaic observation mode, we employed both the ALMA 12~m array and the ACA for band 3 observations, whereas bands 6 and 7 were observed exclusively with the ACA.

The observed areas in all three bands fully cover a 100\arcsec $\times$ 80\arcsec\,rectangular region centered at ($\alpha_{\rm ICRS}$, $\delta_{\rm ICRS}$)=(19$^{\rm h}$11$^{\rm m}$16$\rlap{.}^{\rm s}$87, 09$^\circ$05$\arcmin$21$\rlap{.}\arcsec$7) for W49B east, and a 100\arcsec $\times$ 120\arcsec\,rectangular region centered at ($\alpha_{\rm ICRS}$, $\delta_{\rm ICRS}$)=(19$^{\rm h}$11$^{\rm m}$01$\rlap{.}^{\rm s}$20, 09$^\circ$05$\arcmin$28$\rlap{.}\arcsec$0) for W49B west.
Figure~\ref{fig:w49b_overview} provides an overview of the observed regions toward W49B.
The ALMA 12~m array observations in band 3 were performed in the C-2 configuration, with baseline lengths of 15.1 -- 312.7~m, and the ACA observations in all three bands with baseline lengths of 8.9 -- 48.9~m. 
The original channel width was 122~kHz in band 3, and 488~kHz in bands 6 and 7, corresponding to a velocity spacing of $\sim$0.42~\kms, $\sim$0.56~\kms, and $\sim$0.43~\kms\,for bands 3, 6, and 7, respectively.

Standard calibration and data reduction were performed using the Common Astronomy Software Applications \citep[CASA,][]{2022PASP..134k4501C}. Imaging was performed with CASA version 6.4.1.12. 
Line imaging was performed using the CASA task \texttt{tclean} with the hogbom deconvolver, a briggs weighting with robust of 0.5, and the auto-multithresh algorithm for emission masking \citep{2020PASP..132b4505K}. 
All cleaned images have a pixel size of 0$\rlap{.}\arcsec$81
and were resampled to a velocity spacing of 0.6~\kms.
Primary beam correction was applied to all images.
The observational details are summarized in Table~\ref{Tab:lines}.

To recover the missing flux and diffuse emission, the cleaned ALMA data were combined with the calibrated single-dish data (see details in Sect.~\ref{Sec:sd_obs}) using the CASA task \texttt{feather}.
The clumpy CO and \hco\ emission in W49B west at velocities of $V_{\rm LSR} >$ +75~\kms\ (see Sect.~\ref{subsec:gas_west}) is the focus of this work. These components are characterized by broad line widths indicative of SNR--MC interaction, and we find no significant missing flux associated with them.
To verify the imaging results for these $V_{\rm LSR} >$ +75~\kms\ emission components, we additionally performed manual masking for them in the \texttt{tclean} task.
A comparison between manual and auto-masking shows that the fluxes derived for \hco\ agree within 10\%. For CO, however, manual masking is more reliable, as its residual map does not show significant emission features, and in some regions it recovers about 20--30\% more flux than auto-masking. We therefore adopted the imaging results obtained with manual masking for all lines when analyzing the broad-line features (see Sect.~\ref{Sec:discuss}).

For visualization purposes, the images presented in Sect.~\ref{Sec:results} are shown at the native angular resolution and were used to select the regions of interest. 
For quantitative analyses in Sect.~\ref{Sec:discuss}, based on spectral line information including line ratios, all data were convolved to a common circular beam of 8$\rlap{.}\arcsec$5 to minimize the effects caused by different beam sizes. 
All spectra presented in this paper were extracted from the convolved data cubes.

In this study, we only present the band~3 continuum emission for both regions, as no continuum was detected in bands 6 and 7. Two line-free spectral windows covering 99.7–103.2~GHz were selected, and the 12~m array and ACA data were combined for imaging.
Similar to the line imaging, the continuum imaging was performed with the CASA task \texttt{tclean}, but applying the multi-scale multi-frequency synthesis (mtmfs) deconvolver and manual masking.

\subsection{Single-dish observations and data}\label{Sec:sd_obs}

To recover the large-scale emission, we conducted ALMA total power (TP) observations of \co\,$J$=3--2 for both W49B east and west, as part of the same ALMA project mentioned above.
The half-power beam width (HPBW) is 18$\rlap{.}\arcsec$8. The typical 1-$\sigma$ noise level per 488~kHz (i.e., 0.43~\kms) channel is 0.5~Jy~beam$^{-1}$ for both regions.

We also carried out Atacama Pathfinder Experiment (APEX) mapping observations of \hco $J$=3--2 and 4--3 toward W49B west only (project id: M-0115.F-9505C-2025, PI: W. Yang).
The observations were carried out using the nFLASH230\footnote{\url{https://www.apex-telescope.org/ns/observing-run/observing/the-telescope/instruments/nflash/}} and the Large APEX sub-Millimeter Array (LAsMA\footnote{\url{https://www.apex-telescope.org/ns/observing-run/observing/the-telescope/instruments/lasma-large-apex-sub-millimetre-array/}}) receivers. 
The HPBWs are 24$\rlap{.}\arcsec$8 and 18$\rlap{.}\arcsec$6 for \hco~$J$=3--2, and $J$=4--3 transitions, respectively.
The forward efficiency ($\eta_{\rm fw}$) is 0.99 for all lines. The main beam efficiency ($\eta_{\rm mb}$) is 0.67 for the \hco~$J$=3--2 line, and 0.71 for the \hco~$J$=4--3 line.  
The reduced data cube has a typical rms noise of 20~mK in $T_{\rm A}^*$ scale at a velocity spacing of 0.6~\kms.

We also utilized \hco\,$J$=1--0 data from IRAM 30~m observations for both west and east regions \citep{2022ApJ...931..144Z}.
The data cube has a typical rms noise of 20~mK at a velocity spacing of 0.7~\kms\ and a pixel size of 10\arcsec.

\subsection{Archival ALMA ACA data} \label{Sec:alma_archive}

We utilized archival ALMA ACA data of the $J$=2--1 transitions of \co\ and \ico\ for W49B \citep[project id: 2018.1.01780.S, see][]{2021ApJ...919..123S}.
Their observations covered the entire SNR using both the ACA and TP antennas.
In this work, we re-imaged the data following the procedures described in Sect.~\ref{Sec:alma_obs}, and restricted the analysis to the regions covered by our new observations.
For the manually masked \co~$J=2$--1 data cube at $V_{\rm LSR} >$ +75~\kms, we did not include the TP data because no TP emission was detected in this velocity range.
Besides, the velocity coverage of the two CO lines in this project spans about $-$20 to +135~\kms, narrower than that of our new ALMA and ACA observations.

\subsection{Other data} \label{Sec:nir_archive}

We used the 2.12~$\mu$m H$_2$ and 1.64~$\mu$m [Fe~II] images from \citet{2019AJ....157..123L}, which were obtained as part of the UWISH2 \citep{2011MNRAS.413..480F} and UWIFE \citep{2014MNRAS.443.2650L} surveys, respectively. 
Both surveys were carried out with the Wide-Field Camera (WFCAM) at the United Kingdom Infrared Telescope.
We additionally adopted the Chandra X-ray image in the 0.3--10~keV band from \cite{2018A&A...615A.150Z}.

\section{Results}\label{Sec:results}

\subsection{Compact broad-line molecular emission in W49B west} \label{subsec:gas_west}

\subsubsection{Overview} \label{subsec:west_overview}

\begin{figure*}[ht]
\hspace{127mm}
\mbox{\begin{minipage}[b]{5.5cm}
\includegraphics[width=1\textwidth]{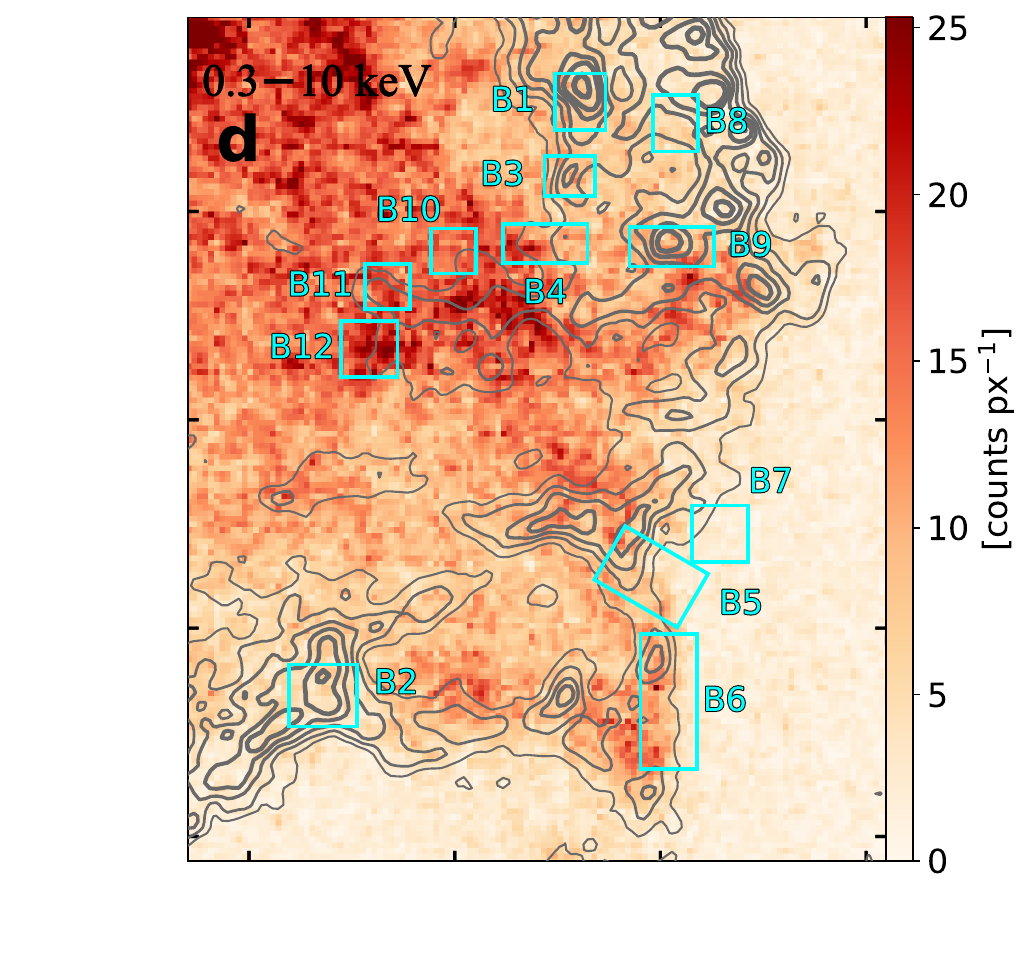}
\end{minipage}
\hspace{-103mm}
\mbox{\begin{minipage}[b]{5.5cm}
\includegraphics[width=1\textwidth]{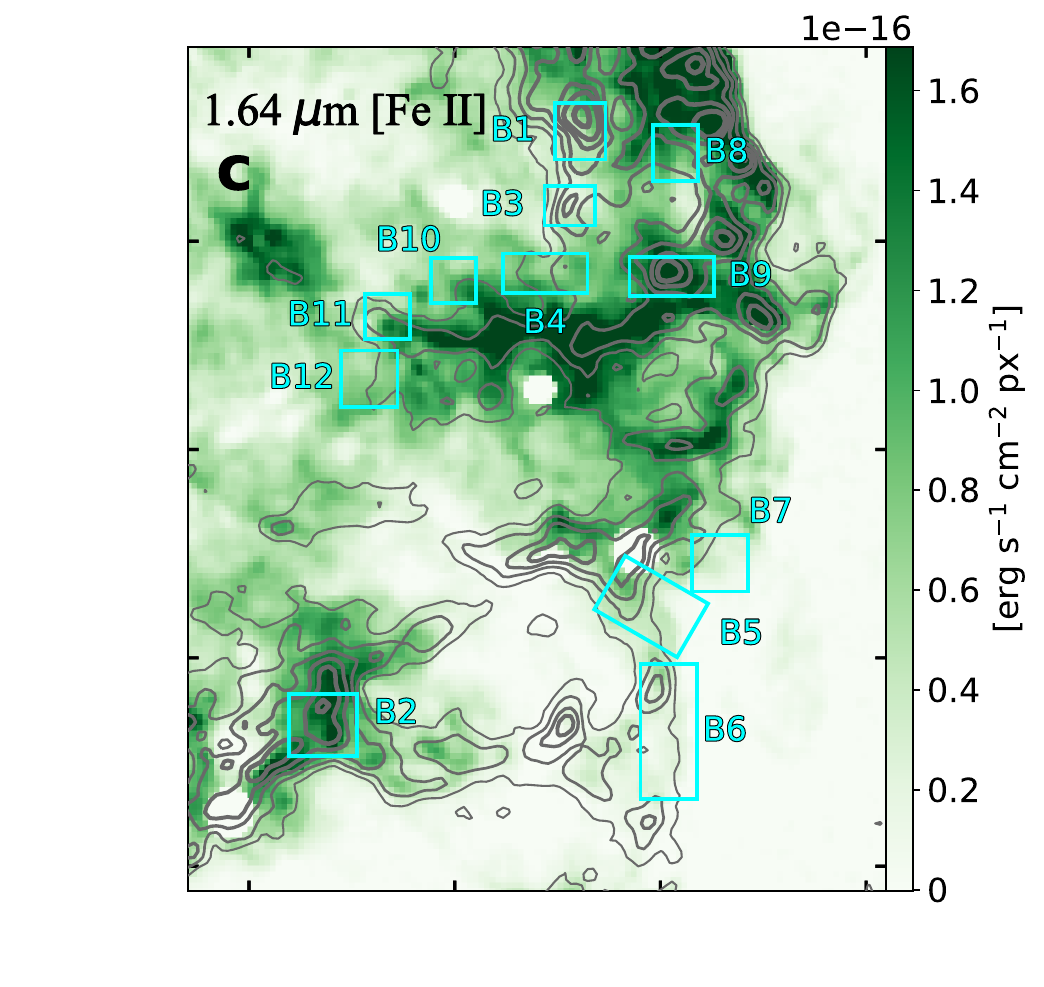}
\end{minipage}}
\hspace{-103mm}
\mbox{\begin{minipage}[b]{5.5cm}
\includegraphics[width=1\textwidth]{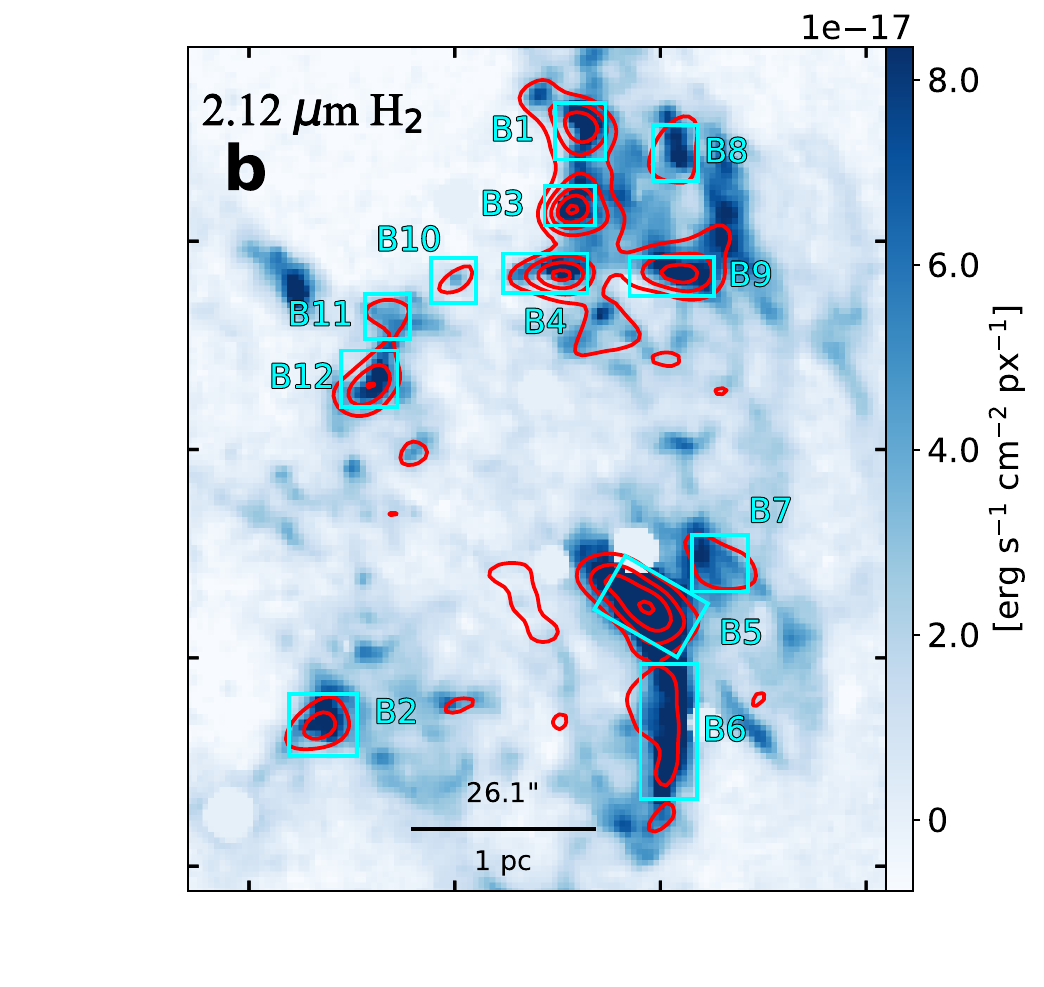}
\end{minipage}}
\hspace{-103mm}
\mbox{\begin{minipage}[b]{5.5cm}
\includegraphics[width=1\textwidth]{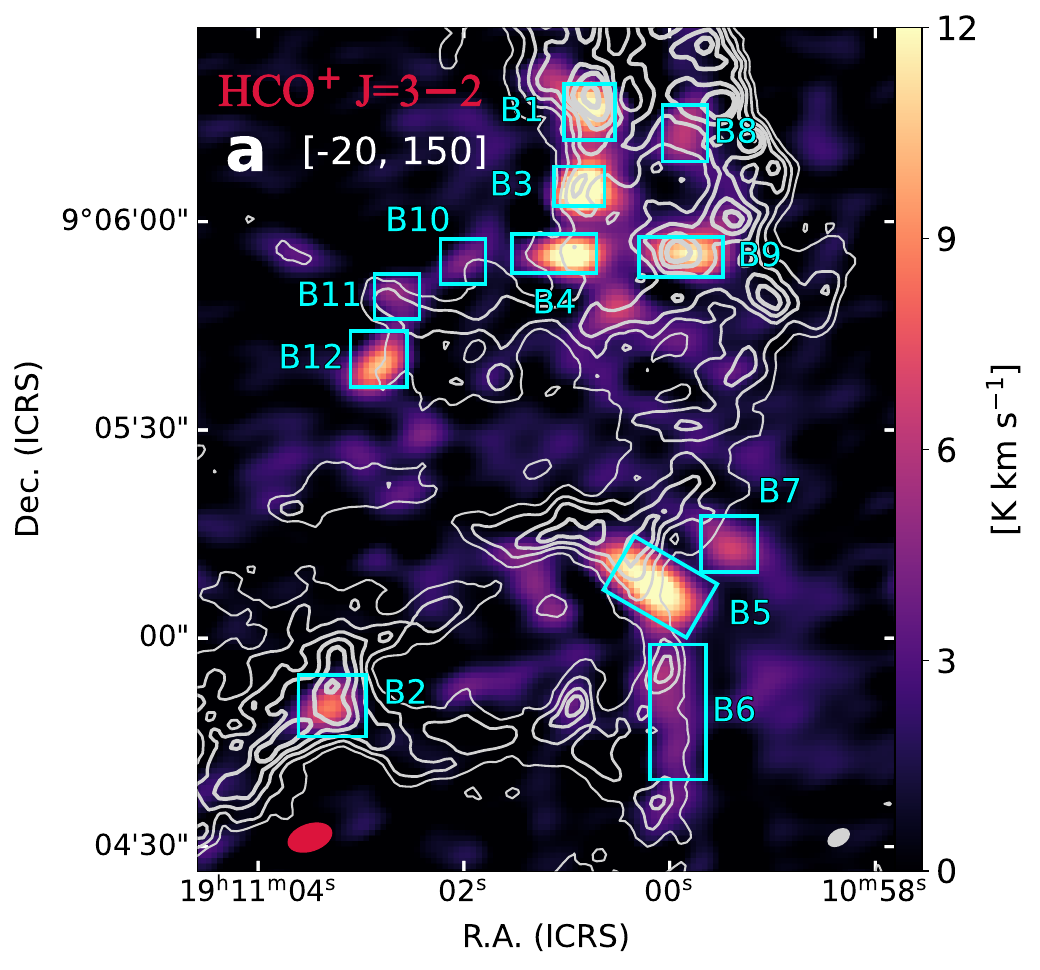}
\end{minipage}}}

\caption{Multi-phase gas in W49B west. The cyan boxes in each panel denote regions where broad  \hco\,emission is detected.
(a) Integrated intensity map of \hco~$J$=3--2 over $-$20 to +150~\kms, overlaid with the light-gray contours of the 3~mm continuum from the SNR. 
Contour levels are [5, 10, 15, 20, 25, 30, 40] $\times \sigma_{\rm cont}$, where 1$\sigma_{\rm cont}$ of 0.06~mJy~beam$^{-1}$, with thicker contours indicating stronger continuum emission. 
Synthesized beams for the line and continuum emission are shown in the lower left and lower right, respectively. 
(b) 2.12~$\mu$m H$_2$ map \citep{2019AJ....157..123L} overlaid with the red contours of the integrated \hco~$J$=3–2 emission. 
Contour levels are [5, 10, 15, 20, 25, 30] $\times \sigma_{I(\rm HCO^+)}$, where $1\sigma_{I(\rm HCO^+)}$ of $7\times10^{-4}$ K~\kms. A horizontal scale bar of 26$\rlap{.}\arcsec$1 corresponds to 1~pc at a distance of 7.9~kpc. 
(c) 1.64~$\mu$m [Fe~II] map \citep{2019AJ....157..123L} and (d) 0.3--10~keV X-ray emission map \citep{2018A&A...615A.150Z} overlaid with the same 3~mm continuum contours as shown in panel (a).
\label{fig:west_inte_map_all}}
\end{figure*}

In W49B west, we detected all the lines listed in Table~\ref{Tab:lines}.
Compared to \hco\,emission, \co\ emission is more spatially extended and exhibits a larger number of components. In this work, we focus on the \co\ emission components exhibiting a spatial and kinematic correspondence with the \hco\ emission.

Broadened molecular lines provide strong evidence that MCs are being disturbed by shocks \citep[e.g.,][]{2010ApJ...712.1147J}.
Broad emission of \hco~$J$=1--0, 3--2, 4--3, and \co~$J$=2--1, 3--2 exhibits multiple clumpy structures in W49B west.
These structures are primarily concentrated in two distinct areas with different velocity ranges (see Fig.~\ref{fig:west_inte_map}): a northern component spanning $\sim$ +25 to +150~\kms\ (corresponding to regions B and C in \citealt{2022ApJ...931..144Z}) and a southwestern component spanning +40 to +100~\kms\ (corresponding to region A).
At an angular resolution of 8$\rlap{.}\arcsec$5, our observations resolve the broad-line emission into compact substructures with a typical full width at half maximum (FWHM) of $\sim$40~\kms.
This suggests that the extremely broad \hco~$J$=1–0 line observed with the IRAM 30~m telescope \citep{2022ApJ...931..144Z}, with a FWHM of 75~\kms, arises from the blending of multiple shocked velocity components.

\begin{figure*}[htb]
\centering
\includegraphics[width=1\textwidth]{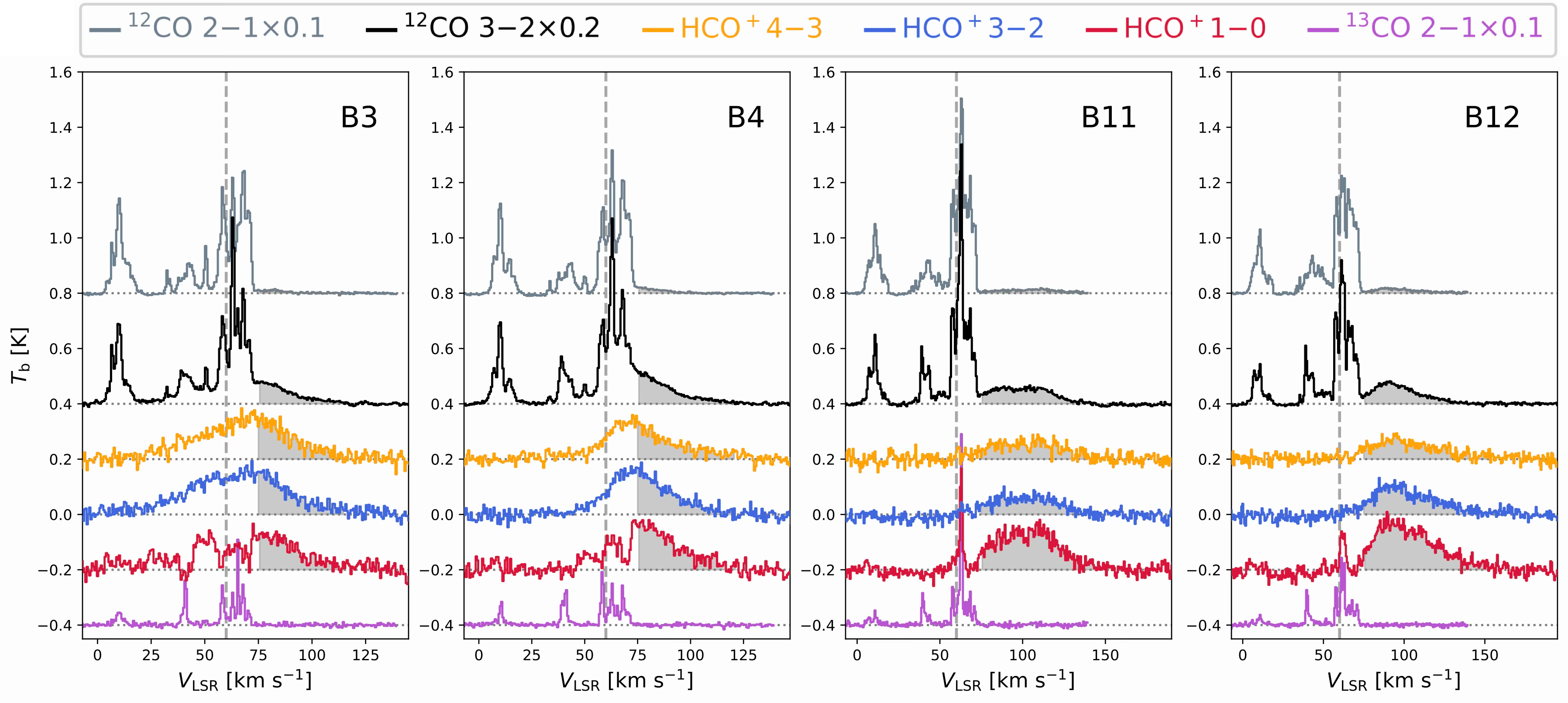}
\caption{Averaged spectra of \hco~$J$=1--0, 3--2, 4--3, \co~$J$=3--2, 2--1, and \ico~$J$=2--1 lines from four representative broad-line emitting regions in W49B west (labeled in Fig.~\ref{fig:west_inte_map_all}). The dashed vertical line indicates +60~\kms. The shaded regions present the velocity ranges ($>$ +75~\kms) used in subsequent analysis in Sect.~\ref{Sec:discuss}. The \co\ and \ico\ $J$=2--1 observations cover velocities from approximately $-20$ to $+135$~\kms, narrower than the velocity coverage of the other transitions.
All spectra are extracted from data cubes produced using auto-masking, then combined with the single-dish data and convolved to a circular beam of 8$\rlap{.}\arcsec$5. Spectra from the other eight regions are shown in Fig.~\ref{fig:west_broad_spec}.
\label{fig:west_broad_spec_example}}
\end{figure*}

We employ the \hco~$J$=3--2 line to identify a total of 12 broad-line emitting regions in W49B west (Fig.~\ref{fig:west_inte_map_all}a), as it is stronger than the \hco~$J$=4--3 line, not affected by the absorption present in the \hco~$J$=1--0 line, and less contaminated by unrelated line-of-sight components than the \co\ emission. 
These broad-line emissions exhibit a compact morphology at high angular resolution ($\sim$5\arcsec).
At this resolution, the shocked molecular gas traced by compact broad CO and \hco\ emission shows no significant spatial offset (Fig.~\ref{fig:west_inte_map}).

\subsubsection{Line profiles} \label{subsec:west_spectra}

The extracted spectra of the 12 broad-line regions are shown in Figures~\ref{fig:west_broad_spec_example} and \ref{fig:west_broad_spec}.
Even at an angular resolution of 8$\rlap{.}\arcsec$5 (i.e., 0.33~pc at a distance of 7.9~kpc), these compact emissions still exhibit very broad line profiles, with FWHM values of 25--60~\kms. 
These FWHMs are significantly broader than typical molecular line widths observed in other SNR-MC interaction regions \citep[$\sim$ 7--35~\kms, e.g.,][]{1997ApJ...485..263K,1999ApJ...511..836R, 2016ApJ...816....1K,2024ApJ...966..178T, 2024ApJ...974..262T}.
In addition, no corresponding \ico~$J$=2--1 counterpart to these broad \hco\ and \co\ components is detected at the current sensitivity.
In some regions, such as B3, B4, and B5, the \hco~$J=1$--0 emission shows absorption dips at velocities $<$ +75~\kms. These dips are aligned in velocity with narrow \ico~$J=2$--1 emission peaks, indicating the presence of cooler foreground \hco\ gas.

Interestingly, we detected broad molecular emission extending beyond +75~\kms. 
The maximum LSR velocity permitted by the Galactic rotation along the direction toward W49B ($l~\sim$ 43~deg) is approximately +75~\kms, i.e., the tangent velocity, as shown in Figure~4 of \citet{2001ApJ...547..792D} and Figure~17 of \citet{2026ApJS..282...65Y}. 
Emission beyond the tangent velocity can be attributed to local dynamical perturbations, similar to the high-velocity H{\sc i} gas detected in W51C \citep{1997ApJ...475..194K}.
At velocities above +75~\kms, line-of-sight contamination in \hco\ and \co\ emission is  substantially reduced for all regions, making this velocity range a particularly clean probe of the shocked molecular gas.
We therefore select six regions (B3, B4, B7, B9, B11, and B12; see their spectra in Fig.~\ref{fig:west_broad_spec_gt75}) exhibiting prominent emission above this velocity for the detailed analysis in Sect.~\ref{Sec:discuss}.

To characterize the high-velocity profiles, Figure~\ref{fig:west_broad_line-ratio} shows the channel-by-channel intensity ratios between the different \hco\ and \co\ transitions for the six regions at $V_{\rm LSR}>+75$~\kms. 
In comparing the \hco\ and \co\ profiles, we adopt the \hco~$J$=1--0 /\co~$J$=3--2 ratio because these two transitions have relatively high signal-to-noise ratios. The spectra are binned over five channels to improve the robustness of the velocity-dependent trends.
Within the uncertainties, most line ratios are generally constant across the high-velocity channels, suggesting that the velocity structures and excitation of these transitions are broadly similar at the current resolution. The main exceptions are regions B7 and B12, where the \hco~$J$=1--0/\co~$J$=3--2 intensity ratio increases toward the high-velocity wings, with a possible but less pronounced trend also seen in B9. This velocity-dependent enhancement is discussed further in Sect.~\ref{Sec:dis_CR_diff_length}.

The velocity ranges and velocity centroids of these broad-line emissions allow us to infer the location of the shocked gas relative to the remnant. The line centroids of the B1 and B2 regions are blue-shifted with respect to the systemic velocity, indicating that shocked gas lies on the near side of the remnant. The B5 and B6 regions show centroid velocities close to the systemic velocity, suggesting shocks propagating roughly perpendicular to the line-of-sight. The remaining regions show red-shifted velocity centroids, indicating that the shocked gas is located on the far side of the remnant.
We note that B11 and B12 are more red-shifted than in the other regions, possibly because the shock propagation direction is more closely aligned with the line of sight.

\subsubsection{Multi-band comparison} \label{subsec:west_multi-band}

In Figure~\ref{fig:west_inte_map_all}b, the shocked molecular gas traced by the broad emission of \hco\, is closely associated with the bright 2.12~$\mu$m H$_2$ knots \citep{2019AJ....157..123L}. 
Similar associations have also been reported in SNR IC~443 and HB~3 \citep{2001ApJ...547..885R, 2021ApJ...917...47R}.
On the other hand, the broad \hco\ emission in most of the regions, except for B2 and B9, is offset from the bright 1.64~$\mu$m [Fe~II] emission (Fig.~\ref{fig:west_inte_map_all}c) and 3~mm continuum peaks.
In Fig.~\ref{fig:west_inte_map_all}d, the broad \hco\ emission regions do not show a clear spatial correlation with the 0.3--10~keV band X-ray emission \citep{2018A&A...615A.150Z}.

The continuum emission\footnote{We estimate a spectral index of about  $-$0.7 to $-$0.6 for W49B west, using VLA 327~MHz continuum \citep{2001ApJ...559..954L}, MeerKAT 1359.7~MHz continuum \citep{2024MNRAS.531..649G}, and our 3 mm continuum data. The three datasets show nearly identical intensity distributions, indicating that the 3~mm continuum emission is likely dominated by synchrotron radiation. In this work, we use the 3~mm continuum emission to define the boundary of the SNR because it has the highest angular resolution among these datasets.} 
from the SNR originates from synchrotron radiation of relativistic electrons 
accelerated at the shock front.
The [Fe~II] emission closely follows the SNR shell delineated by the 3~mm continuum morphology (Fig.~\ref{fig:west_inte_map_all}c), suggesting that the [Fe~II]-emitting gas traces fast shocks along the shell.
In contrast, the molecular gas is spatially offset from the [Fe~II] and continuum emission, suggesting that these tracers probe different shock components.

\subsection{Narrow-line molecular emission in W49B east} \label{subsec:gas_east}

\begin{figure*}[!tbp]
\mbox{
\hspace{129mm}
\mbox{\begin{minipage}[b]{5.45cm}
\includegraphics[width=1\textwidth]{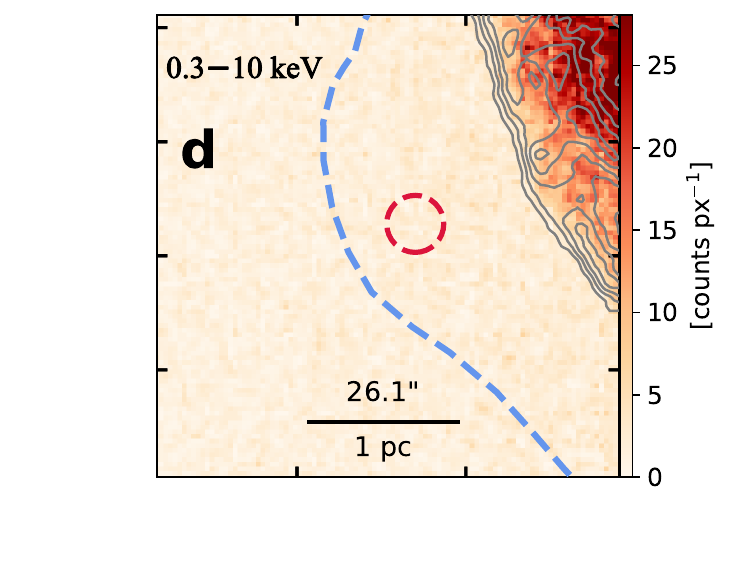}
\end{minipage}}
\hspace{-100mm}
\mbox{\begin{minipage}[b]{5.5cm}
\includegraphics[width=1\textwidth]{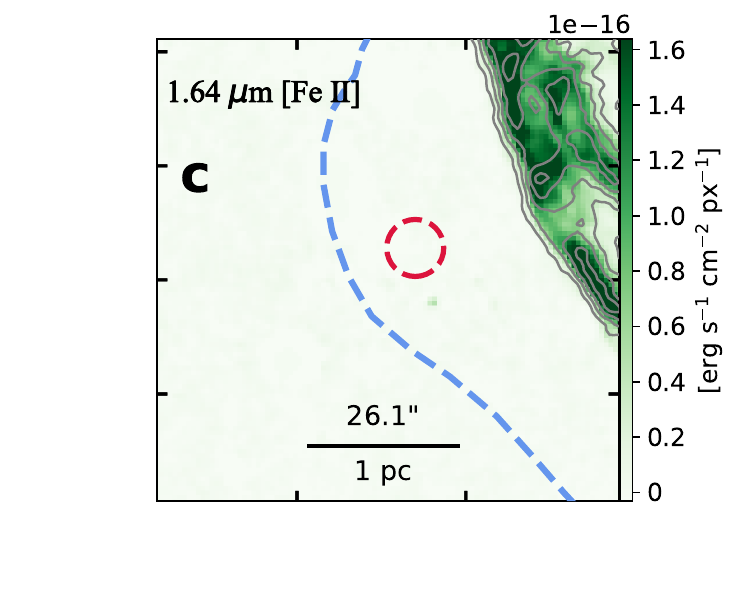}
\end{minipage}}
\hspace{-100mm}
\mbox{\begin{minipage}[b]{5.4cm}
\includegraphics[width=1\textwidth]{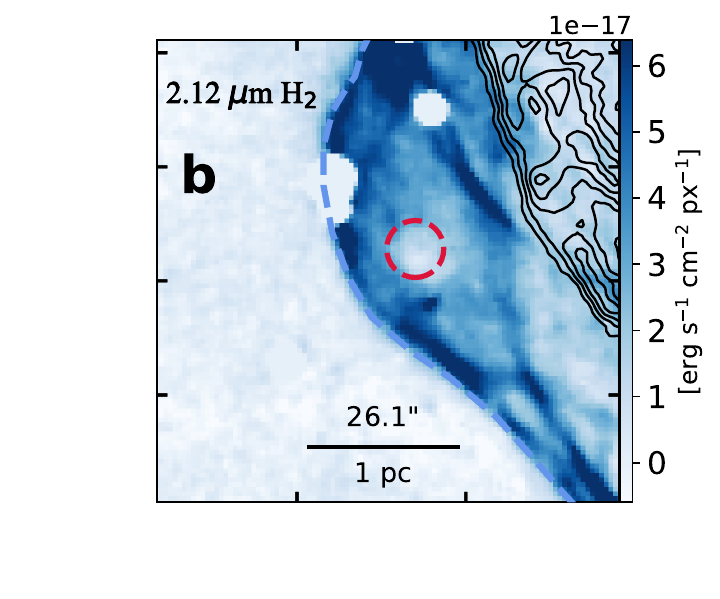}
\end{minipage}}
\hspace{-100mm}
\mbox{\begin{minipage}[b]{5.55cm}
\includegraphics[width=1\textwidth]{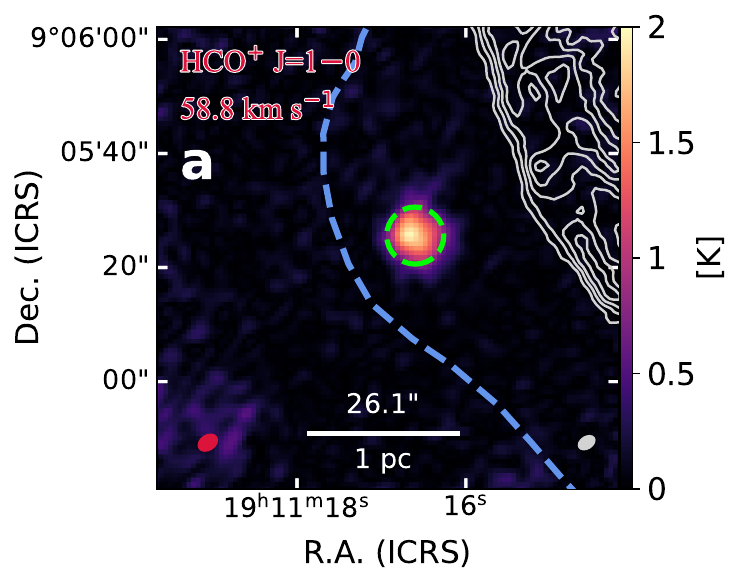}
\end{minipage}}
}
\caption{Bright \hco~$J$=1--0 blob-like component at +58.8~\kms\,and its multi-wavelength counterparts in W49B east.
(a) \hco~$J$=1--0, (b) 2.12~$\mu$m H$_2$ and (b) 1.64~$\mu$m [Fe~II] images from \citet{2019AJ....157..123L}. (d) 0.3--10~keV X-ray map \citep{2018A&A...615A.150Z}.
The contours in each panel show the 3~mm continuum emission from the SNR with levels of [5, 10, 15, 20, 25] $\times \sigma_{\rm cont}$, where 1$\sigma_{\rm cont}$ of 0.06~mJy~beam$^{-1}$.
Synthesized beams for the lines and continuum are shown in the bottom left and right corners, respectively. 
The dashed circle (10\arcsec\ in diameter) in each panel outlines the extent of the \hco~$J$=1--0 emission.
The blue, dashed curve delineates the eastern boundary of the H$_2$ filament. A horizontal scale bar of 26$\rlap{.}\arcsec$1 corresponding to 1~pc at a distance of 7.9~kpc. 
\label{fig:east_blob}}
\end{figure*}

\begin{figure}[htbp]
\centering
\includegraphics[width=0.49\textwidth]{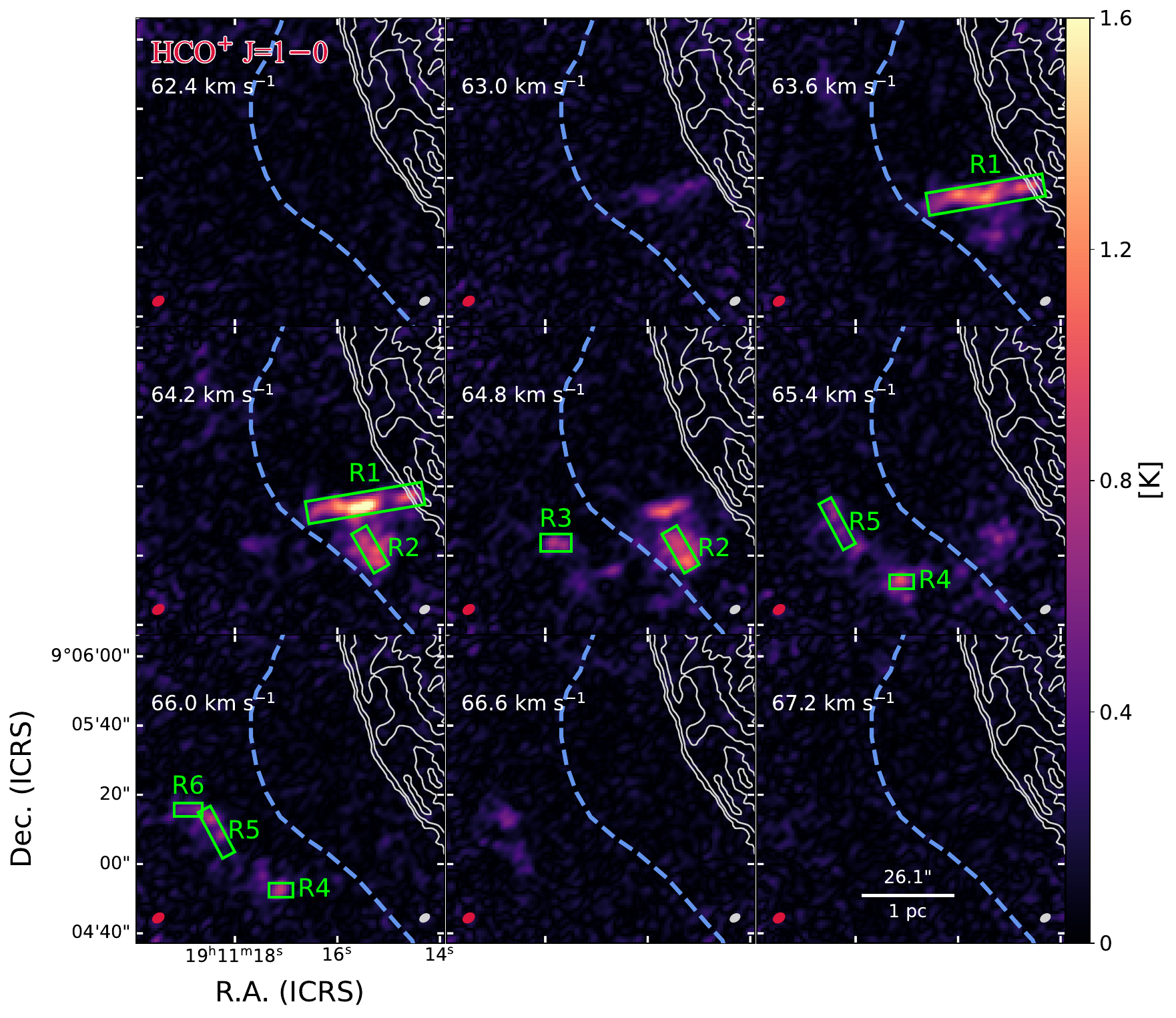}
\caption{Channel maps of \hco~$J$=1--0 emission in W49B east from +62.4 to +67.2~\kms\ in steps of 0.6~\kms. The green boxes denote the regions of interest showing bright \hco~$J$=1--0 emission.
The blue dashed curves, synthesized beams, 3~mm continuum contour levels and scale bar are identical to those in Fig.~\ref{fig:east_blob}.
\label{fig:east_channel_map}}
\end{figure}

In W49B east, we detected emission from \hco~$J$=1--0. 
Bright \hco~$J$=1--0 emission ($>$0.4~K)
is mainly found in the velocity range from +58.8 to +66.6~\kms. Based on the morphology of the emission at different velocities, this emission can be divided into a blob-like component (Fig.~\ref{fig:east_blob}) and several filamentary and clumpy structures (Fig.~\ref{fig:east_channel_map}). All of these features show narrow line profiles.

The \hco~$J$=3--2 emission
is only marginally detected toward the blob-like component (Fig.~\ref{fig:east_blob_appendix}a) and a filamentary structure (labeled as R1 in Fig.~\ref{fig:east_channel_map}). Its intensity is roughly one-tenth that of the \hco~$J$=1--0 emission. 
No \hco~$J$=4--3 emission is detected in W49B east.
The weak or absent higher-$J$ \hco\ emission suggests low-excitation conditions, likely due to low gas density and/or low kinetic temperature.

As shown in Fig.~\ref{fig:east_blob}, the blob-like structure at +58.8~\kms\, is located outside the 3~mm continuum, [Fe~II], and X-ray emission, but appears to coincide with a local dip in the 2.12~$\mu$m H$_2$ emission.
This feature, centered at (19$^{\rm h}$11$^{\rm m}$16$\rlap{.}^{\rm s}$60, 09$^\circ$05$\arcmin$25$\rlap{.}\arcsec$6), has an angular size of about 10\arcsec, corresponding to a diameter of 0.38~pc at a distance of 7.9~kpc, and is located about 21\arcsec\  (0.8~pc) from the eastern edge of the SNR.
This feature has a \ico~$J$=2--1 counterpart, while the \co~$J$=2--1 and 3--2 emission appears more extended (see Figs.~\ref{fig:east_blob_appendix}b--d).
Notably, the line profiles of all transitions mentioned above are narrow, with detectable emission mostly confined to a single channel (i.e., 0.6~\kms; see Fig.~\ref{fig:east_blob_appendix}e). 

A filamentary structure is prominent between +63.6 and +64.2~\kms\ (R1 in Fig.~\ref{fig:east_channel_map}) in the channel map of \hco~$J$=1--0 emission. It has a length of about 34\arcsec, corresponding to 1.3~pc, and is inclined by about 70$^\circ$ with respect to the continuum edge of the SNR.
Between +64.8 and +66.6~\kms, the \hco~$J$=1--0 emission exhibits several clumpy features (R2--R5 in Fig.~\ref{fig:east_channel_map}), all located outside the SNR. With increasing velocity, these structures appear to be further away from the SNR, and many of them do not show clear spatial correspondence with the 2.12~$\mu$m H$_2$ emission. 
All these features have counterparts in \co~$J$=3--2 and 2--1, but no corresponding \ico~$J$=2--1 emission is detected (see Fig.~\ref{fig:east_channel_map_co}), in contrast to the blob-like structure at +58.8~\kms. 

However, the limited number of detected transitions in W49B east, together with the lack of clear spatial association with the remnant, prevents a reliable excitation and column density analysis. 
If local thermodynamics equilibrium (LTE) conditions are assumed, the inferred result would be generally consistent with that reported by \citet{2022ApJ...931..144Z}. 
We therefore briefly describe the detected emission and provide a simple analysis of the gas in Appendix~\ref{sec:appendix-A}.

We note the presence of faint, diffuse \hco\ emission at about +58 to +62~\kms\ within our field of view, with counterparts in \co\ and \ico. This emission is primarily recovered by the IRAM 30~m data \citep{2022ApJ...931..144Z}, as the interferometric observations are insensitive to 
 such extended emission. Owing to the limited signal-to-noise ratio, we do not further investigate this emission in this work.

\subsection{Narrow-line molecular emission in W49B west} \label{subsec:gas_narrow_west}

Narrow \hco~$J$=1--0 and 3--2 lines are detected in the W49B west region, while no \hco~$J$=4--3 narrow emission is found. The absence of narrow \hco~$J=4$--3 emission suggests that the narrow component is relatively weakly excited, with densities and/or temperatures insufficient to populate the higher rotational levels efficiently. In this work, we focus on the \hco\ emission near +60~\kms, which spans a velocity range from +61 to +66~\kms. More details are given in Appendix \ref{sec:appendix-C}.




\section{Discussion}\label{Sec:discuss}

\subsection{Physical conditions of the shocked gas}\label{Sec:dis_phy_conditions}

To derive the physical conditions of the shocked gas, it is necessary to decompose the broad-line feature from the observed spectra (Figs.~\ref{fig:west_broad_spec_example} and \ref{fig:west_broad_spec}). For \hco~$J$=3--2 and 4--3 spectra, the broad emission is generally well described by a single Gaussian profile. In contrast, for \hco~$J$=1--0 and both \co\ spectra, the broad-line component is blended with multiple narrow features at $V_{\rm LSR} <$ +75~\kms, which hampers a reliable decomposition.
Therefore, we select six regions which exhibit clear broad emission at $V_{\rm LSR}>$ +75~\kms\ without contamination (see Fig.~\ref{fig:west_broad_spec_gt75}), and perform a 
non-LTE analysis using RADEX \citep{2007A&A...468..627V}.

For the RADEX calculations, we adopt the plane-parallel slab geometry, with the escape probability $\beta$=(1$-$e$^{-3\tau}$)/3$\tau$, where $\tau$ is optical depth. 
The cosmic microwave background is the only external radiation field included in the calculations.
The molecular data are taken from the Leiden Atomic and Molecular Database \citep[LAMDA;][]{2005A&A...432..369S}. 

We adopt the peak intensity of the \hco~$J$=3--2 line (see Table~\ref{Tab:2hcop_line_para_west_broad}), and the line ratios derived from the $>$ +75~\kms\,component of three \hco\ and two \co\ transitions (see Table~\ref{Tab:line_ratio_west}), as the observational inputs.
The linewidth is set to 40~\kms\, for RADEX calculations, as the typical FWHM of the lines in these regions.
From our observations, we do not find significant spatial deviation in the broad \hco\ and \co\ emission. Thus, the filling factor of all emission is assumed to be 0.2, estimated from the ratio of the smallest synthesized beam size to the convolved beam size.
We note that varying the filling factor from 0.1 to 1 changes the inferred gas density by less than 50\% and the column density ratio between the two molecules by less than 10\%. Therefore, the adopted filling factor does not strongly affect our results.

We construct RADEX model grids, with the parameter ranges of 10--200~K for gas temperature ($T_{\rm kin}$), 10$^3$--10$^7$~cm$^{-3}$ for H$_2$ density ($n_{\rm H_2}$), 10$^{10}$--10$^{16}$~cm$^{-2}$ for \hco\ column density ($N_{\rm HCO^+}$), and 10$^{14}$--10$^{19}$~cm$^{-2}$ for \co\ column density ($N_{\rm CO}$).
We take 200~K as the upper limit, as the collisional rate coefficients for \hco\ are available up to this temperature \citep{2020MNRAS.497.4276D}.
Then, to constrain these physical parameters of shocked gas in six regions, we use the emcee code \citep{2013PASP..125..306F} to perform Markov chain Monte Carlo (MCMC) sampling with an affine-invariant ensemble sampler \citep{2010CAMCS...5...65G}. We adopt a linear-uniform prior on $T_{\rm kin}$ and log-uniform priors on $n_{\rm H_2}$, $N_{\rm HCO^+}$ and $N_{\rm CO}$.
We found that $T_{\rm kin}$ converges only in regions B11 and B12, where the integrated intensity ratio $R_{\rm CO}$ between \co~$J$=3--2 and \co~$J$=2--1 is relatively low (see Table~\ref{Tab:line_ratio_west}). 
For the other regions (B3, B4, B7 and B9), $T_{\rm kin}$ does not converge and instead approaches the upper limit of 200~K.
This behavior is expected because these regions show much stronger \co~$J=3$--2 emission relative to \co~$J=2$--1, suggesting that higher kinetic temperatures, possibly above 200~K, are required to reproduce the excitation.

To investigate whether $T_{\rm kin}$ significantly affects the $N_{\rm HCO^+}$/$N_{\rm CO}$ ratio, which is used as a proxy for the abundance ratio, we also derive the physical parameters by fixing $T_{\rm kin}$ at 100~K and 200~K.
Table~\ref{Tab:physical_para_west_broad} lists the physical parameters of six broad-line regions, with $T_{\rm kin}$ either fixed or treated as a free parameter.
We found that $N_{\rm HCO^+}$ is well-constrained by three \hco\, lines, while a higher $T_{\rm kin}$ leads to a slightly larger $N_{\rm CO}$ and consequently a slightly lower $N_{\rm HCO^+}$/$N_{\rm CO}$ ratio.
Thus, we conclude that the derived $N_{\rm HCO^+}/N_{\rm CO}$ ratio is not strongly affected by the assumed $T_{\rm kin}$, varying by less than a factor of two.
For the B3, B4, B7, and B9 regions, we adopt the results obtained with $T_{\rm kin}$ fixed at 200~K.
Among the six regions, the $N_{\rm HCO^+}$/$N_{\rm CO}$ ratios are in the range of (4.2--14.5)$\times $10$^{-4}$. The highest value is found in region B7, which is spatially separated from the other five regions, whereas the lowest value is measured in region B4.
It is worth noting that our RADEX calculations indicate that all these broad-line emissions are optically thin. The derived abundance ratios therefore likely trace conditions throughout a substantial fraction of the shocked gas, rather than being restricted to its surface layers.

Compared to other SNRs, the $N_{\rm HCO^+}$/$N_{\rm CO}$ ratios of (4.2--14.5)$\times$ 10$^{-4}$ found in W49B west are remarkably high. 
For comparison, abundance ratios are $\sim$2$\times$10$^{-4}$ in W28  \citep{2024ApJ...966..178T}, (1--4)$\times$ 10$^{-4}$ in W51C \citep{1997ApJ...485..263K,2025A&A...693A.222T}, and (6--9)$\times$ 10$^{-5}$ in clump B and 1.7$\times$10$^{-4}$ in clump G of IC~443 \citep{1989ApJ...341..857Z}. All of these remnants show clear evidence of shock interaction with MCs.
In contrast, \citet{2025ApJ...978...83T} found that SNRs without clear evidence of interactions with MCs typically show lower abundance ratios of $\sim$10$^{-5}$.



\subsection{Enhanced cosmic-ray ionization as the origin of the high \texorpdfstring{\hco/CO}{HCO+/CO} abundance ratio}\label{Sec:dis_crir}

\begin{figure*}[htbp]
\centering
\includegraphics[width=0.4\textwidth]{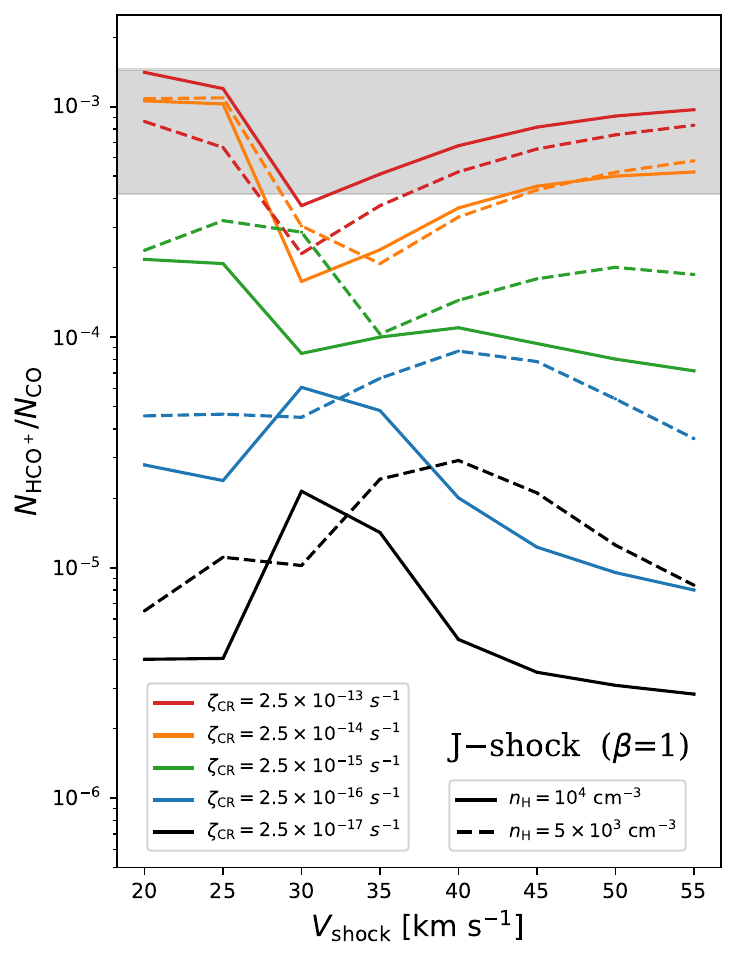}
\includegraphics[width=0.4\textwidth]{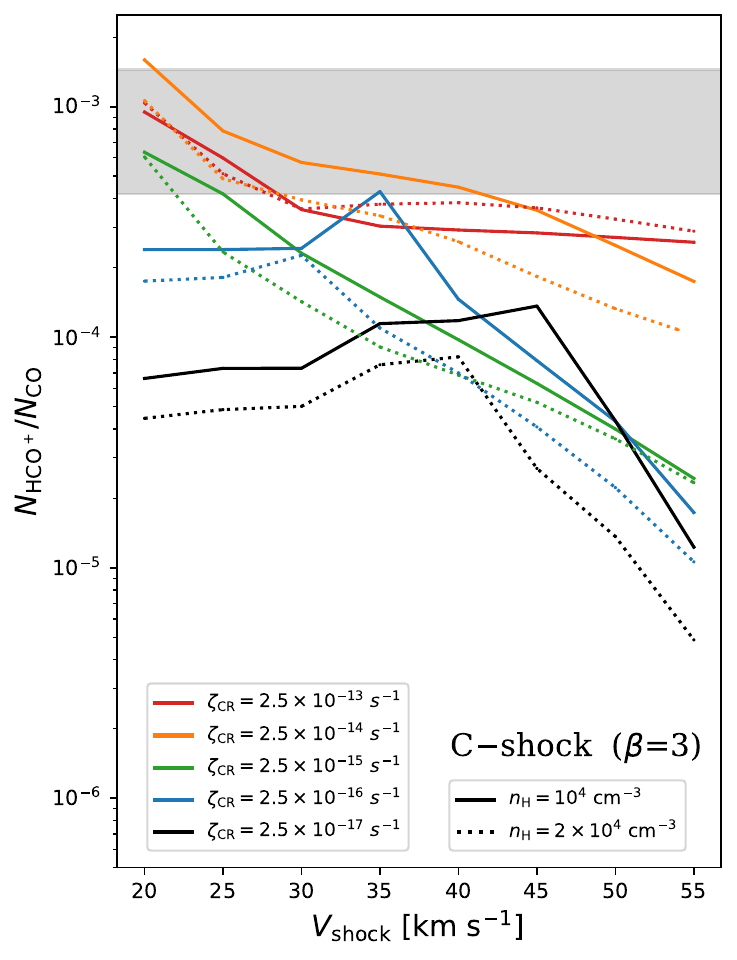}
\caption{The modeled $N_{\rm HCO^+}$/$N_{\rm CO}$ ratio as a function of shock velocity, for J-shock (left) and C-shock (right), with different pre-shock densities $n_{\rm H}$ and CR ionization rates $\zeta_{\rm CR}$.
The gray shaded region marks the observed abundance ratio range of (4.2 -- 14.5)$\times$10$^{-4}$.
\label{fig:west_shock_mhd_code}}
\end{figure*}

\citet{2022ApJ...931..144Z} used the Paris-Durham magnetohydrodynamic (MHD) shock code \citep[][]{2003MNRAS.343..390F, 2019A&A...622A.100G} to interpret the high \hco~$J=1$--0/\co~$J=1$--0 intensity ratios observed in the broad-line regions of W49B west. They found that these ratios can be explained by chemistry driven by low-energy CRs in shocked MCs, with ionization rates 1–2 orders of magnitude above the standard Galactic level. 
In this work, we used multiple \hco\ and \co\ lines to better constrain post-shock gas density, and to derive more reliable $N_{\rm HCO^+}$/$N_{\rm CO}$ ratios in six regions, confirming that these abundance ratios are indeed enhanced. We therefore revisit the role of cosmic rays in producing the enhanced abundance ratios observed in the clumpy shocked gas of W49B west, using an updated Paris-Durham code (V1.1\_rev122).
This code is built to simulate the dynamical, thermal, and chemical evolution of the ISM in a steady-state plane-parallel shock, and includes 140 species linked by $\sim$3000 chemical reactions.

Running a shock model involves two steps. We first run the code to derive the thermo-chemical pre-shock conditions. These conditions are then used as inputs for the shock model, which follows the cloud’s dynamical and chemical evolution.
The shock age is fixed at 1~kyr, following the setup in \cite{2022ApJ...931..144Z}. This is well below the estimated age of SNR W49B \citep[5--6~kyr;][]{2018A&A...615A.150Z, 2020ApJ...893...90S}, but is consistent with the recombination age of the cooler X-ray plasma in the southwestern region \citep[$\sim$1~kyr;][]{2020ApJ...903..108H}, where hot plasma interacts with cold and dense gas.

The varied parameters include the shock velocity ($V_{\rm shock}$), the pre-shock gas density ($n_{\rm H}$ = $n({\rm H})+2n({\rm H_2})+n({\rm H^+})$), the CR ionization rate per H$_2$ molecule (CRIR; $\zeta_{\rm CR}$), and the magnetic scaling factor ($\beta$), which determines a magnetic field strength of $B = \beta\sqrt{n_{\rm H}/\mathrm{cm^{-3}}}~\mu\mathrm{G}$.
The broad molecular line profiles ($\Delta V = 30-50$~\kms) indicate strongly disturbed gas. However, owing to projection effects, unresolved kinematic components, and velocity gradients in the post-shock gas, the observed linewidths do not directly measure the shock velocity. We therefore explore $V_{\rm shock}$ of 20--55~\kms, as a plausible range encompassing the characteristic velocities of the molecular gas.
The transition between C (continuous) and J (jump) type shocks is often strongly influenced by the Alfv\'en speed of the charged fluid, $V_{\rm Alfv\acute{e}n} = B/\sqrt{4\pi\rho_c} \simeq 21\beta$~\kms\ \citep{2024ApJ...977..149R}, where $\rho_c$ is the mass density of charged fluid including charged dust grains \citep{2007A&A...476..263G}. Shocks faster than $V_{\rm Alfv\acute{e}n}$ tend to become J-type, while slower shocks may remain C-type or develop a mixed CJ structure \citep[e.g.,][]{1993ARA&A..31..373D}.
J-type shocks compress molecular gas more efficiently than C-type shocks. According to \cite{2024ApJ...977..149R}, the compression is a factor of 30--170 for J-type shocks, compared to 5--10 for C-type shocks.
We therefore explore both regimes in the Paris-Durham shock code, adopting $\beta=1$ for J-type shocks and $\beta=3$ for C-type shocks, although the actual shock structure remains uncertain because the magnetic field strength is not well constrained.
We adopt pre-shock densities of $n_{\rm H}=5\times10^3$ and $10^4~{\rm cm^{-3}}$ for J-type shocks, and $10^4$ and $2\times10^4~{\rm cm^{-3}}$ for C-type shocks, yielding post-shock densities reasonable for our observational constraints.
We explore shock velocities from 20 to 55~\kms\ in steps of 5~\kms. For each model, we vary the cosmic-ray ionization rate, $\zeta_{\rm CR}$, from the standard Galactic value of $2.5\times10^{-17}~{\rm s^{-1}}$ \citep{2000A&A...358L..79V} to values enhanced by several orders of magnitude.

Figure~\ref{fig:west_shock_mhd_code} shows the abundance ratio as a function of shock velocity for J-type and C-type shocks under different pre-shock densities and CR ionization rates.
Across the adopted parameter space of pre-shock density and magnetic scaling factor, the J-type shock models generally reproduce the observed abundance ratios better than the C-type models.
More importantly, regardless of the shock type, an enhanced CRIR appears necessary to explain the observed abundance ratios, with a required CRIR of about a few $10^{-14}$--$10^{-13}$~s$^{-1}$.
We note that such high CRIR in W49B is distinct among other SNRs (e.g., $\sim$10$^{-15}$~s$^{-1}$ in W28, W44, and W51C; \citealt{2011ApJ...740L...4C, 2014A&A...568A..50V,2024ApJ...966..178T}), and it could be comparable or even higher than that in the Galactic Center \citep[$\sim$10$^{-14}$~s$^{-1}$, e.g.,][]{2013ApJ...768L..34C,2019ApJ...883...54O}.

This interpretation should be treated with caution. 
The present MHD shock models include neither UV radiation from the SNR, nor self-generated UV radiation from fast J-type shocks, both of which may modify the post-shock chemistry \citep[e.g.,][]{2020A&A...643A.101L}.
W49B may also be a luminous UV source, as inferred from its soft X-ray emission \citep{2022ApJ...931..144Z}. The chemical effect of UV irradiation depends on the UV field strength, extinction, gas density \citep[e.g.,][]{2023ApJ...946...91L}. Since the UV spectrum of W49B and the extinction of the shocked gas are poorly constrained, the net effect of UV irradiation on the modeled abundance ratios cannot be reliably quantified.

Although X-ray radiation is not considered in the MHD code, we roughly estimate its contribution to the ionization rate.
We derive an unabsorbed X-ray spectrum in the 0.3--10~keV band from the observations of \citet{2018A&A...615A.150Z}, and then convert the photon flux density $N_{\rm ph}(E)$, in units of photon~cm$^{-2}$~s$^{-1}$~keV$^{-1}$, to energy flux density via $F(E)$ = $N_{\rm ph}(E) \times E \times 1.602 \times 10^{-9}$, where $F(E)$ is in units of erg~cm$^{-2}$~s$^{-1}$~keV$^{-1}$.
Assuming a point-source geometry for the X-ray emission, the flux at a clump located at a distance $r$ from the X-ray–emitting region (i.e., the center of W49B) is scaled as $F_{\rm clump}(E)$ = $F(E) \times (D/r)^2$, where $D$=7.9~kpc is the distance to W49B. 
Using the photoelectric absorption cross-section $\sigma_{\rm pa}$ from \cite{1983ApJ...270..119M}, the local X-ray energy deposition rate per H nucleus, $H_{\rm X}$, at an attenuating column density, $N_{\rm H}^{\rm att}$, is calculated following \cite{1996ApJ...466..561M}:
\begin{equation}
H_{\rm X} = \int^{E_{\rm max}}_{E_{\rm min}} \sigma_{\rm pa}(E)\,F_{\rm clump}(E)\,\exp\!\left[-\sigma_{\rm pa}(E)\,N_{\rm H}^{\rm att}\right] \mathrm{d}E,
\end{equation}
The corresponding ionization rate per H nucleus $\zeta_{\rm X}$
can be derived using Eq.~(A10) in \cite{1996ApJ...466..561M}\footnote{For comparisons with previous studies, we note that throughout this work, $\zeta_{\rm X}$ is defined per H nucleus, whereas $\zeta_{\rm CR}$ is given per H$_2$ molecule. Assuming fully molecular gas, $\zeta{\rm X}$ should therefore be multiplied by a factor of two when compared with $\zeta_{\rm CR}$, that is, $\zeta({\rm H_2})\simeq2\zeta({\rm H})$.}:
\begin{equation}
\zeta_{\rm X} \simeq 1.83 \times 10^{10} \frac{f_{\rm i}}{0.4} H_{\rm X},
\end{equation}
where $f_{\rm i} \sim 0.4$ is the fraction of the primary photoelectron energy that goes into ionization.



\begin{figure}[htbp]
\centering
\includegraphics[width=0.49\textwidth]{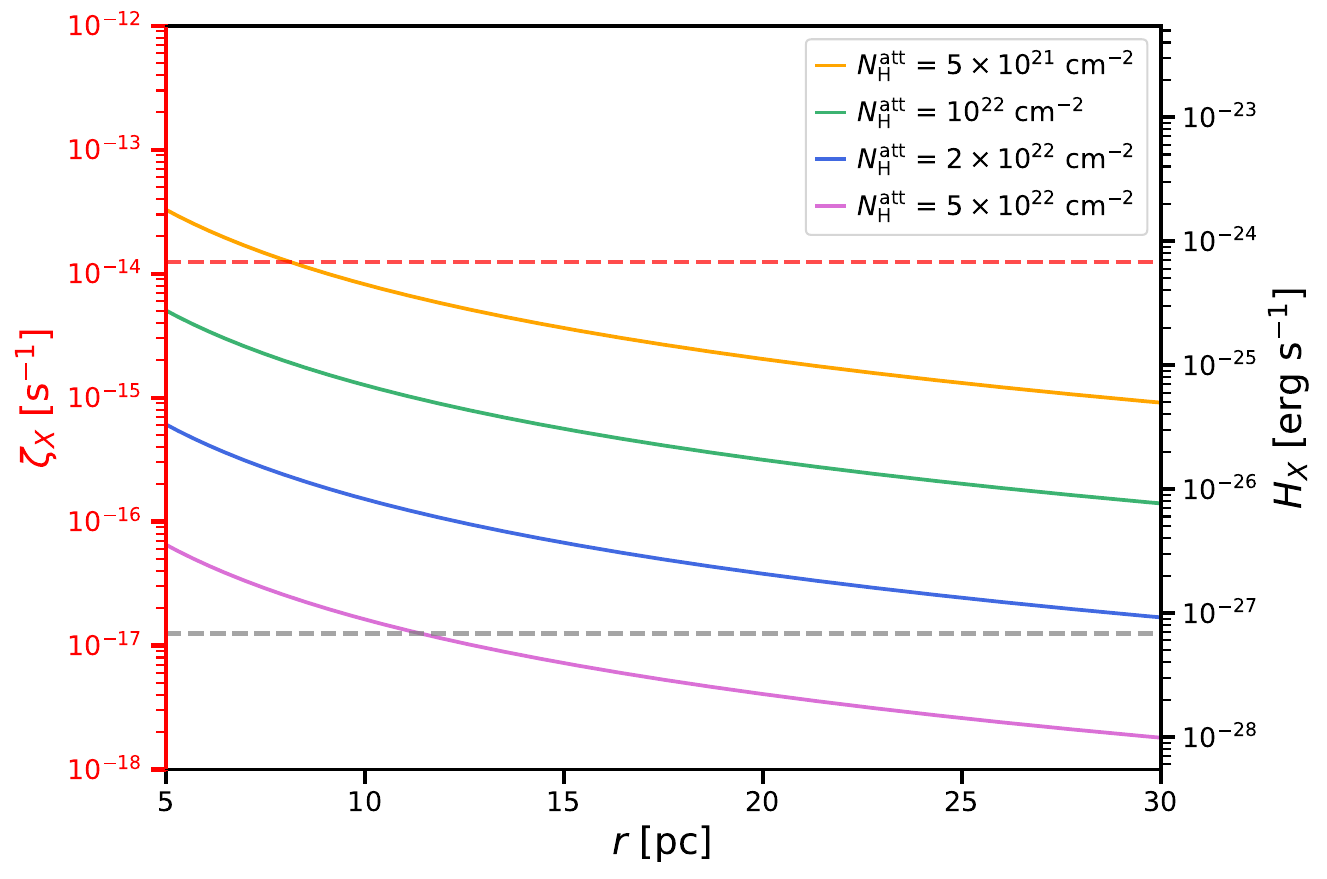}
\caption{ X-ray ionization rate per H nucleus, $\zeta_{\rm X}$, and X-ray energy deposition rate per H nucleus, $H_{\rm X}$, as functions of distance from the center of W49B for different attenuating hydrogen column densities, $N_{\rm H}^{\rm att}$. The left and right axes show $\zeta_{\rm X}$ and $H_{\rm X}$, respectively. The gray and red dashed lines mark the standard Galactic CRIR per H nucleus and a value enhanced by a factor of 1000, respectively.
\label{fig:X-ray_IR}}
\end{figure}

Figure~\ref{fig:X-ray_IR} shows that the X-ray ionization rate $\zeta_{\rm X}$ decreases with increasing attenuating column density $N_{\rm H}^{\rm att}$ and distance from SNR.
Given the radius of W49B is $\sim$2\arcmin, corresponding to 4.6~pc at a distance of 7.9~kpc, $r$ = 5~pc can be regarded as closely located on the boundary of W49B. 
According to X-ray dominated region (XDR) models \citep{2005A&A...436..397M,2022ARA&A..60..247W}, CO can be efficiently destroyed in XDR gas, especially at column densities of $N_{\rm H} < 10^{22}$~cm$^{-2}$. 
The CO and \hco\, detections imply that these molecular emitting regions are sufficiently shielded from the intense X-ray flux. 
Consequently, the attenuating column density $N_{\rm H}^{\rm att}$ between the X-ray source and the molecular clumps is inferred to be at least a few $10^{22}$ cm$^{-2}$. Even at the SNR boundary with $N_{\rm H}^{\rm att} = 10^{22}$~cm$^{-2}$, $\zeta_{\rm X}$ remains below 5$\times 10^{-15}$~s$^{-1}$ and decreases further at higher $N_{\rm H}^{\rm att}$.  
The contribution from X-ray ionization is therefore at most $\sim40\%$ of the CRIRs inferred from the molecular abundance ratios. X-ray ionization is thus likely to partially influence the molecular chemistry but does not play a dominant role. Instead, the unusually high HCO$^+$/CO abundance ratios in the shocked gas of W49B are most naturally explained by enhanced cosmic-ray ionization.





\subsection{
Velocity-dependent \texorpdfstring{\hco/CO}{HCO+/CO} intensity ratio in line wings}\label{Sec:dis_CR_diff_length}

\begin{figure*}[htbp]
\centering
\includegraphics[width=0.442\textwidth]{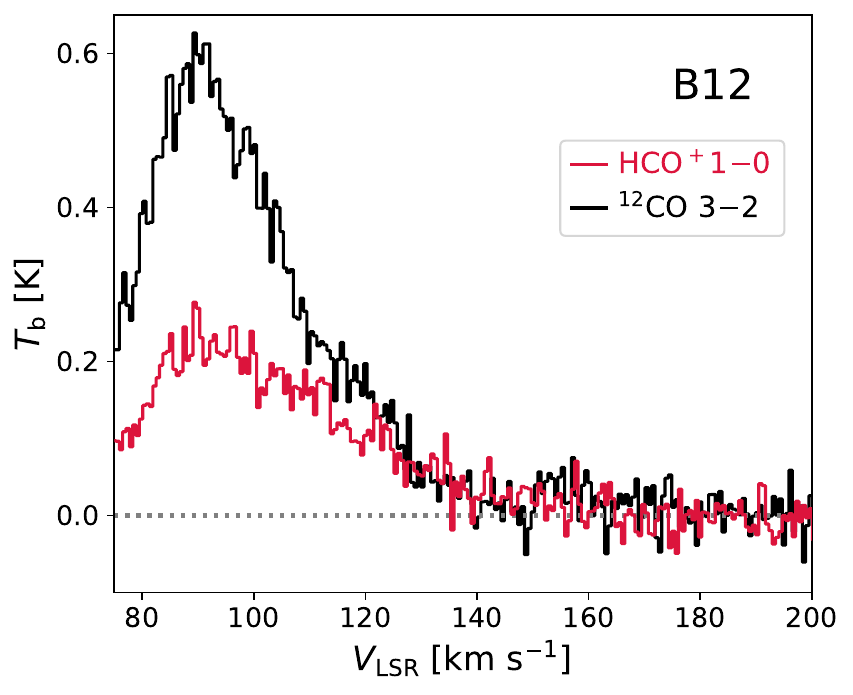}
\includegraphics[width=0.52\textwidth]{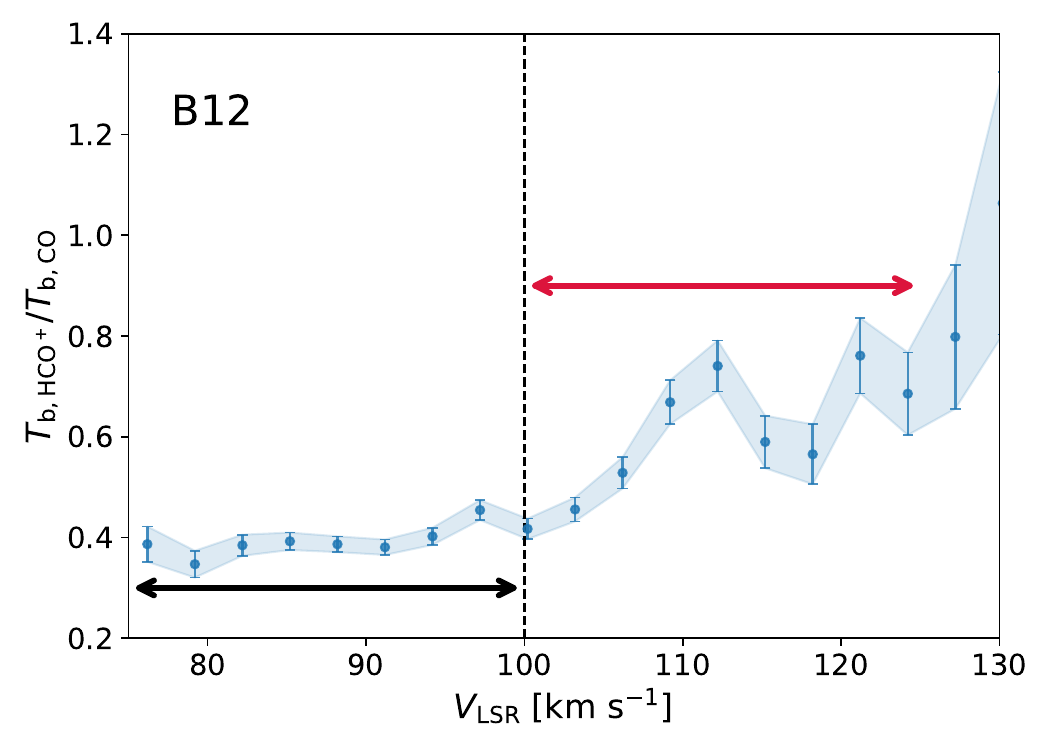}
\caption{(Left) Spectra of \hco~$J$=1--0 and \co~$J$=3--2 in region B12, both convolved to a circular beam of 5$\rlap{.}\arcsec$7.
(Right) The corresponding channel-by-channel \hco/\co\ intensity ratios versus $V_{\rm LSR}$, after binning over five channels.
The vertical dashed line marks $V_{\rm LSR}$=100~\kms, and the two double-headed arrows indicate the velocity intervals used for the non-LTE analysis: 75--100~\kms\ and 100--125~\kms.
\label{fig:line-ratio_channel}}
\end{figure*}

The current observations do not resolve spatial differences between the clumpy shocked \hco\ and \co\ emission. 
Nevertheless, the line profiles can be used to investigate variations among different kinematic components. 
Figure~\ref{fig:west_broad_line-ratio} shows the evident increase in the \hco~$J$=1--0/\co~$J$=3--2 intensity ratio toward the high-velocity line wings in regions B7 and B12.
To minimize the possible effects of beam dilution, we convolved the original data cubes to a common circular beam of 5$\rlap{.}\arcsec$7 for the two transitions.
The increase toward the high-velocity wings remains evident only in regions B7 and B12.
Figure~\ref{fig:line-ratio_channel} shows the spectra and intensity ratio as a function of $V_{\rm LSR}$ for region B12 (5$\rlap{.}\arcsec$7 resolution). The intensity ratio remains nearly constant between +75 and +100~\kms, but increases from +100 to +125~\kms.

To assess whether variations in excitation conditions of gas can account for the change in the \hco~$J$=1--0/\co~$J$=3--2 line ratio from +75 to +100~\kms\ and from +100 to +125~\kms, we applied the same RADEX modeling and MCMC fitting procedure described in Sect.~\ref{Sec:dis_phy_conditions}. 
The RADEX setups, including the geometry, linewidth, beam-filling factor, and model grids, are identical to those described in Sect.~\ref{Sec:dis_phy_conditions}. The observational inputs are the integrated intensities of the three \hco\ and two \co\ transitions measured over the two velocity intervals in region B12 (8$\rlap{.}\arcsec$5 resolution).

Figure~\ref{fig:west_B12_RADEX+MCMC} presents the resulting posterior distributions from the MCMC modeling for the two velocity intervals. The inferred temperatures and densities of the gas in the two velocity intervals are comparable. 
In contrast, the beam-averaged $N_{\rm HCO^+}$/$N_{\rm CO}$ ratio increases from (5.8$\pm$0.4)$\times$10$^{-4}$ in +75 to +100~\kms\ interval to (8.5$\pm$0.8)$\times$10$^{-4}$ in +100 to +125~\kms\ interval, corresponding to an enhancement of approximately 50\%.
Both column densities decrease toward higher velocities, but $N_{\rm CO}$ decreases more rapidly than  $N_{\rm HCO^+}$. These results suggest that the velocity dependence of the \hco/\co\ intensity ratio in the line wings is unlikely to be driven primarily by excitation effects, but instead reflects an enhanced column-density ratio in the higher-velocity gas.

One possible interpretation of the elevated \hco/\co\ intensity ratios in the high-velocity wings is that broad line profiles arise from multiple shocks propagating at different velocities within the beam. 
Since real MCs comprise a range of densities, the inhomogeneity naturally gives rise to a distribution of shock propagation velocities.
The higher-velocity emission preferentially trace gas affected by faster shocks, the elevated \hco/\co\ ratios in the line wings are consistent with the increase in $N_{\rm HCO^+}/N_{\rm CO}$ at $V_{\rm shock}>30$~\kms\ under high-CRIR conditions in the J-shock models (Fig.~\ref{fig:west_shock_mhd_code}).
In addition, a spatially inhomogeneous CRIR provides another, not mutually exclusive, interpretation. Different mixtures of these components may contribute to the velocity dependence of the \hco/\co\ ratio.

For the other regions, such as B11, there is no clear evidence that the line ratios vary systematically with velocity (Fig.~\ref{fig:west_broad_line-ratio}). 
The present data therefore do not require a velocity-dependent CRIR, and a single CRIR provides an adequate description of the beam-averaged abundance ratio.

\subsection{Remarks on ionizing cosmic rays}\label{Sec:CR_diff}

Given the absence of clear velocity-dependent variations in the line ratios toward region B11, we consider a common CRIR for all gas components to be compatible with the observations in this region.
In the Paris–Durham shock models, the \hco/\co\ abundance ratios vary across a thin post-shock layer, over a characteristic scale of $l_{\rm layer}\sim10^{16}$~cm (i.e., 0.003~pc). This scale is substantially smaller than the spatial resolution of the present observations. 
Although the layer is spatially unresolved, its predicted thickness provides a characteristic scale for constraining CR transport. We therefore explore the propagation of low-energy CRs through the post-shock layer.

Cosmic rays are believed to be accelerated by SNR shocks, gaining energy through repeated crossings of the shock front \citep{1983RPPh...46..973D}. 
Fresh acceleration of CRs at the local shock front is possible, and alternatively, the ionization may be supplied by a more spatially extended SNR-associated CR population, including particles accelerated during earlier stages of the remnant's evolution and subsequently retained within the SNR before entering the molecular gas \citep[e.g.,][]{2019PASJ...71...78M}.
While, irrespective of CR origin, low-energy CRs experience rapid ionization which could be particularly important at MeV to sub-GeV energies, limiting their lifetimes and penetration depths \citep{1996A&A...309.1002O, padovani09,2018MNRAS.480.5167P}.
 Studies of SNR-MC interactions further suggest that MeV CR protons can remain confined near the shock front before entering adjacent molecular clouds \citep{2019PASJ...71...78M,2019PASJ...71..115N}.

An order-of-magnitude estimation of the timescale of proton energy loss $t_{\rm loss}$ is given by,
\begin{equation}
 t_{\rm loss}(E)\sim
 \frac{E}
 {L_{\rm p}(E)n_{\rm H_2}v(E)},
\end{equation}
where $E$ is the proton kinetic energy, $L_{\rm p}(E)$ is the proton energy loss per unit molecular-hydrogen column density \citep[see Eq.~20 and Fig.~7 in][]{padovani09}.
The velocity of the CR proton is $v(E)=c\left[1-\left(1+\frac{E}{m_{\rm p}c^2}\right)^{-2}\right]^{1/2}$, where $m_{\rm p}$ is the proton rest mass and $c$ is the speed of light. For the gas density inferred from region B11, $n_{\rm H_2}=
6.3\times 10^4\,\rm cm^{-3}$, we find that $t_{\rm loss}
\sim 1.2\,$yr for 10\,MeV protons and $\sim$ 39\,yr for 100\,MeV protons.
Taking the shock age of 1~kyr as a fiducial proxy for the duration of enhanced CR illumination, \(t_{\rm CR}\), both energy-loss timescales are substantially shorter than \(t_{\rm CR}\).
This implies that individual low-energy protons cannot retain their initial energies for a significant fraction of the shock–cloud interaction time. A sustained ionizing population would therefore require continuous replenishment.

We define the effective transport time of CR protons in the molecular gas as
\(t_{\rm eff}(E)=\min \left[t_{\rm CR},t_{\rm loss}(E)\right].\)
Given $t_{\rm loss}(E) \ll t_{\rm CR}$ for the ionizing low-energy protons considered here, the effective transport time is determined by energy losses, i.e., \(t_{\rm eff}(E)\simeq t_{\rm loss}(E)\).
The characteristic advection length $l_{\rm adv}(E)$ can be given by \(l_{\rm adv}(E)=u_{\rm adv}t_{\rm eff}(E)\), where \(u_{\rm adv}=V_{\rm shock}/r\) is the velocity of the downstream particles advected way from the shock, and \(r\) is the shock compression ratio \citep[e.g.,][]{2006A&A...453..387P}. 
For an order-of-magnitude estimation, we adopt \(u_{\rm adv}\sim10~{\rm km\,s^{-1}}\) and \(t_{\rm eff}(E)\) ranging from approximately one to a few tens of years, yielding \(l_{\rm adv}(E)\sim10^{-5} - 10^{-4}\)~pc. This suggests that the particles are unlikely to traverse the entire post-shock layer by advection alone.

 We therefore consider a scenario in which ionizing CRs enter the post-shock layer at or near one of its boundaries and are subsequently transported across the layer predominantly by diffusion, without necessarily being freshly accelerated at the current shock front.
The characteristic diffusion length can be estimated as $l_{\rm diff} \sim  \sqrt{D(E)t_{\rm eff}(E)}$, where $D(E)$ is the low-energy CR diffusion coefficient, which describes transport through scattering by magnetic turbulence \citep[e.g.,][]{2006A&A...453..387P,2012A&ARv..20...49V,2014A&A...568A..50V}. 
For diffusion to distribute the protons across the full thickness of the post-shock layer, we require $l_{\rm diff}> 10^{16}$~cm.
Combining the above estimation of $t_{\rm eff}$, we may obtain the following approximate lower limits on the diffusion coefficients of 10 and 100~MeV CR protons:

\begingroup
\fontsize{8.5pt}{10pt}\selectfont
\begin{align}
D(10\,{\rm MeV}) &>
2.8\times10^{24}
\left(\frac{l_{\rm diff}}{10^{16}{\rm cm}}\right)^2
\left(\frac{n({\rm H_2})}
{6.3\times10^4{\rm cm^{-3}}}\right)
{\rm cm^2\,s^{-1}},\\
D(100\,{\rm MeV}) &>
8.0\times10^{22}
\left(\frac{l_{\rm diff}}{10^{16}{\rm cm}}\right)^2
\left(\frac{n({\rm H_2})}
{6.3\times10^4{\rm cm^{-3}}}\right)
{\rm cm^2\,s^{-1}}.
\end{align}
\endgroup

The above estimate assumes diffusive transport, but the diffusion coefficient may itself be regulated by CR--wave interactions. If CRs stream along the magnetic field faster than the local Alfv\'en speed, their anisotropic distribution can excite resonant Alfv\'en waves through the CR streaming instability \citep{Skilling1975,Lucek2000}. Scattering by these self-generated waves can reduce the CR drift speed and suppress the effective parallel diffusion coefficient. In the strong self-confinement limit, the net CR drift approaches the Alfv\'en speed. For the $10-100\,$MeV ionizing CRs, particularly $\sim10$ MeV protons, such Alfv\'enic transport would cover a distance much smaller than $l_{\rm layer}$ during their short ionization-loss time. Therefore, under the interpretation that a nearly uniform CRIR is considered across this layer, very strong self-confinement of the lowest-energy CRs would be disfavored. Indeed, in partially ionized shocked molecular gas, ion--neutral collisions can damp the self-generated Alfv\'en waves, weakening CR self-confinement and allowing a larger effective parallel diffusion coefficient \citep{Morlino2015,Plotnikov2021}.
So inclusion of such transport would not significantly alter the diffusion-dominated lower limit derived above. 


\subsection{Energy partition among feedback channels}\label{Sec:dis_energy}

To compare the contributions of different feedback channels (i.e., shocks, CRs, and X-ray radiation), we estimate the energy budget of a representative shock clump selected from region B11 in our ALMA observations.
The physical properties of the clump are derived from data at an angular resolution of $8\rlap{.}\arcsec5$ (Sects.~\ref{Sec:dis_phy_conditions} and \ref{Sec:dis_crir}), corresponding to a linear scale of 0.33~pc. As the shocked clumps are unresolved at this resolution, we adopt this scale as a characteristic clump diameter, which should be regarded as an upper limit. Assuming a spherical geometry, this gives a clump volume of $V_{\rm clump}\sim5.5\times10^{53}$~cm$^{3}$.


Cosmic ray ionization can also heat molecular gas. The heating rate can be expressed as $\Gamma_{\rm CR}=\zeta_{\rm CR}n_{\rm H_2} \Delta Q_{\rm cr}$ \citep{1978ApJ...222..881G}, where $\Delta Q_{\rm cr}$ is the mean heat per ionization event\footnote{$\Delta Q_{\rm cr}$ varies with CR energy (17--26 eV for 10--100~MeV protons), as well as gas composition and density \citep{1978ApJ...222..881G,2012ApJ...756..157G}.}, for which we use $\Delta Q_{\rm cr}$=13~eV. Adopting $\zeta_{\rm CR} = 10^{-13}$~s$^{-1}$ yields a heating rate of $\Gamma_{\rm CR} = 1.3\times10^{-19}$~erg~cm$^{-3}$~s$^{-1}$.
The cumulative CR-deposited energy is $E_{\rm CR} = \Gamma_{\rm CR} V_{\rm clump} t_{\rm CR} \sim 2.3\times10^{45}(t_{\rm CR}/1~{\rm kyr})$~erg, where $t_{\rm CR}$ is duration of exposure to the enhanced CR field.
Since the enhanced CRIR could be closely related to CR acceleration by the SNR shocks, we use the adopted shock age of 1~kyr as a fiducial approximation for $t_{\rm CR}$, yielding $E_{\rm CR}\simeq2.3\times10^{45}$~erg. 
This estimate applies primarily to low-energy CRs, which dominate ionization and heating in molecular gas. In contrast, $>$GeV-energy protons have relatively small cross sections and can hardly produce efficiently MC heating \citep{padovani09,gabici22}.

To estimate the X-ray heating rate per unit volume, we adopt $\Gamma_{\rm X} \approx 3 \times 10^4 (f_{\rm h}/0.3) (n_{\rm H}/10^5~{\rm cm^{-3}}) H_{\rm X}$ \citep[see Eq.~A11 in][]{1996ApJ...466..561M}, where $f_{\rm h} \sim 0.3$ represents the fraction of energy converted into heat.
We adopt the upper limit of $H_{\rm X}$ by assuming that the clump is located at the SNR boundary with $N_{\rm H}^{\rm att} = 10^{22}$~cm$^{-2}$, corresponding to $\zeta_{\rm X} \sim 5\times10^{-15}$~s$^{-1}$. This yields a heating rate of $\Gamma_{\rm X} = 1.0\times10^{-20}$~erg~cm$^{-3}$~s$^{-1}$.
The cumulative X-ray heating energy is $E_{\rm X} = \Gamma_{\rm X} V_{\rm clump} t_{\rm X} \sim 1.7\times10^{44} (t_{\rm X}/1~{\rm kyr})$~erg, where $t_{\rm X}$ is exposure duration to the X-ray radiation field. 
The X-ray emission from W49B arises from shock-heated plasma and is therefore linked to the evolution of the SNR shocks \citep{2006A&A...453..567M,2020ApJ...893...90S}. 
However, the duration of X-ray irradiation of an individual clump does not necessarily coincide with its local shock age. In the absence of an independent constraint on \(t_{\rm X}\), we adopt the local shock age of 1~kyr as an order-of-magnitude estimate of $t_{\rm X}$, yielding $E_{\rm X} \sim 1.7\times10^{44}$~erg.


For the mechanical input from shocks, we estimate the kinetic energy associated with the bulk motion of the clump as
$E_{\rm kin}\sim \frac{1}{2}M_{\rm clump}V_{\rm exp}^2$, where $M_{\rm clump}=\mu m_{\rm H} n_{\rm H_2} V_{\rm clump}$, $\mu$ =2.8 is the mean molecular weight, $m_{\rm H}$ is the mass of a hydrogen atom. Here $V_{\rm exp} = 40$~km~s$^{-1}$ is the outward bulk velocity of the clump, estimated from the velocity offset between the shocked gas centroid velocity (about +100~\kms) and the systemic velocity (about +60~\kms). This yields $E_{\rm kin}$ $\sim 1.3\times10^{48}~{\rm erg}$.

The above estimates indicate that the bulk kinetic-energy reservoir associated with the shocks substantially exceeds the time-integrated energy inputs from CR and X-ray heating. 
Considering $E_{\rm CR}$ and $E_{\rm X}$ depend on different and uncertain exposure durations, the resulting ratio, $E_{\rm kin}:E_{\rm CR}:E_{\rm X} \sim 10^4:10:1$, provides an illustrative comparison.
Because our analysis focuses on compact shocked gas, it may preferentially emphasize the mechanical contribution. Furthermore, the shock term represents the current kinetic-energy reservoir, whereas the CR and X-ray terms are time-integrated heating inputs. Their comparison should therefore not be interpreted as a direct energy partition over a common timescale.
Nevertheless, shocks dominate the local energy budget, while cosmic rays provide more heating than X-rays.


\section{Summary}\label{Sec:sum}
We performed observations of the \hco~$J$=1--0, 3--2, and 4--3 transitions, as well as the \co~$J$=3--2 line, toward the eastern and western regions of SNR W49B using ALMA, ACA, and APEX. 

Toward W49B east, the \hco~$J=1$--0 emission reveals a compact blob-like component, together with several filamentary and clumpy structures located outside the SNR.
These features exhibit narrow line profiles. 
The higher-$J$ \hco\ transitions are significantly weaker, with $J$=3--2 only marginally detected and $J$=4--3 undetected.

Toward W49B west, we conduct a detailed investigation of the shocked gas in the SNR-MC interaction region. The main results are summarized as follows:

\begin{itemize}

\item[1.] We identify 12 regions exhibiting broad \hco\ and \co\ line profiles, with FWHMs of 25--60~km~s$^{-1}$. These emissions show compact morphologies (at a resolution of $\sim$5\arcsec) and are spatially associated with the 2.12~$\mu$m H$_2$ emission, but are generally offset from the [Fe~II] emission and continuum peaks.

\item[2.]  Using three \hco\ and two \co\ transitions, we performed non-LTE analyses to derive the physical conditions for six broad-line regions. 
The shocked gas has densities of (0.5--1.6)$\times$ 10$^{5}$~cm$^{-3}$. The derived $N_{\rm HCO^+}$/$N_{\rm CO}$ abundance ratios $(4.2$--$14.5) \times10^{-4}$ are remarkably higher than those reported for other SNRs. 

\item[3.]  Comparing the observational results with the MHD shock models, we find that reproducing the high $N_{\rm HCO^+}$/$N_{\rm CO}$ abundance ratios in shocked MCs requires a high CRIR of a few 10$^{-14}$--10$^{-13}$~s$^{-1}$, corresponding to an enhancement of about 3--4 orders of magnitude above the standard Galactic value.
We also analyzed the possible contribution of X-ray photons to the ionization and find that CRs are likely the dominant driver of the chemical enhancement.

\item[4.] We find elevated \hco/\co\ intensity ratios in high-velocity line wings in regions B7 and B12. A non-LTE analysis of the two velocity intervals in B12 indicates comparable excitation conditions, but a $\sim$50\% higher $N_{\rm HCO^+}/N_{\rm CO}$ ratio in the high-velocity gas.
This enhancement may result from multiple unresolved shocks with different velocities arising from the range of gas densities in MCs, an inhomogeneous CRIR, or both.

\item[5.] In region B11, the nearly velocity-independent HCO$^+$/CO line ratios are compatible with a common CRIR across different gas components. The short energy-loss timescales of low-energy protons imply continuous replenishment and suggest that advection alone is insufficient. Under the scenario in which particles are accelerated at the shock front, we therefore explore the diffusion coefficients required for them to propagate across the post-shock layer.

\item[6.]  We further estimate the energy budget within a representative shocked clump. The shock-associated kinetic-energy reservoir substantially exceeds the estimated cumulative CR and X-ray heating inputs, with CR heating exceeding X-ray heating.

\end{itemize}

\begin{acknowledgments}
This paper makes use of the following ALMA data: ADS/JAO.ALMA\#2022.1.01608.S, and ADS/JAO.ALMA\#2018.1.01780.S. ALMA is a partnership of ESO (representing its member states), NSF (USA) and NINS (Japan), together with NRC (Canada), MOST and ASIAA (Taiwan), and KASI (Republic of Korea), in cooperation with the Republic of Chile. The Joint ALMA Observatory is operated by ESO, AUI/NRAO and NAOJ. 
We thank the APEX staff for their assistance with our observations.
APEX is a collaboration between the Max-Planck-Institut f{\"u}r Radioastronomie, the European Southern Observatory, and the Onsala Space Observatory. 
We acknowledge the use of ChatGPT (OpenAI) for assistance with language polishing, and script development.
WY and PZ acknowledge the NSFC funding under grants 12273010, and 12403027, and the Fundamental Research Funds for the Central Universities with grant No.\ KG202502.
WY acknowledges the support from China Postdoctoral Science Foundation (2024M751376), and Jiangsu Funding Program for Excellent Postdoctoral Talent (2024ZB347). 
YG was supported by the Ministry of Science and Technology of China under the National Key R\&D Program (Grant No. 2023YFA1608200), the National Natural Science Foundation of China (Grant No. 12427901), and the Strategic Priority Research Program of the Chinese Academy of Sciences (Grant No. XDB0800301).
HS was supported by the JSPS KAKENHI grant Nos. 21H01136, 24H00246 and 26K21722. HS was also supported by NAOJ ALMA Scientific Research Grant Code 2023-25A.
YC acknowledges the NSFC funding under grants 12573047, 12121003, and 12393852.

\end{acknowledgments}




%
\facilities{ALMA, ACA, APEX}

\software{CASA \citep{2022PASP..134k4501C}, GILDAS/CLASS \citep{2005sf2a.conf..721P},
astropy \citep{2022ApJ...935..167A}, 
matplotlib \citep{Hunter:2007}, 
scipy \citep{jones2001scipy}, and 
numpy \citep{5725236}}.


\clearpage
\appendix
\onecolumngrid

\section{Broad emission toward the W49B west region}\label{sec:appendix-B}

This section presents images, spectra and line parameters of the broad-line emission detected in W49B west.

\restartappendixnumbering
\begin{figure*}[ht!]
\center
\includegraphics[width=0.945\textwidth]{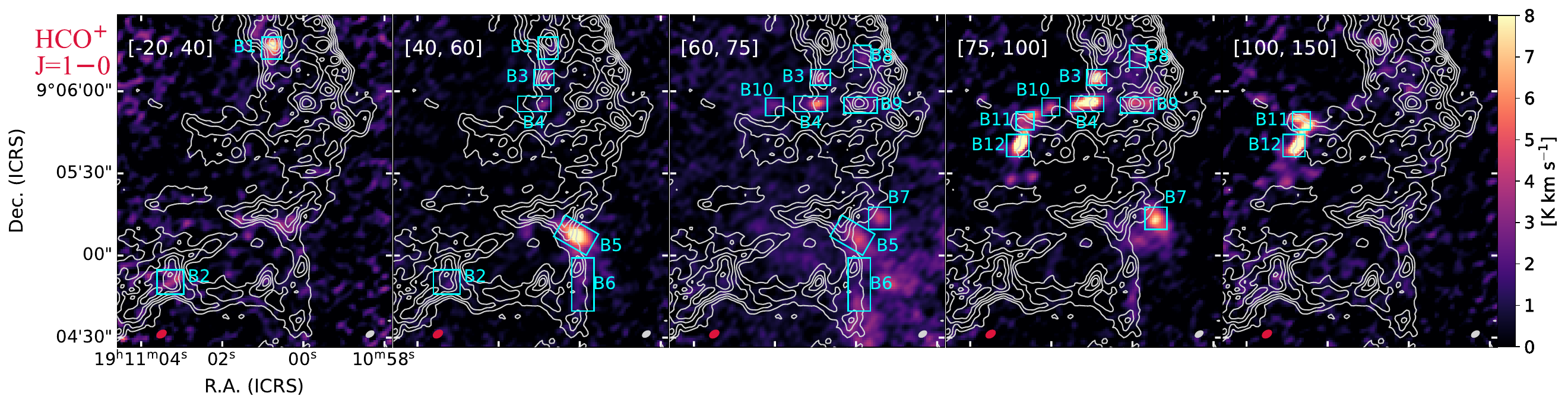}
\includegraphics[width=0.94\textwidth]{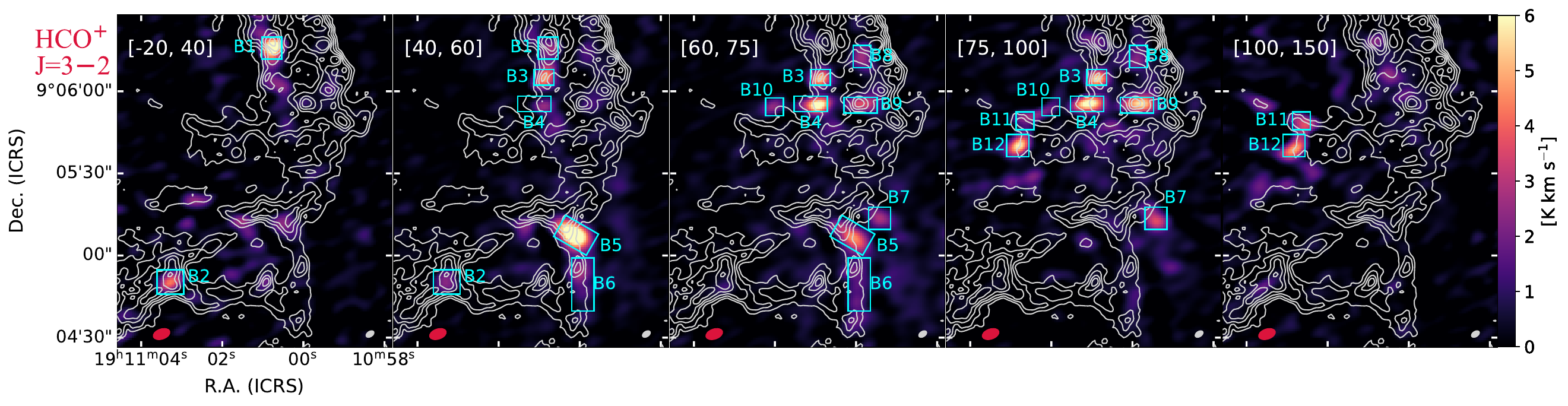}
\includegraphics[width=0.94\textwidth]{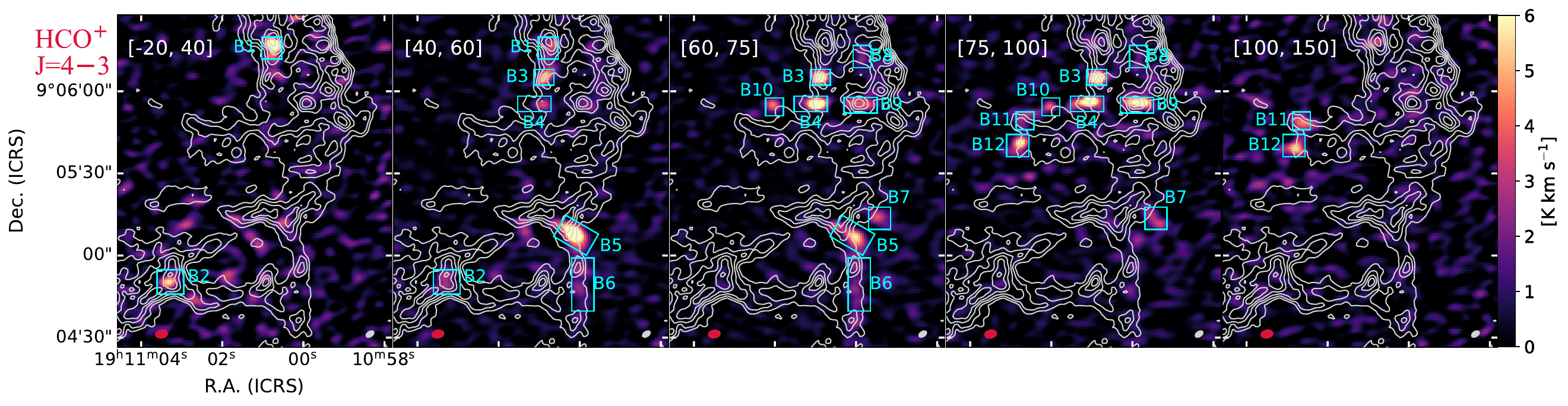}
\includegraphics[width=0.94\textwidth]{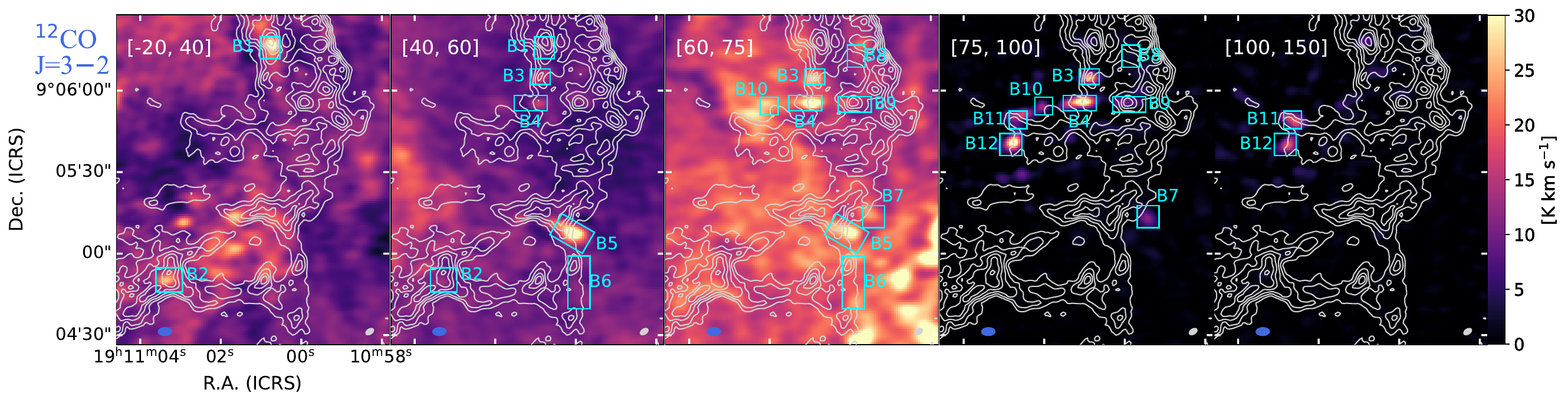}
\caption{From top to bottom, integrated intensity maps of \hco~$J$=1--0, \hco~$J$=3--2, \hco~$J$=4--3 and \co~$J$=3--2 over different velocity intervals, to better characterize the distribution of selected regions.
The synthesized beams, 3~mm continuum contour levels and cyan boxes are identical to those shown in Fig.~\ref{fig:west_inte_map_all}. 
\label{fig:west_inte_map}}
\end{figure*}

\begin{figure*}[htb]
\centering
\includegraphics[width=0.246\textwidth]{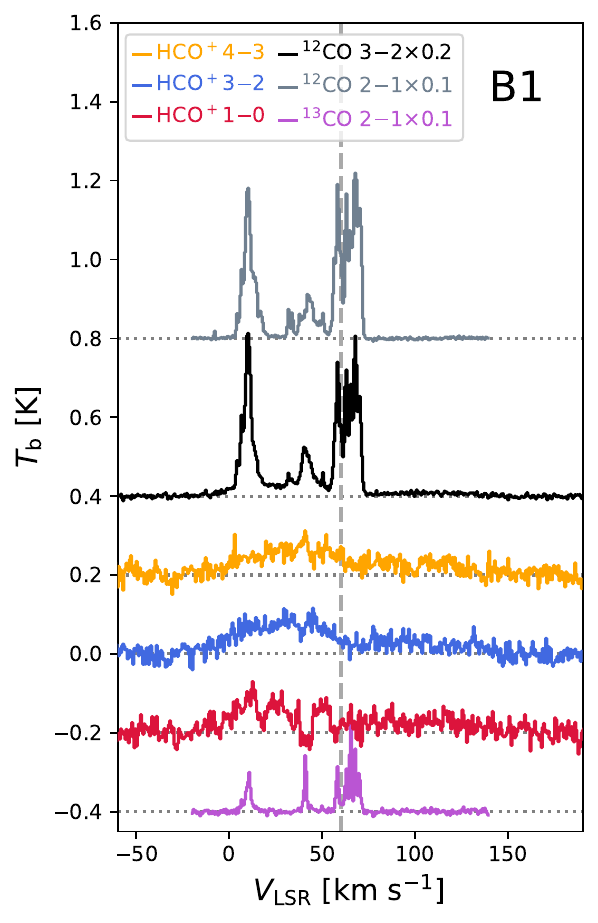}
\includegraphics[width=0.246\textwidth]{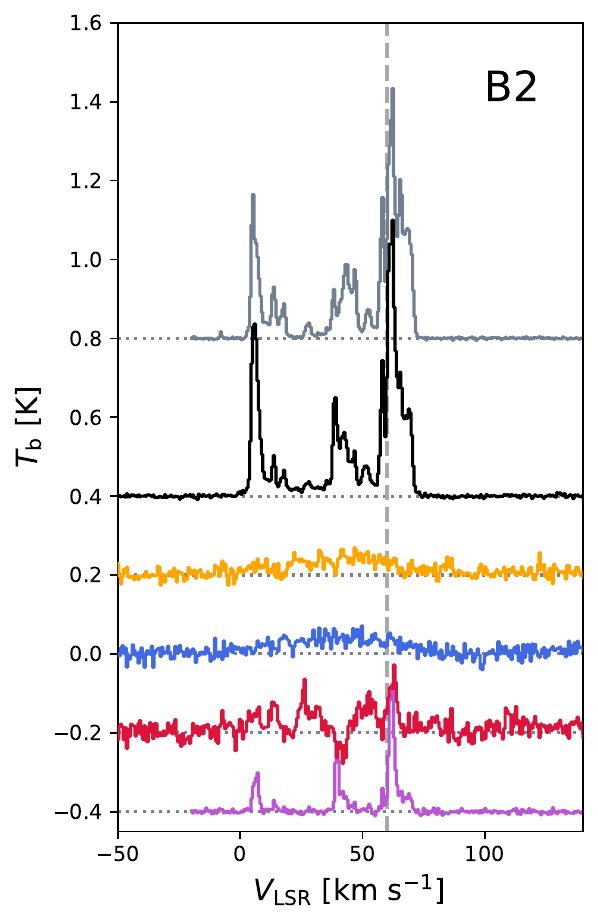}
\includegraphics[width=0.246\textwidth]{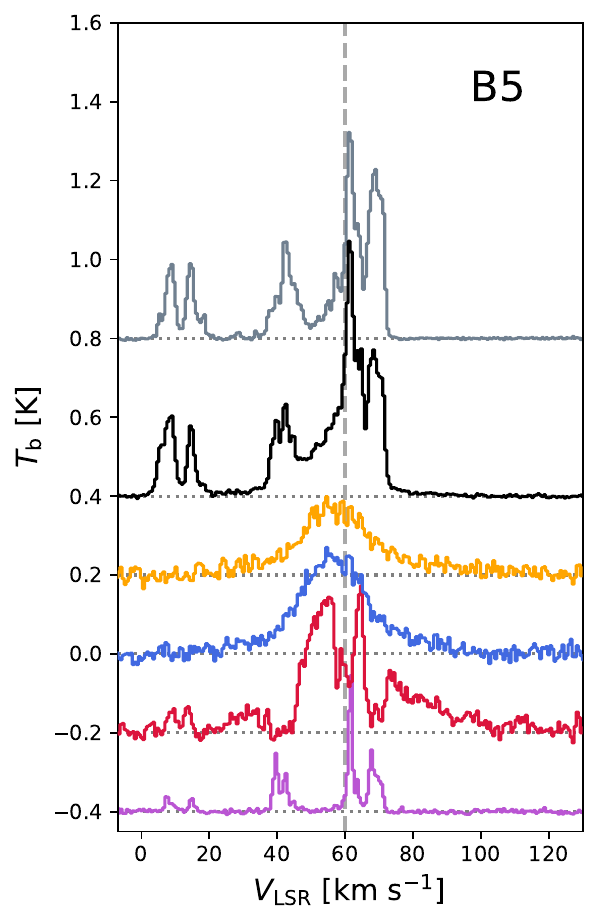}
\includegraphics[width=0.246\textwidth]{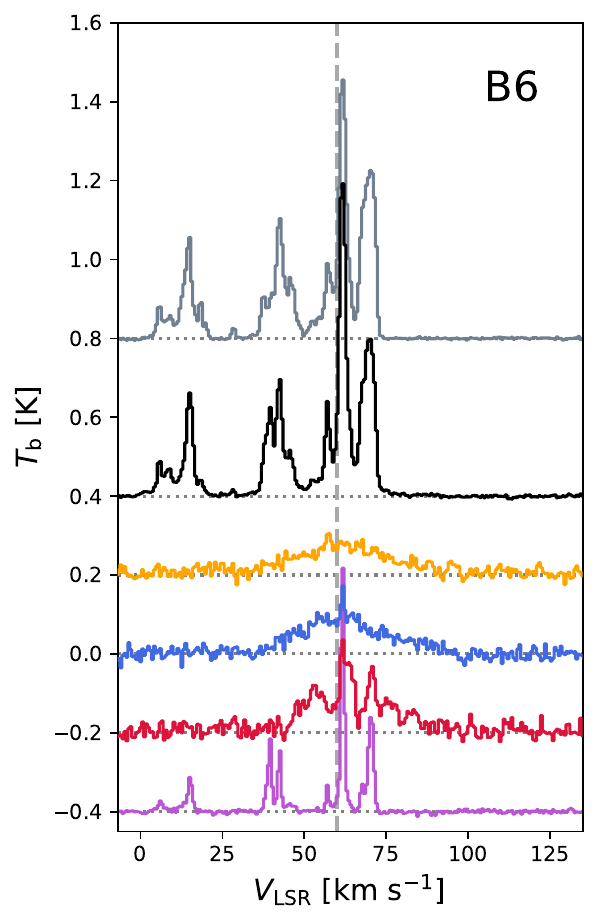}

\vspace{-1mm}
\centering
\includegraphics[width=0.246\textwidth]{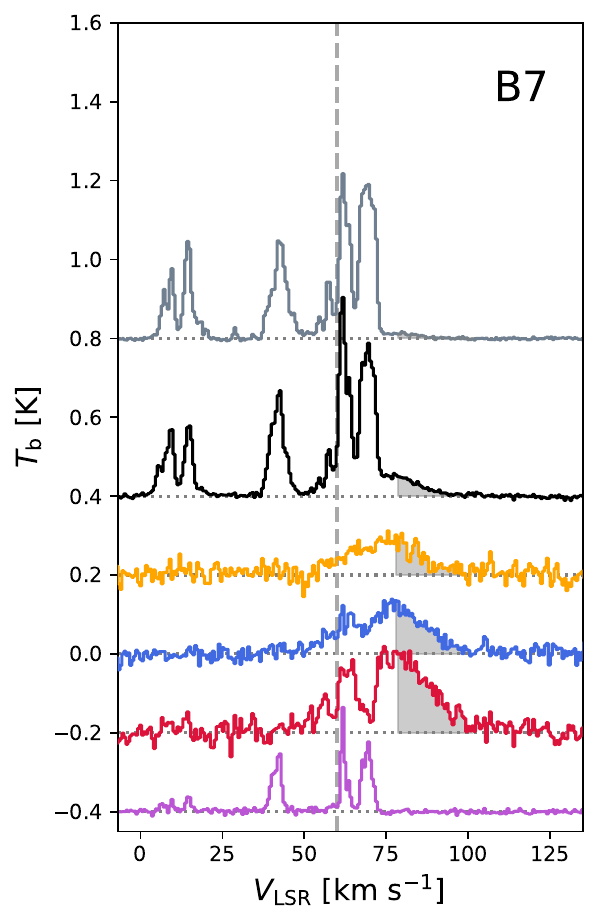}
\includegraphics[width=0.246\textwidth]{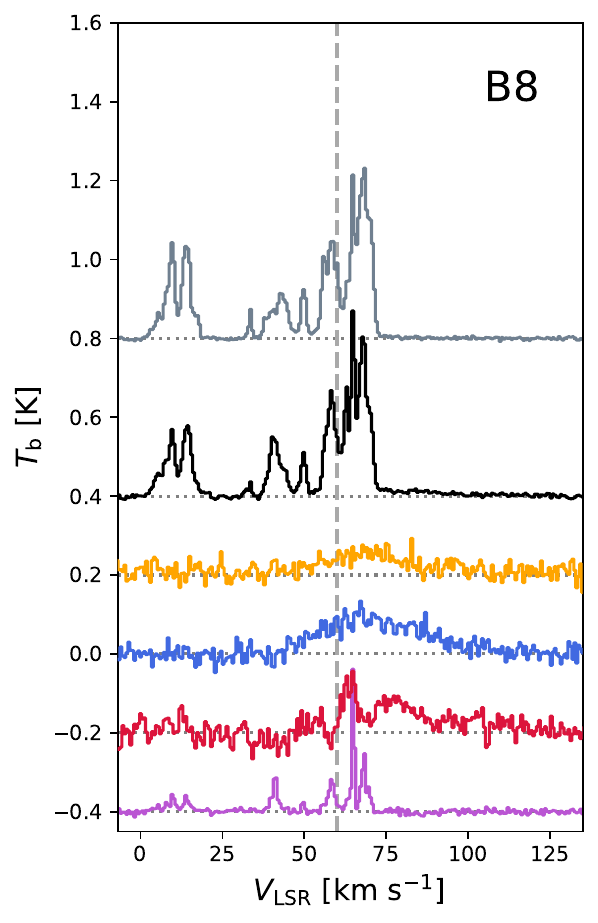}
\includegraphics[width=0.246\textwidth]{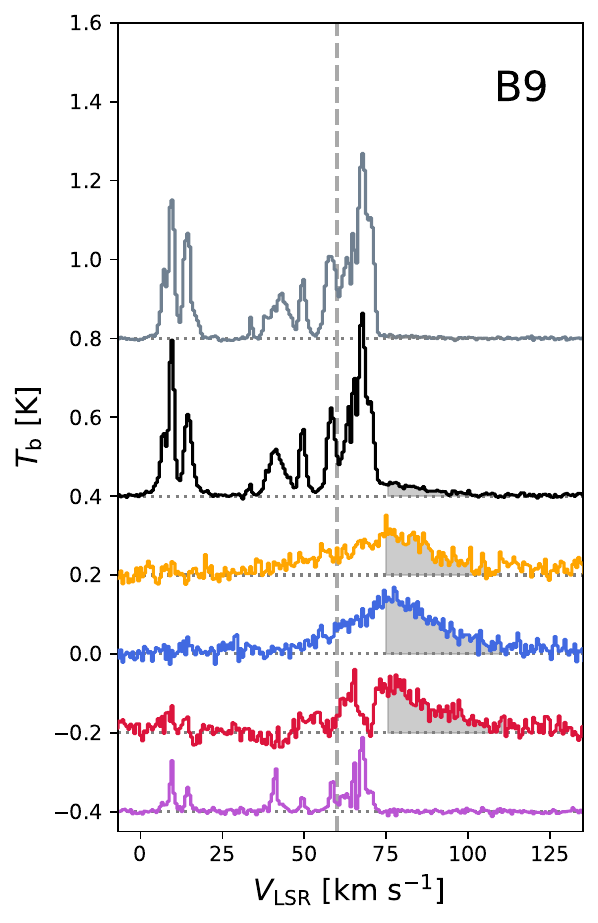}
\includegraphics[width=0.246\textwidth]{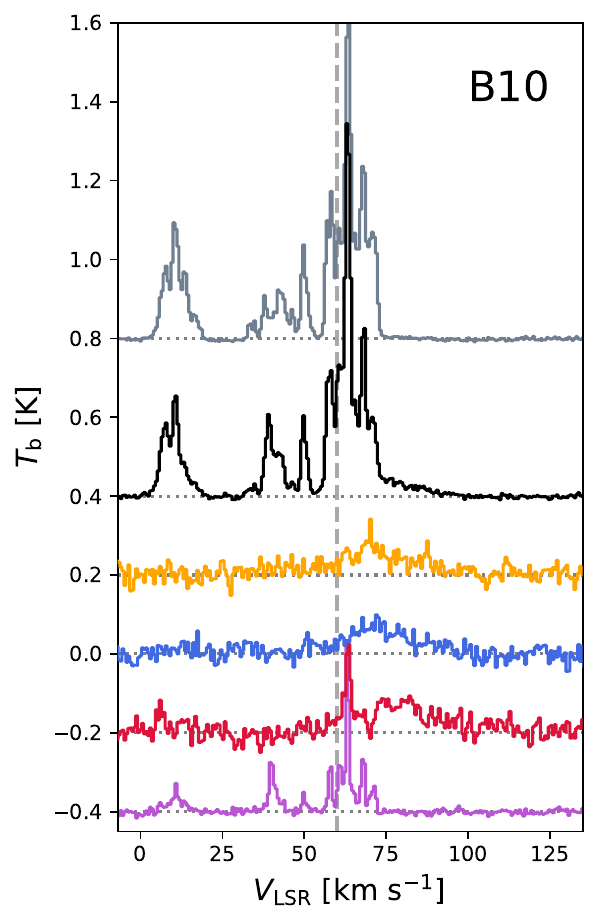}
\caption{Same as Fig.~\ref{fig:west_broad_spec_example}, for the other eight regions showing broad line emission in W49B west (labeled in Fig.~\ref{fig:west_inte_map_all}).
\label{fig:west_broad_spec}}
\end{figure*}

\begin{figure*}[htb]
\centering
\includegraphics[width=0.24\textwidth]{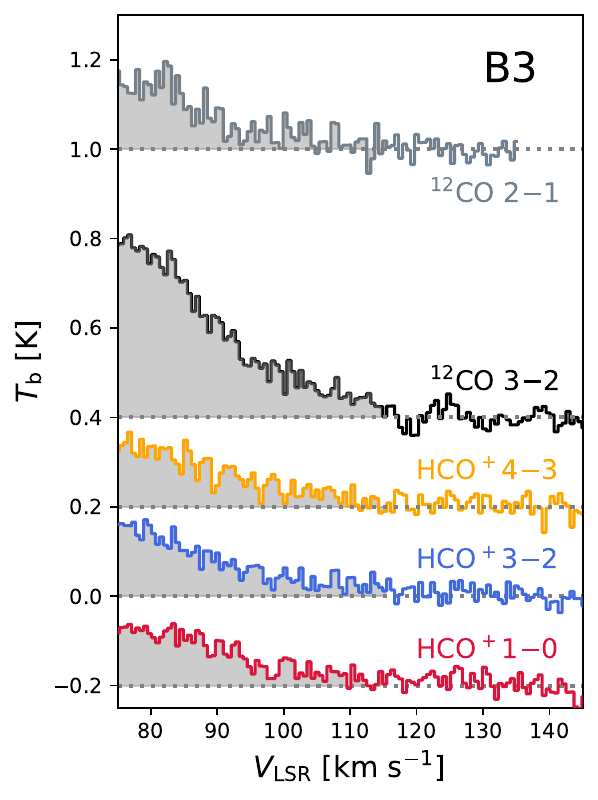}
\includegraphics[width=0.24\textwidth]{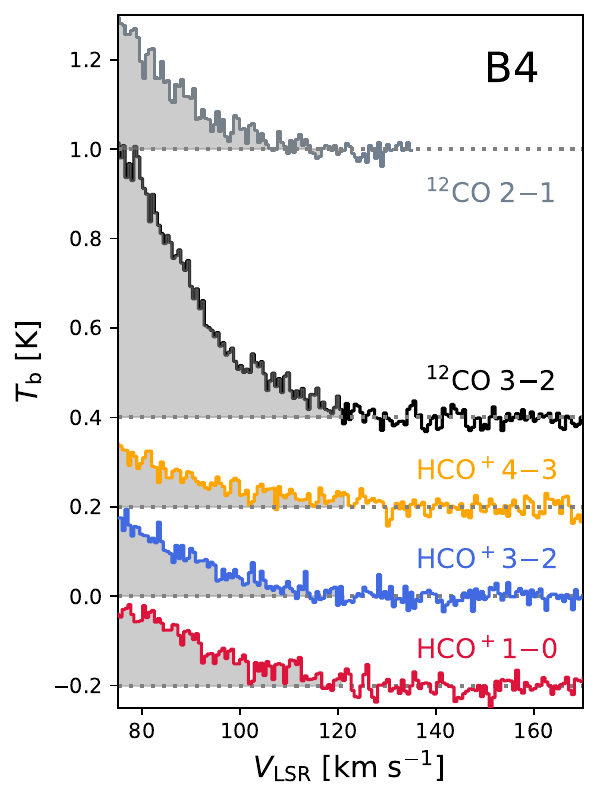}
\includegraphics[width=0.24\textwidth]{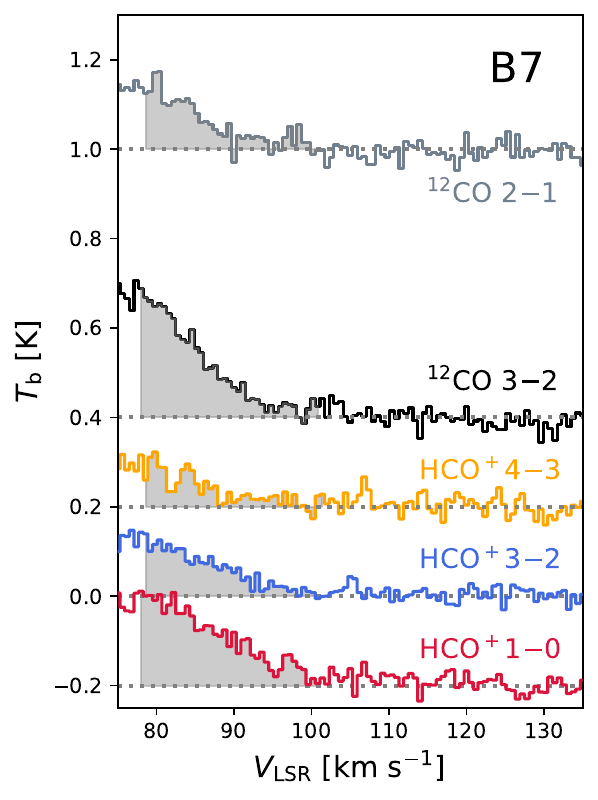}

\includegraphics[width=0.24\textwidth]{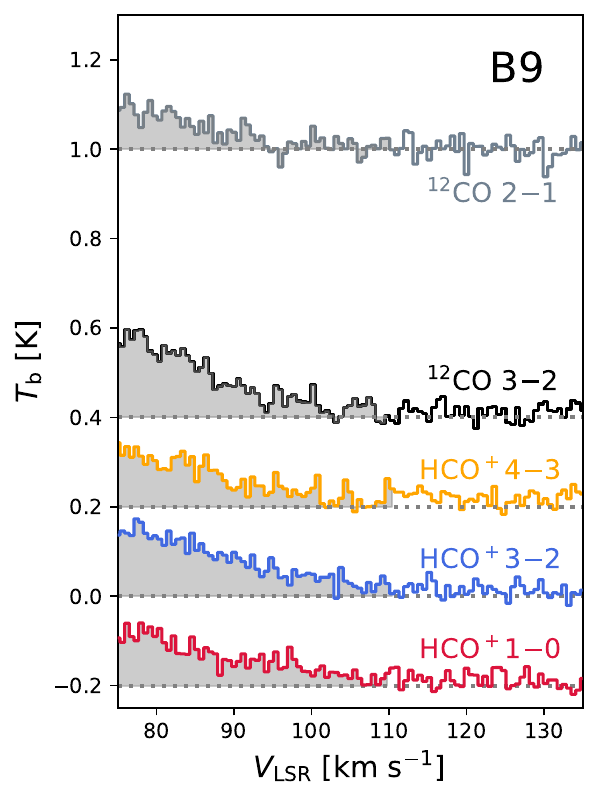}
\includegraphics[width=0.24\textwidth]{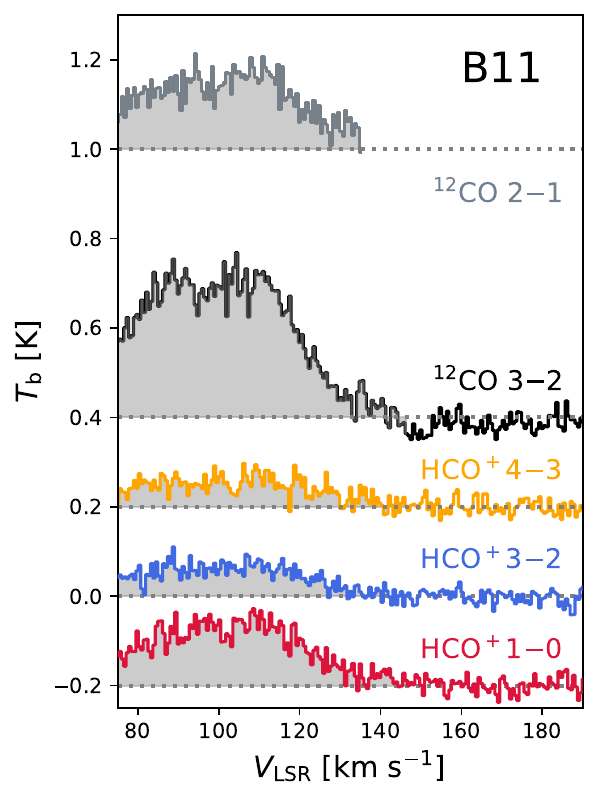}
\includegraphics[width=0.24\textwidth]{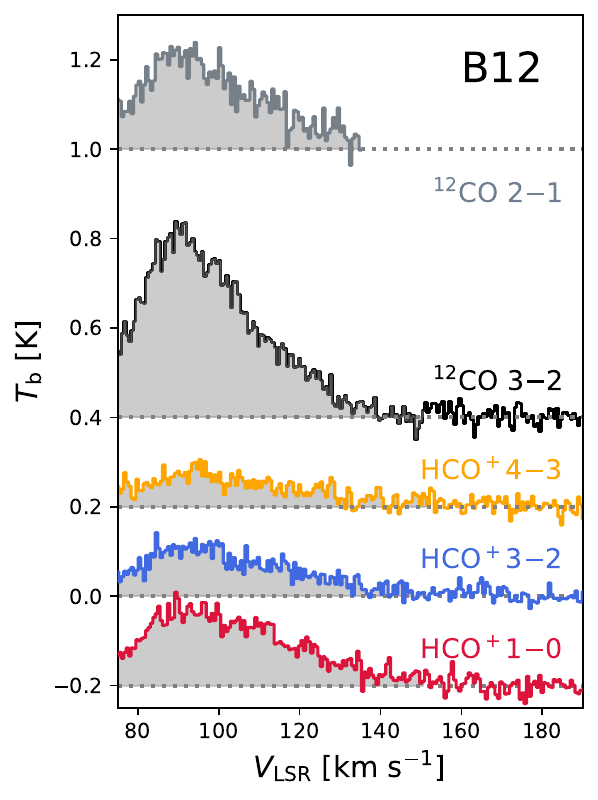}
\caption{Averaged spectra of \hco~$J$=1--0, 3--2, 4--3, and \co~$J$=3--2, 2--1 from the six regions in W49B west (labeled in Fig.~\ref{fig:west_inte_map_all}a). All spectra starting from +75~\kms\, are generally free from narrow-line contamination.
The shaded regions indicate the velocity ranges used to derive the integrated intensities in our subsequent analysis.
All spectra are extracted from data cubes produced using manual masking, with the combination of single-dish data in all cases except for \co~$J$=2--1, and convolved to a circular beam of 8$\rlap{.}\arcsec$5.
\label{fig:west_broad_spec_gt75}}
\end{figure*}

\begin{figure*}[ht!]
\center
\includegraphics[width=1\textwidth]{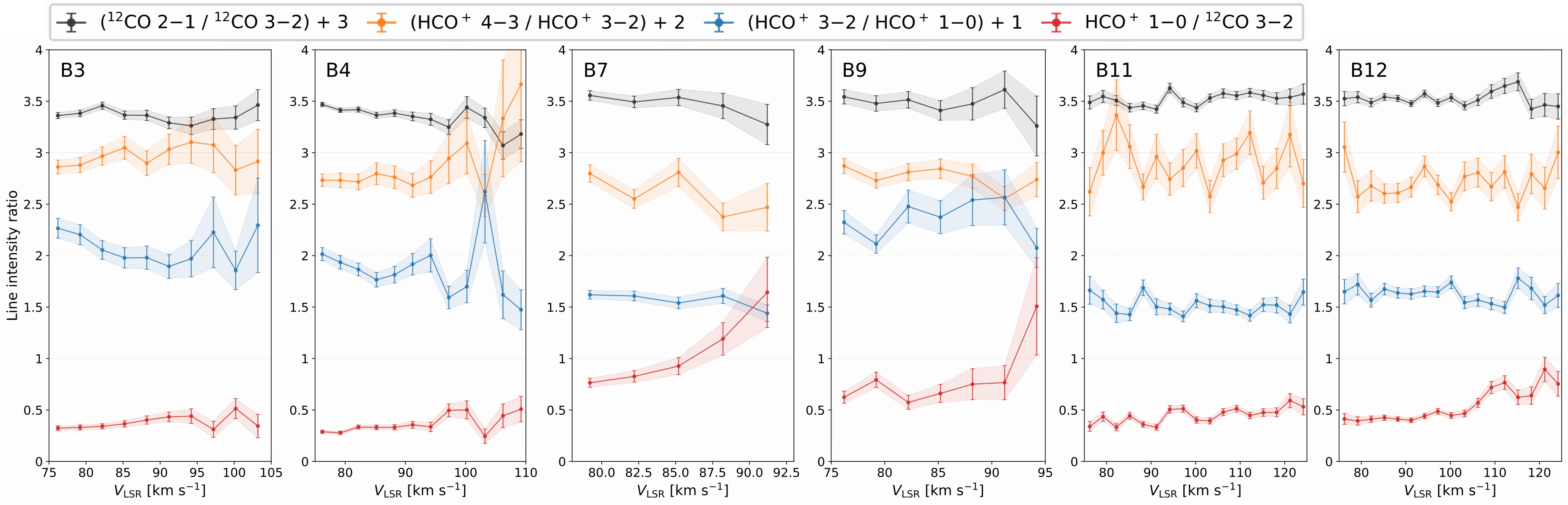}
\caption{Channel-by-channel line intensity ratios versus $V_{\rm LSR}$, after binning over five channels. 
The used spectra are shown in Fig.~\ref{fig:west_broad_spec_gt75}.
\label{fig:west_broad_line-ratio}}
\end{figure*}


\begin{figure*}[ht!]
\center
\includegraphics[width=0.49\textwidth]{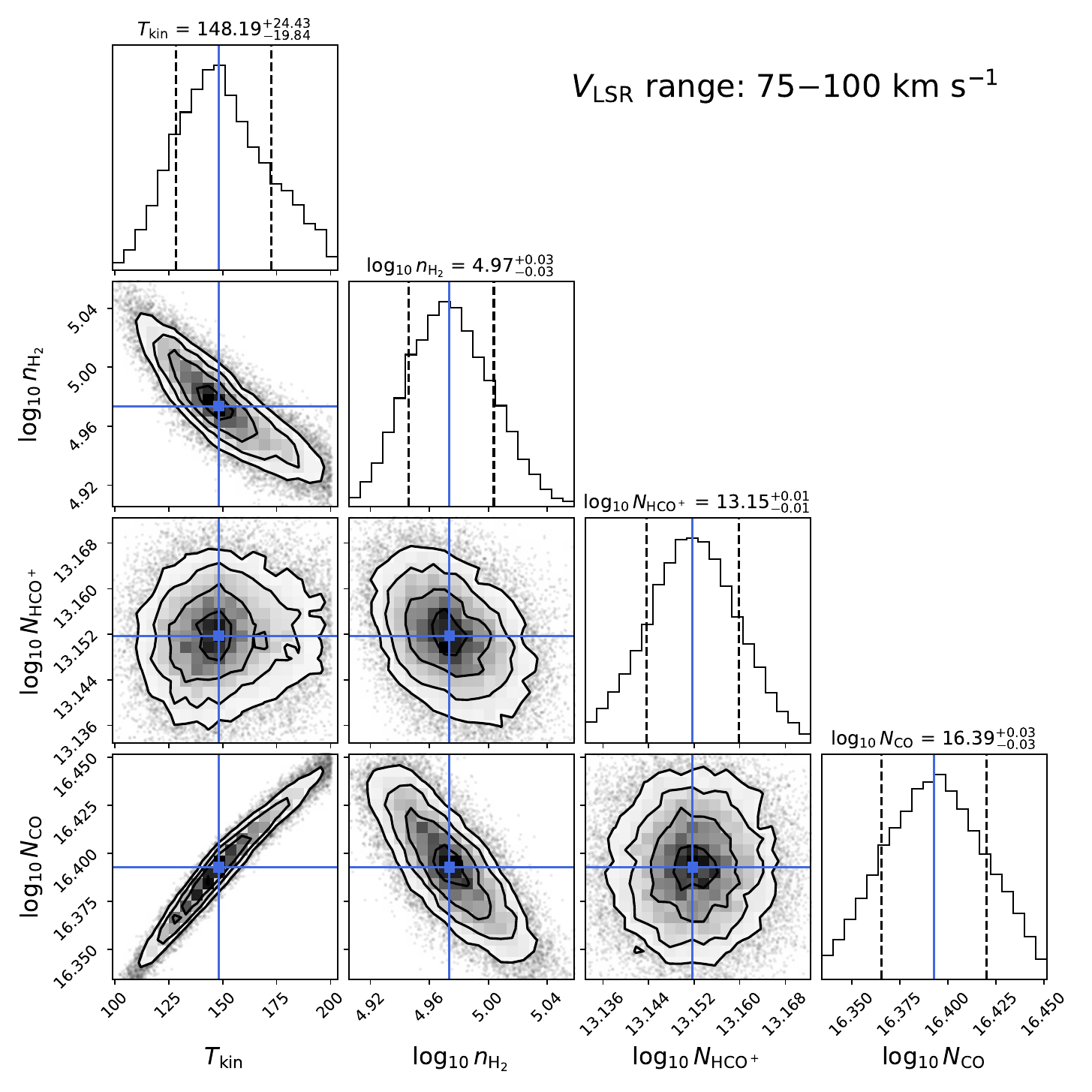}
\includegraphics[width=0.49\textwidth]{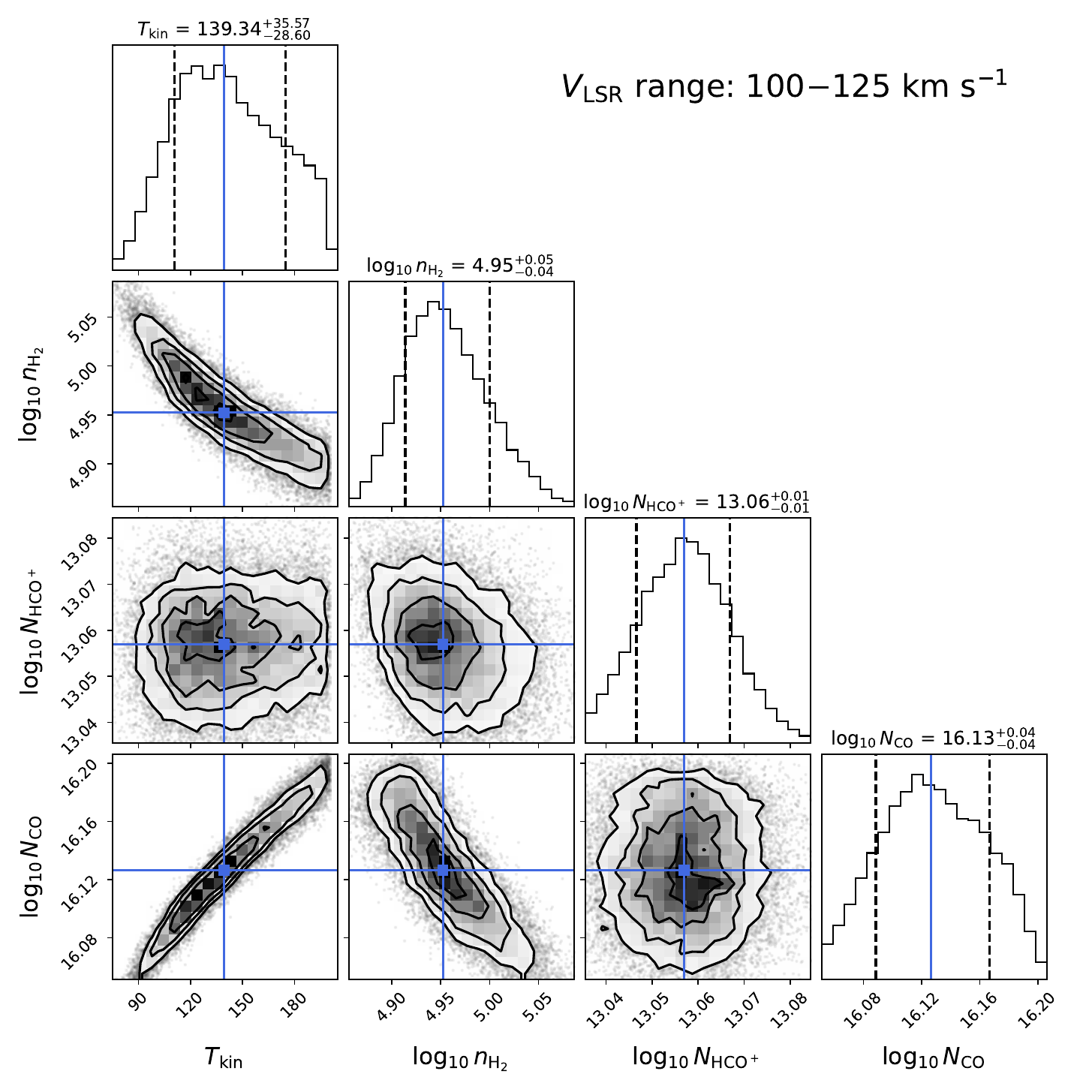}
\caption{Corner plot for region B12 over $V_{\rm LSR}=+75$ to +100~\kms (left) and +100 to +125~\kms (right). 
\label{fig:west_B12_RADEX+MCMC}}
\end{figure*}

\restartappendixnumbering
\begin{deluxetable*}{lllrrrccr}[htb]
\tablecaption{Parameters of the broad \hco~$J$=3--2 and 4--3 emission toward the 12 broad-line regions in W49B west.} \label{Tab:2hcop_line_para_west_broad}
\centering
\tablecolumns{9}
\centering
\tablehead{
\colhead{Region}  & \colhead{Coordinates} & \colhead{Transition} &  \multicolumn{3}{c}{Gaussian fit}  & \colhead{$V_{\rm range}$}  & \colhead{$\int T_{\rm b}$d$v$} & \colhead{$R_{\rm 4-3/3-2}$} \\
\vspace{-1.8em} \\
\cmidrule(lr){4-6} 
\vspace{-1.8em} \\
&   &  & \colhead{$V_{\rm LSR}$} & \colhead{$\Delta V$}  & \colhead{$T_{\rm b}$}  &  \\
\vspace{-1.8em} \\
& \colhead{(ICRS)} &  & \colhead{(\kms)} & \colhead{(\kms)}  & \colhead{(K)} & \colhead{(\kms)} & \colhead{(K~\kms)} }
\startdata
B1  & 19:11:00.78   & $J$=3--2     & 29.80$\pm$0.89 & 57.14$\pm$2.28 & 0.082$\pm$0.003 & [$-$22, 75] & 4.72   & 0.86\\
    & 09:06:15.8    & $J$=4--3     & 35.76$\pm$1.20 & 61.52$\pm$3.21 & 0.067$\pm$0.003 & [$-$22, 75] & 4.04   & ... \\
B2  & 19:11:03.28   & $J$=3--2     & 39.79$\pm$1.25 & 54.21$\pm$3.11 & 0.042$\pm$0.002 & [$-$5, 95]  & 2.31   & 1.00\\
    & 09:04:50.3    & $J$=4--3     & 43.08$\pm$1.43 & 53.42$\pm$3.52 & 0.041$\pm$0.002 & [$-$5, 95]  & 2.30   & ... \\
B3  & 19:11:00.88   & $J$=3--2     & 66.79$\pm$0.44 & 52.31$\pm$1.06 & 0.149$\pm$0.003 & [5, 120]    & 8.30   & 0.88\\
    & 09:06:05.1    & $J$=4--3     & 68.60$\pm$0.55 & 49.59$\pm$1.30 & 0.139$\pm$0.003 & [5, 120]    & 7.30   & ... \\
B4  & 19:11:01.12   & $J$=3--2     & 74.84$\pm$0.32 & 32.73$\pm$0.75 & 0.154$\pm$0.003 & [40, 120]   & 5.44   & 0.88\\
    & 09:05:55.4    & $J$=4--3$^*$ & 73.65$\pm$0.46 & 34.26$\pm$1.11 & 0.125$\pm$0.003 & [40, 120]   & 4.77   & ... \\
B5  & 19:11:00.09   & $J$=3--2     & 56.59$\pm$0.21 & 24.92$\pm$0.49 & 0.237$\pm$0.004 & [20, 100]   & 6.73   & 0.76\\
    & 09:05:07.4    & $J$=4--3     & 57.33$\pm$0.34 & 27.55$\pm$0.81 & 0.160$\pm$0.004 & [20, 100]   & 5.12   & ... \\
B6  & 19:10:59.92   & $J$=3--2     & 61.35$\pm$0.50 & 30.03$\pm$1.19 & 0.093$\pm$0.003 & [30, 100]   & 2.93   & 0.94\\
    & 09:04:49.4    & $J$=4--3     & 61.72$\pm$0.53 & 35.28$\pm$1.30 & 0.074$\pm$0.002 & [30, 100]   & 2.75   & ... \\
B7  & 19:10:59.42   & $J$=3--2$^*$ & 75.07$\pm$0.49 & 29.21$\pm$1.20 & 0.108$\pm$0.004 & [45, 102]   & 3.27  & 0.68\\
    & 09:05:13.6    & $J$=4--3     & 74.50$\pm$0.53 & 24.53$\pm$1.27 & 0.087$\pm$0.004 & [45, 102]   & 2.21   & ... \\
B8  & 19:10:59.85   & $J$=3--2     & 70.84$\pm$0.69 & 36.55$\pm$1.70 & 0.090$\pm$0.003 & [40, 110]   & 3.39   & 0.59\\
    & 09:06:12.7    & $J$=4--3     & 69.80$\pm$1.14 & 37.78$\pm$2.85 & 0.051$\pm$0.003 & [40, 110]   & 2.01   & ... \\
B9  & 19:10:59.89   & $J$=3--2     & 77.34$\pm$0.38 & 34.24$\pm$0.92 & 0.128$\pm$0.003 & [40, 110]   & 4.64   & 0.82\\
    & 09:05:54.9    & $J$=4--3     & 74.98$\pm$0.67 & 39.81$\pm$1.67 & 0.092$\pm$0.003 & [40, 110]   & 3.80   & ... \\
B10 & 19:11:02.01   & $J$=3--2     & 72.38$\pm$0.71 & 25.75$\pm$1.70 & 0.064$\pm$0.004 & [45, 102]   & 1.77   & 1.07\\
    & 09:05:54.3    & $J$=4--3     & 71.60$\pm$1.02 & 30.58$\pm$2.52 & 0.059$\pm$0.004 & [45, 102]   & 1.90   & ... \\
B11 & 19:11:02.65   & $J$=3--2     & 100.67$\pm$0.88 & 45.70$\pm$2.15 & 0.064$\pm$0.003 & [60, 150]  & 2.99   & 1.06\\
    & 09:05:49.2    & $J$=4--3     & 100.61$\pm$1.16 & 52.64$\pm$2.98 & 0.060$\pm$0.003 & [60, 150]  & 3.17   & ... \\
B12 & 19:11:02.83   & $J$=3--2$^*$ & 97.86$\pm$0.66  & 43.15$\pm$1.56 & 0.094$\pm$0.003 & [60, 152]  & 4.31   & 0.79\\
    & 09:05:40.2    & $J$=4--3$^*$ & 98.81$\pm$0.86  & 48.81$\pm$2.06 & 0.065$\pm$0.002 & [60, 152]  & 3.39   & ... \\
\enddata
\tablecomments{The asterisks in the transition column indicate that the Gaussian fit does not well reproduce the corresponding line profile. The last column, $R_{\rm 4-3/3-2}$, represents the integrated intensity ratio of \hco~$J$=4--3 to $J$=3--2.}
\end{deluxetable*}

\begin{deluxetable*}{lrccccccccr}[htb] 
\tablecaption{Integrated intensities and line ratios measured from the less-contaminated velocity intervals toward six broad-line regions in W49B west.} \label{Tab:line_ratio_west}
\centering
\tabletypesize{\small}
\tablecolumns{11}
\centering
\setlength{\tabcolsep}{1pt}
\tablehead{
\colhead{Region}  & \colhead{$V_{\rm range}$}  & \colhead{$I_{\rm HCO^+ (1-0)}$} & \colhead{$I_{\rm HCO^+ (3-2)}$}  & \colhead{$I_{\rm HCO^+ (4-3)}$}  & \colhead{$I_{\rm CO(3-2)}$} & \colhead{$I_{\rm CO(2-1)}$} & \colhead{$R_{\rm 3-2/1-0}$$^{(a)}$} & \colhead{$R_{\rm 4-3/3-2}$$^{(b)}$}  & \colhead{$R^{3-2}_{\rm HCO^+/CO}$$^{(c)}$} & \colhead{$R_{\rm CO}$$^{(d)}$}  \\  
& \colhead{(\kms)} & \colhead{(K~\kms)}  & \colhead{(K~\kms)} & \colhead{(K~\kms)} & \colhead{(K~\kms)} & \colhead{(K~\kms)} & }
\startdata
B3  & [75.0, 115.2] & 2.66$\pm$0.09 & 2.93$\pm$0.07 & 2.69$\pm$0.10 &  7.21$\pm$0.12 & 2.61$\pm$0.12            & 1.10$\pm$0.05 & 0.92$\pm$0.04 & 0.41$\pm$0.01 & 2.76$\pm$0.14\\
B4  & [75.0, 121.2] & 3.56$\pm$0.09 & 2.98$\pm$0.08 & 2.50$\pm$0.11 & 10.42$\pm$0.13 & 3.85$\pm$0.13            & 0.84$\pm$0.03 & 0.84$\pm$0.04 & 0.29$\pm$0.01 & 2.71$\pm$0.10\\
B7  & [78.0, 101.4] & 2.52$\pm$0.07 & 1.31$\pm$0.06 & 0.84$\pm$0.07 &  2.53$\pm$0.09 & 1.23$\pm$0.09            & 0.52$\pm$0.03 & 0.64$\pm$0.06 & 0.52$\pm$0.03 & 2.06$\pm$0.17\\
B9  & [75.0, 110.4] & 2.13$\pm$0.08 & 2.78$\pm$0.07 & 2.12$\pm$0.09 &  2.74$\pm$0.11 & 1.28$\pm$0.11            & 1.31$\pm$0.06 & 0.76$\pm$0.04 & 1.01$\pm$0.05 & 2.14$\pm$0.20\\
B11 & [75.0, 147.0] & 6.32$\pm$0.12 & 3.11$\pm$0.10 & 2.93$\pm$0.13 & 13.78$\pm$0.16 & 7.25$\pm$0.16$^\dagger$  & 0.49$\pm$0.02 & 0.94$\pm$0.05 & 0.23$\pm$0.01 & 1.90$\pm$0.05\\
B12 & [75.0, 151.2] & 7.20$\pm$0.12 & 4.50$\pm$0.10 & 3.25$\pm$0.14 & 13.33$\pm$0.16 & 7.09$\pm$0.16$^\dagger$  & 0.62$\pm$0.02 & 0.72$\pm$0.04 & 0.34$\pm$0.01 & 1.88$\pm$0.05\\
\enddata
\tablecomments{$^{(a)}$ $R_{\rm 3-2/1-0}$ refers to the integrated intensity ratio of \hco~$J$=3--2 to $J$=1--0, measured over the same velocity range for each region. \\
$^{(b)}$ $R_{\rm 4-3/3-2}$ refers to the integrated intensity ratio of \hco~$J$=4--3 to $J$=3--2. \\
$^{(c)}$ $R^{3-2}_{\rm HCO^+/CO}$ refers to the integrated intensity ratio of \hco~$J$=3--2 to \co~$J$=3--2.\\
$^{(d)}$ $R_{\rm CO}$ refers to the integrated intensity ratio of \co~$J$=3--2 to \co~$J$=2--1.\\
$\dagger$ denotes that the observations did not cover possible \co~$J$=2--1 emission above 135~\kms, the intensities of each channel are estimated using a noise level of 24~mK in the box region. Thus the $R_{\rm CO}$ values of B11 and B12 regions is an upper limit.
}
\end{deluxetable*}

\begin{deluxetable*}{llcccc}[htb]
\tablecaption{Physical parameters derived from RADEX toward six broad-line regions in W49B west.} \label{Tab:physical_para_west_broad}
\centering
\renewcommand{\arraystretch}{1.12}
\tablecolumns{6}
\centering
\tablehead{
\colhead{Region}    & \colhead{$T_{\rm kin}$} & \colhead{$n_{\rm H_2}$} & \colhead{$N_{\rm HCO^+}$} & \colhead{$N_{\rm CO}$} & \colhead{$N_{\rm HCO^+}$/$N_{\rm CO}$}  \\  
\vspace{-1.8em} \\
&  \colhead{(K)}  &  \colhead{($\times 10^{5}$ cm$^{-3}$)}  & \colhead{($\times 10^{13}$ cm$^{-2}$)} & \colhead{($\times 10^{16}$ cm$^{-2}$)} & \colhead{($\times 10^{-4}$)}}
\startdata
B3  & 100 (fixed)             & 1.78$^{+0.08}_{-0.08}$  & 2.51$^{+0.12}_{-0.11}$ & 4.37$^{+0.10}_{-0.10}$ & 5.8$^{+0.3}_{-0.3}$ \\ 
    & 200 (fixed)             & 1.32$^{+0.06}_{-0.06}$  & 2.51$^{+0.12}_{-0.11}$ & 5.89$^{+0.14}_{-0.13}$ & 4.3$^{+0.2}_{-0.2}$ \\ 
    & 192.9$^{+5.2}_{-10.7}$  & 1.35$^{+0.06}_{-0.06}$  & 2.51$^{+0.12}_{-0.11}$ & 5.75$^{+0.27}_{-0.26}$ & 4.4$^{+0.3}_{-0.3}$ \\ 
B4  & 100 (fixed)             & 1.23$^{+0.06}_{-0.06}$  & 3.31$^{+0.08}_{-0.08}$ & 6.03$^{+0.28}_{-0.27}$ & 5.5$^{+0.3}_{-0.3}$ \\ 
    & 200 (fixed)             & 0.91$^{+0.04}_{-0.04}$  & 3.31$^{+0.16}_{-0.08}$ & 7.94$^{+0.37}_{-0.36}$ & 4.2$^{+0.3}_{-0.2}$ \\ 
    & 194.8$^{+3.8}_{-7.8}$   & 0.93$^{+0.04}_{-0.04}$  & 3.31$^{+0.16}_{-0.08}$ & 7.94$^{+0.37}_{-0.36}$ & 4.2$^{+0.3}_{-0.2}$ \\ 
B7  & 100 (fixed)             & 0.69$^{+0.05}_{-0.05}$  & 3.72$^{+0.27}_{-0.25}$ & 2.09$^{+0.15}_{-0.14}$ & 17.8$^{+1.8}_{-1.7}$\\ 
    & 200 (fixed)             & 0.52$^{+0.04}_{-0.05}$  & 3.80$^{+0.27}_{-0.25}$ & 2.63$^{+0.19}_{-0.18}$ & 14.5$^{+1.5}_{-1.3}$\\ 
    & 170.7$^{+21.1}_{-31.7}$ & 0.55$^{+0.07}_{-0.05}$  & 3.72$^{+0.36}_{-0.25}$ & 2.45$^{+0.24}_{-0.22}$ & 15.1$^{+2.1}_{-1.6}$\\ 
B9  & 100 (fixed)             & 2.14$^{+0.10}_{-0.10}$  & 1.95$^{+0.09}_{-0.09}$ & 1.55$^{+0.07}_{-0.07}$ & 12.6$^{+0.8}_{-0.8}$\\
    & 200 (fixed)             & 1.58$^{+0.07}_{-0.07}$  & 1.95$^{+0.09}_{-0.09}$ & 2.14$^{+0.15}_{-0.10}$ & 9.1$^{+0.6}_{-0.7}$ \\ 
    & 184.2$^{+11.6}_{-22.2}$ & 1.62$^{+0.12}_{-0.07}$  & 1.95$^{+0.09}_{-0.09}$ & 2.04$^{+0.15}_{-0.14}$ & 9.5$^{+0.8}_{-0.8}$ \\ 
B11$^*$ & 100 (fixed)         & 0.76$^{+0.04}_{-0.03}$  & 2.14$^{+0.15}_{-0.14}$ & 2.82$^{+0.20}_{-0.19}$ & 7.6$^{+0.8}_{-0.7}$ \\ 
    & 200 (fixed)             & 0.56$^{+0.03}_{-0.03}$  & 2.19$^{+0.16}_{-0.15}$ & 3.55$^{+0.25}_{-0.24}$ & 6.2$^{+0.6}_{-0.6}$ \\ 
    & 140.2$^{+22.4}_{-18.7}$ & 0.63$^{+0.05}_{-0.06}$  & 2.19$^{+0.16}_{-0.15}$ & 3.16$^{+0.23}_{-0.21}$ & 6.9$^{+0.7}_{-0.6}$ \\ 
B12$^*$ & 100 (fixed)         & 0.91$^{+0.04}_{-0.04}$  & 2.63$^{+0.12}_{-0.12}$ & 2.95$^{+0.14}_{-0.13}$ & 8.9$^{+0.6}_{-0.6}$ \\ 
    & 200 (fixed)             & 0.69$^{+0.03}_{-0.03}$  & 2.63$^{+0.12}_{-0.12}$ & 3.80$^{+0.18}_{-0.17}$ & 6.9$^{+0.5}_{-0.4}$ \\ 
    & 128.1$^{+21.0}_{-17.4}$ & 0.81$^{+0.06}_{-0.05}$  & 2.63$^{+0.12}_{-0.12}$ & 3.24$^{+0.23}_{-0.22}$ & 8.1$^{+0.7}_{-0.6}$ \\ 
\enddata
\tablecomments{$^*$ In RADEX calculations, the gas kinetic temperature ($T_{\rm kin}$) reaches convergence only in the B11 and B12 regions.
}
\end{deluxetable*}

\clearpage
\onecolumngrid
\section{Narrow-line emission in W49B east}\label{sec:appendix-A}

\begin{figure*}[htbp]
\includegraphics[width=1\textwidth]{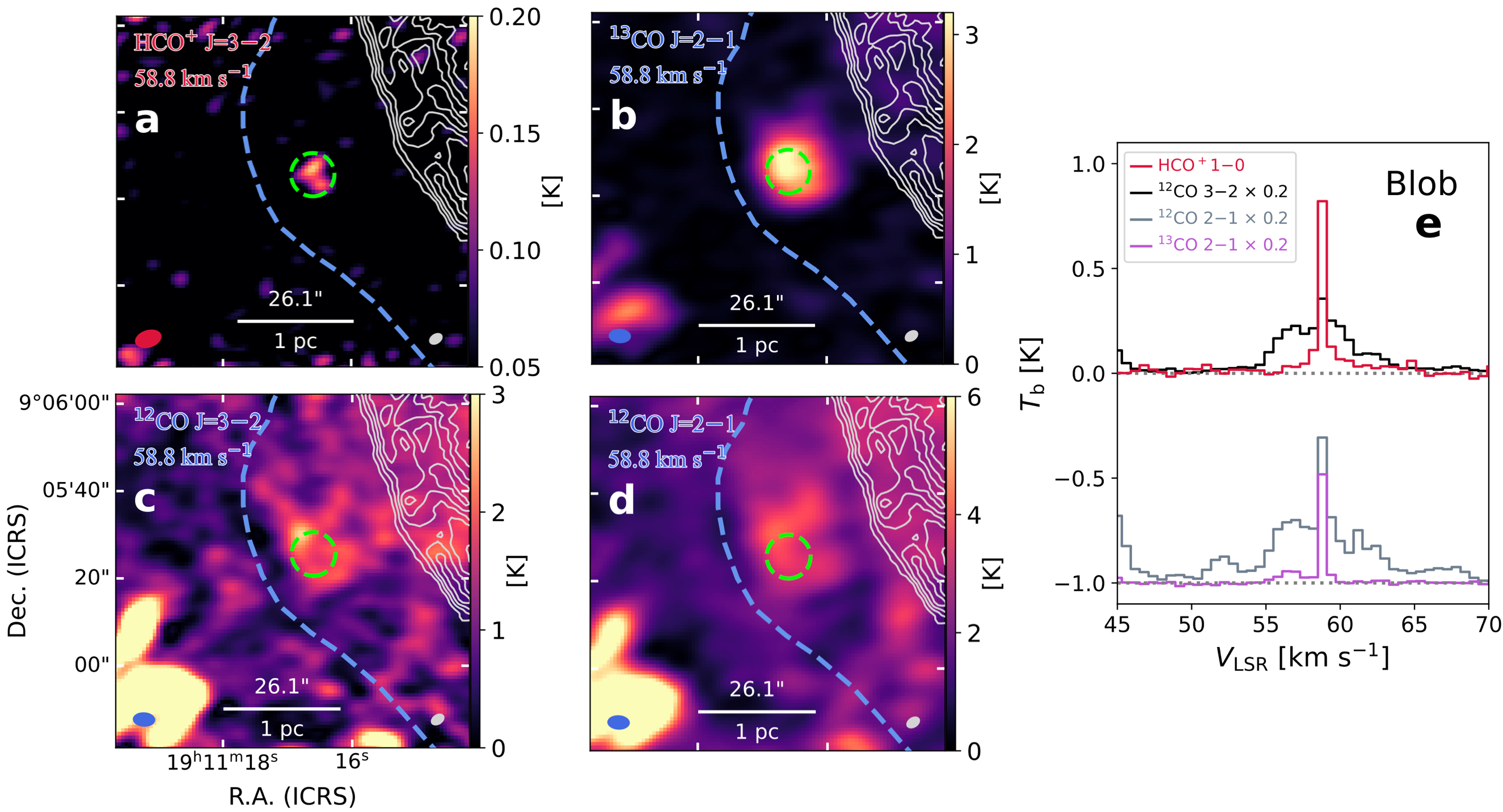}
\caption{Molecular emission counterparts for the bright \hco~$J$=1--0 blob-like component in W49B east.
Panels (a)--(d) show the channel emission at +58.8~\kms\ of (a) \hco~$J$=3--2 (ACA data only), (b) \ico~$J$=2--1, (c) \co~$J$=3--2, and (d) \co~$J$=2--1, respectively.
The green dashed circle, blue dashed curves, synthesized beams, 3~mm continuum contour levels and scale bar are identical to those in Fig.~\ref{fig:east_blob}.
Panel (e) shows the averaged spectra of \hco~$J$=1--0, \co~$J$=3--2, \co~$J$=2--1, and \ico~$J$=2--1 toward the blob-like component. All spectra are convolved to a circular beam of 8$\rlap{.}\arcsec$5.
\label{fig:east_blob_appendix}}
\end{figure*}
\begin{figure}[htbp]
\centering
\includegraphics[width=1\textwidth]{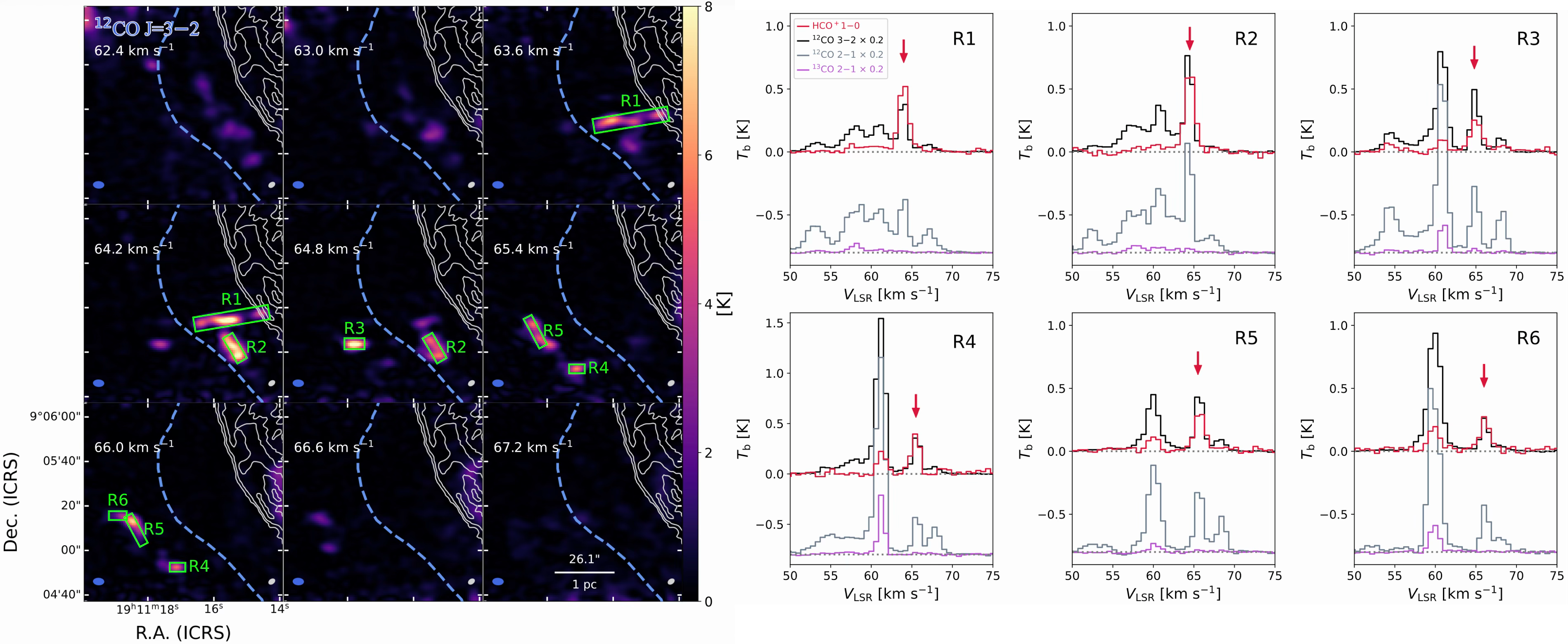}
\caption{Molecular emission from +62.4 to +67.2~\kms\ in W49B east. The left panel shows channel maps of \co~$J$=3--2 emission from +62.4 to +67.2~\kms\ in steps of 0.6~\kms, with green boxes marking six regions of interest with bright \hco~$J$=1--0 emission. The six panels on the right show the averaged spectra of \hco~$J$=1--0, \co~$J$=3--2, \co~$J$=2--1, and \ico~$J$=2--1 toward these regions. All spectra are convolved to a circular beam of 8$\rlap{.}\arcsec$5.
\label{fig:east_channel_map_co}}
\end{figure}

This section presents the images and spectra of the blob-like structure and the six clumpy regions  R1--R6 in W49B east, as shown in Figs.~\ref{fig:east_blob_appendix} and \ref{fig:east_channel_map_co}, respectively.
The weak or absent higher-$J$ \hco\ emission in all these components suggests low-excitation conditions. 

Toward the blob-like structure, the \co~$J$=2--1 emission appears weaker than the \ico~$J$=2--1 emission at +58.8~\kms\,(Fig.~\ref{fig:east_blob_appendix}e), after accounting for contamination from other components estimated from the average emission in the adjacent velocity channels. This suggests severe optical-depth effects, and self-absorption in the \co\ emission. 
As a result, the \co\ intensity may not reliably trace the CO column density, and using the \hco/\co\ intensity ratio as a proxy for the abundance ratio may lead to an overestimate.

In regions R1--R6, the good spatial and velocity correspondence between the \hco~$J$=1--0 and \co\ emission suggests that they likely originate from the same layer of molecular gas. Given their narrow linewidths, these components are more likely associated with low-excitation, relatively quiescent molecular emission. The absence of \ico~$J$=2--1 and higher-$J$ \hco\ emission also suggests that its column density, temperature, and/or density are limited. 
We also attempted RADEX calculations and MCMC modeling, similar to that in Sect.~\ref{Sec:dis_phy_conditions}, by assuming that the \co\ and \hco~$J$=1--0 emission arise from the same gas component. However, the available transitions are insufficient to robustly constrain $T_{\rm kin}$, $n({\rm H_2})$, $N({\rm HCO^+})$, and $N({\rm CO})$. The solutions either fail to converge or yield large uncertainties, so we do not derive reliable \hco/\co\ abundance ratios for these components.
Table~\ref{Tab:line_para_narrow} lists the line parameters of these narrow-line components.

\restartappendixnumbering

%

\begin{deluxetable*}{lclccr}[b]
\tablecaption{Parameters of the narrow emission in the selected regions of W49B east.} \label{Tab:line_para_narrow}
\centering
\tablecolumns{6}
\centering
\tablehead{
\colhead{Region}  & \colhead{Coordinates} & \colhead{$V_{\rm range}$}  & \colhead{$I_{\rm HCO^+ (1-0)}$} & \colhead{$I_{\rm ^{12}CO(2-1)}$} & \colhead{$I_{\rm ^{12}CO(3-2)}$} \\
\vspace{-1.8em} \\
 & \colhead{(ICRS)} &  \colhead{(\kms)} & \colhead{(K~\kms)} & \colhead{(K~\kms)} & \colhead{(K~\kms)} }
\startdata
R1 & 19:11:15.46, 9:05:15.2 & [63.6, 64.2] & 0.58$\pm$0.013 & 2.28$\pm$0.020 & 2.13$\pm$0.014 \\
R2 & 19:11:15.36, 9:05:01.8 & [64.2, 64.8] & 0.71$\pm$0.013 & 4.46$\pm$0.020 & 3.90$\pm$0.014 \\
R3 & 19:11:17.79, 9:05:03.8 & [64.8, 65.4] & 0.28$\pm$0.013 & 2.54$\pm$0.020 & 2.33$\pm$0.014 \\
R4 & 19:11:17.10, 9:04:52.5 & [65.4, 66.0] & 0.40$\pm$0.013 & 2.00$\pm$0.020 & 1.89$\pm$0.014 \\
R5 & 19:11:18.36, 9:05:09.2 & [65.4, 66.0] & 0.34$\pm$0.013 & 2.69$\pm$0.020 & 2.44$\pm$0.014 \\
R6 & 19:11:18.91, 9:05:15.7 & [66,0, 66.6] & 0.27$\pm$0.013 & 1.83$\pm$0.020 & 1.24$\pm$0.014 \\
\enddata
\end{deluxetable*}

\clearpage
\onecolumngrid
\section{Narrow-line emission in W49B west}\label{sec:appendix-C}

\begin{figure}[htb]
\includegraphics[width=0.5\textwidth]{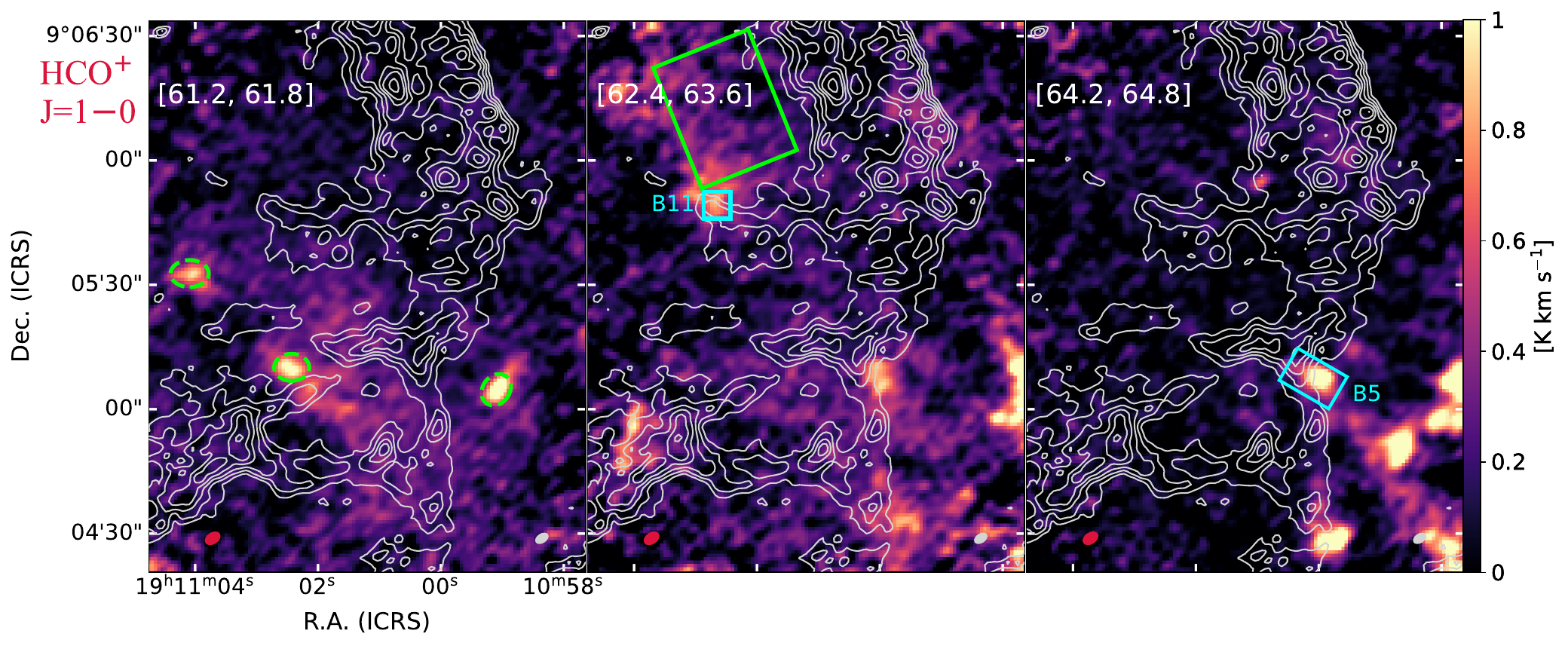}
\includegraphics[width=0.5\textwidth]{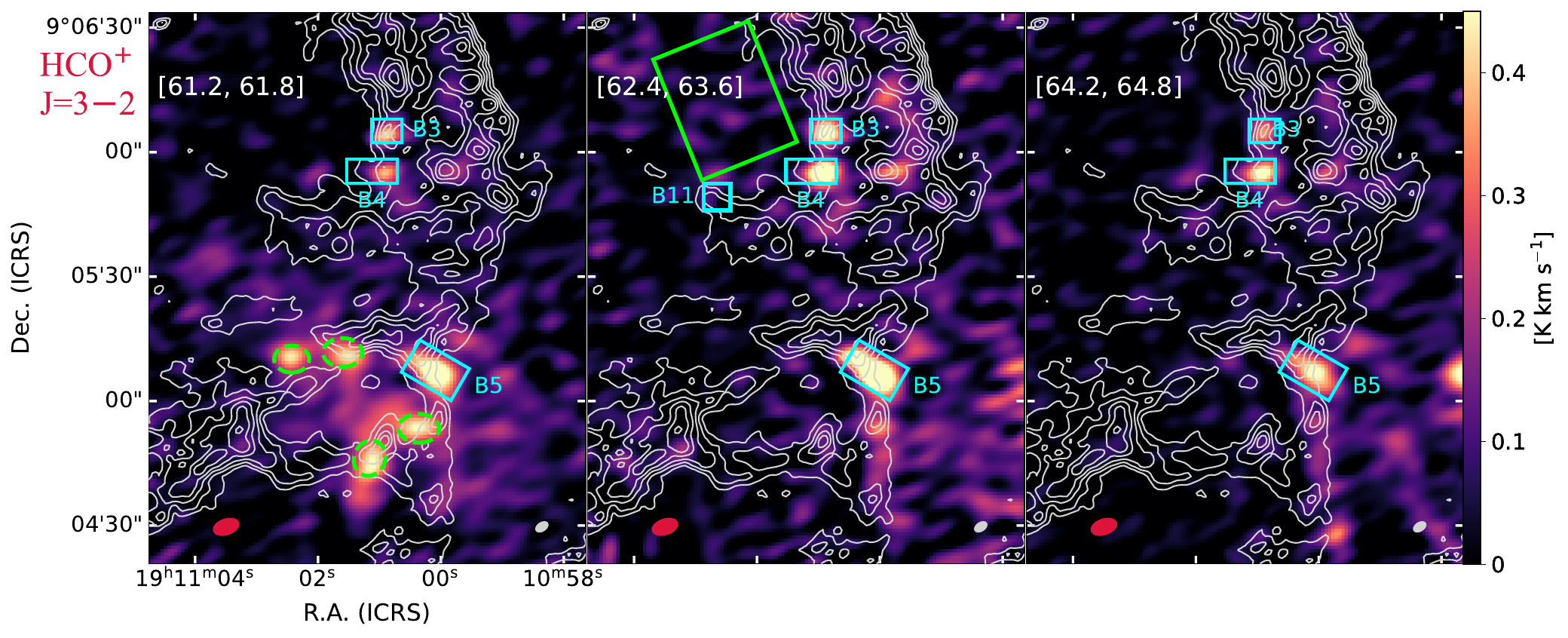}
\includegraphics[width=0.5\textwidth]{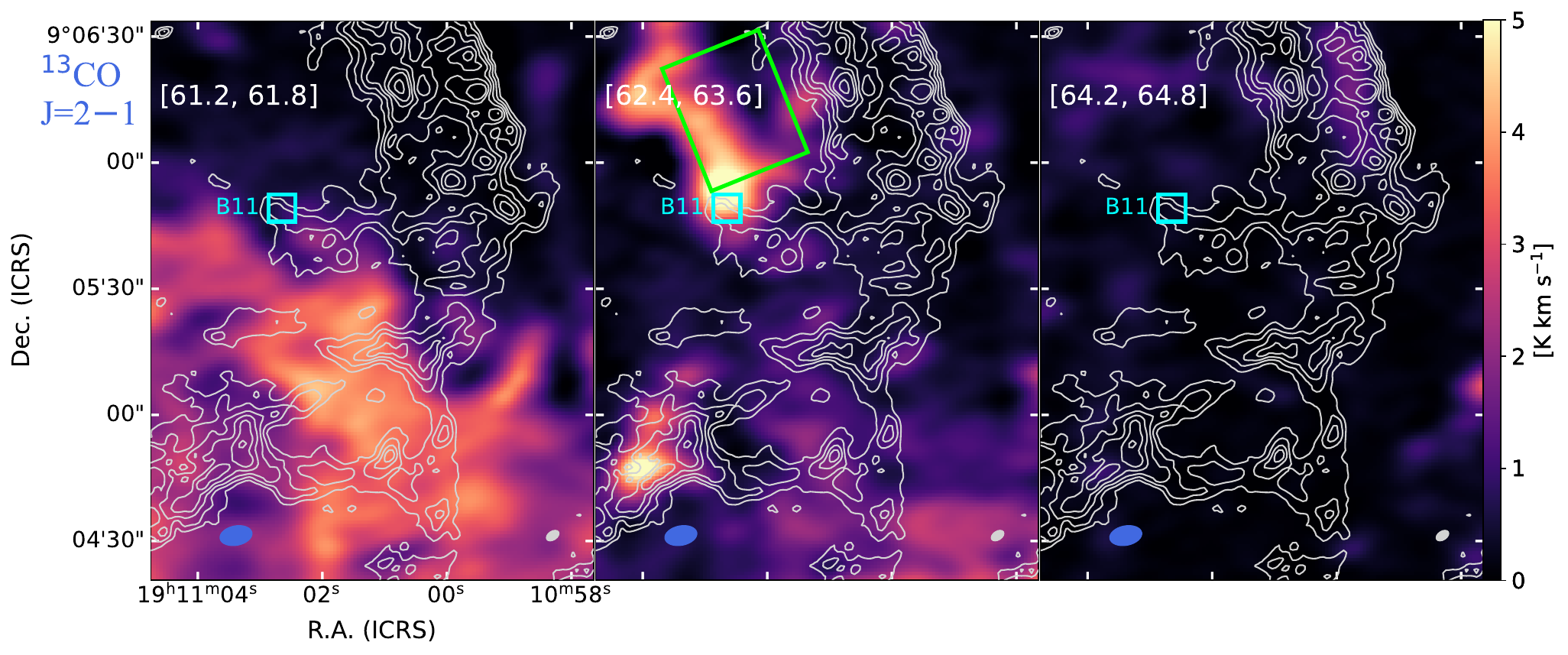}
\includegraphics[width=0.5\textwidth]{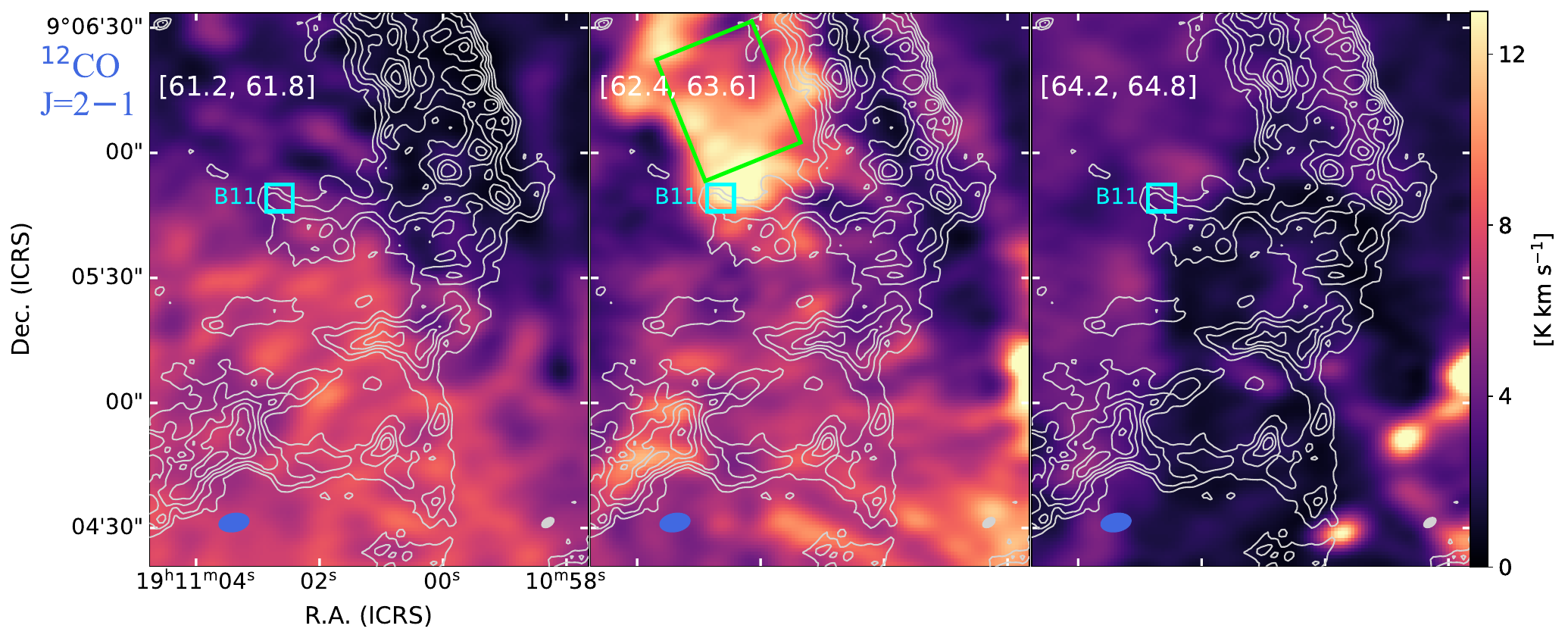}
\includegraphics[width=0.5\textwidth]{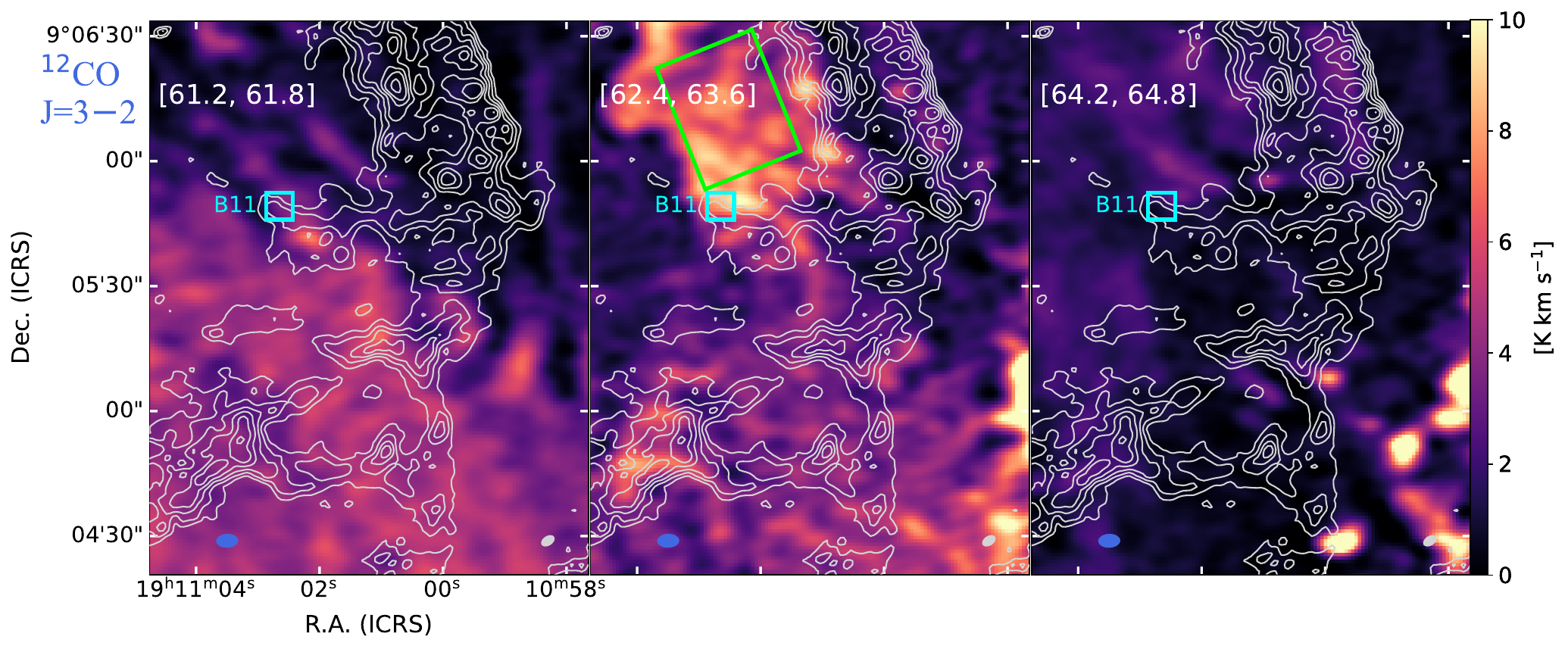}
\caption{Integrated intensity maps of \hco~$J$=1--0, \hco~$J$=3--2, \ico~$J$=2--1, \co~$J$=2--1 and \co~$J$=3--2 over different velocity intervals in W49B west. The cyan box marks the broad-line region B11, while the green box indicates a possible pre-shock gas region.
The synthesized beams, and 3~mm continuum contour levels are identical to those shown in Fig.~\ref{fig:west_inte_map_all}. 
\label{fig:west_narrow_image}}
\end{figure}

This section presents the image (Fig.~\ref{fig:west_narrow_image}) of the narrow-line emission detected in W49B west.
Both compact and faint extended components are shown in \hco~$J$=1--0 and 3--2 maps from +61.2 to +61.8~\kms.
The \hco~$J$=1--0 line is generally brighter than $J$=3--2 line, and the bright clumpy emission of the two transitions is not fully co-spatial.
In these regions, \co\ and \ico\, emission exhibits a more extended distribution.
We note that owing to projection effects, it is difficult to determine whether these clumpy \hco\ emission components are associated with the SNR.

In region B11, the \hco~$J$=1--0 line exhibits a narrow component at about +63~\kms\ and a broad component spanning about +70 to +150~\kms\ (Fig.~\ref{fig:west_broad_spec_example}). The \hco\ $J$=1--0 map from +62.4 to +63.6~\kms\ reveals faint and diffusive emission (see green boxed region in Fig.~\ref{fig:west_narrow_image}) in and around B11. This component has brighter counterparts in \co\ and \ico, but no \hco~$J$=3--2 emission is detected.
Compared to the \hco~$J$=1--0 transition, the $J$=3--2 line requires higher critical densities and energies \citep{2015PASP..127..299S}. Therefore, the absence of \hco~$J$=3--2 emission generally indicates low excitation conditions, such as low gas density and/or kinetic temperature.
Possibly due to projection effects, it appears to lie within the SNR, but is bounded by continuum emission on the western and southern sides.
All these observations suggest that this emission likely traces pre-shock gas.

Bright \hco~$J$=1--0 emission above +64.2~\kms\ mainly arises from the southwestern side of the SNR, as well as its vicinity and regions outside the remnant. As shown in Fig.~4 in \cite{2022ApJ...931..144Z}, a large MC is shown in the southwestern part of SNR, where the \hco~$J$=1--0/\co~$J$=1--0 intensity ratio is relatively low, suggesting that this emission may not be associated with the SNR.

\restartappendixnumbering

\bibliography{ALMA_W49B}{}
\bibliographystyle{aasjournalv7}



\end{document}